\documentclass{aa}  

\usepackage[utf8]{inputenc}   
\usepackage[T1]{fontenc}      

\usepackage{graphicx}
\usepackage{newtxtext,newtxmath}
\usepackage{xcolor}
\usepackage[hidelinks,colorlinks=true,linkcolor=blue,citec
olor=blue]{hyperref}
\usepackage{graphicx}	
\usepackage{amsmath}	
\usepackage{amssymb}	
\usepackage{textgreek}
\usepackage{ulem}

\newcommand{\hi}{\ion{H}{i}}

\titlerunning{H{\sc i}-to-Halo mass relation}
\authorrunning{M. Korsaga et al.}

\begin{document}


\title {The galactic H{\sc i}-to-halo mass relation from isolated galaxies to cosmological hydrodynamic simulations}

\author{
M. Korsaga\inst{1,2}%
\and
J. Freundlich\inst{3}
\and
B. Famaey\inst{3}
\and
L. Verdes-Montenegro\inst{1}
\and
A. Sorgho\inst{1}
\and
F. Baraggioni\inst{4,5}
\and
A. V. Macci\`o\inst{4,5}
\and
K. Kraljic\inst{3}
\and
L. Chemin\inst{3}
\and
P. Kamphuis\inst{1}
\and
P. Amram\inst{6}
\and
B. Epinat\inst{6}
\and
R. Ianjamasimanana\inst{1}
\and
M. Bureau\inst{7}
\and
P. Dominiak\inst{7}
\and
B. Namumba\inst{1}
\and
J. Garrido\inst{1}
\and
S. Sanchez-Exp\'osito\inst{1}
}

\institute{
Instituto de Astrof\'isica de Andaluc\'ia (CSIC), PO Box 3004, 18080 Granada, Spain\\
\email{mkorsaga@iaa.es}
\and
Laboratoire de Physique et de Chimie de l'Environnement, Universit\'e Joseph Ki-Zerbo, 03 BP 7021, Ouaga 03, Burkina Faso
\and
Universit\'e de Strasbourg, CNRS, Observatoire astronomique de Strasbourg, UMR 7550, F-67000 Strasbourg, France
\and
Center for Astrophysics and Space Science (CASS), New York University Abu Dhabi, PO Box 129188, Abu Dhabi, UAE
\and
New York University Abu Dhabi, PO Box 129188, Abu Dhabi, UAE
\and
Aix-Marseille Université, CNRS, CNES, LAM, Marseille, France
\and 
Sub-department of Astrophysics, Department of Physics, University of Oxford, Denys Wilkinson Building, Keble Road, Oxford, OX1 3RH, UK
}

\abstract
{The relation between H{\sc i} and dark matter halo masses provides key insights into gas accretion and the regulation of galaxy formation. In a previous study based on the {\it Spitzer} Photometry and Accurate Rotation Curves (SPARC) and the `Local Irregulars That Trace Luminosity Extremes, The H{\sc i} Nearby Galaxy Survey' (LITTLE THINGS) samples, we showed that the ratio between H{\sc i} mass and dark matter halo mass remains approximately constant with stellar mass for nearby disc galaxies ($\rm log(M_{\hi}/M_{200}) = -1.90$ with a 1$\sigma$ scatter of 0.37 dex). In this work, we extend that analysis by incorporating galaxies from the Analysis of the interstellar Medium in Isolated GAlaxies (AMIGA) sample and from the Gassendi HAlpha survey of SPirals (GHASP). The AMIGA sample provides the most rigorously selected sample of isolated galaxies in the local Universe, well-suited for testing whether the relation found in the previous samples holds in interaction-free galaxies. We construct individual mass models from high-resolution rotation curves and infrared photometry, and derive dark matter halo parameters. We confirm the proportionality between H{\sc i} and dark halo masses and find a nearly constant H{\sc i}-mass-to-halo-mass ratio of $\rm log(M_{\hi}/M_{200}) = -1.90$ over nearly five orders of magnitude in stellar mass ($\rm 7 \leq log (M_{\star}/M_\odot) \leq 11.5$), with a 1$\sigma$ scatter of 0.36 dex. The AMIGA and GHASP samples are statistically consistent with the relation previously found for the SPARC and LITTLES THINGS samples, indicating that it is robust across galaxies spanning a broad range of isolation levels. In contrast, cosmological hydrodynamic simulations such as SIMBA, IllustrisTNG, and NIHAO predict a dependence on stellar mass, with a break at the high stellar mass end for disc galaxies ($\rm M_{\star}> 10^{10}~M_{\sun}$). Our results thus further demonstrate that the H{\sc i}-mass-to-halo-mass ratio is remarkably self-similar across rotationally-supported disc galaxies, hinting at mass-independent self-regulation mechanisms that are not yet understood in current theoretical models.}

\keywords{galaxies: isolated galaxies -- dynamics -- kinematics -- spiral and irregular -- dark matter -- simulations: NIHAO -- IllustrisTNG and SIMBA}

\maketitle


\section{Introduction}

Over recent decades, detailed studies of galactic dynamics and of the distribution of baryons within galaxies have led to the establishment of scaling relations connecting a galaxy's light, emitted by stars and gas, to its total gravitational potential, including dark matter (DM). These relations, such as the baryonic Tully Fisher relation \citep{McGaugh2000, McGaugh2005, McGaugh2012,  Lelli2016b, Lelli2019} and the Radial Acceleration Relation \citep{McGaugh2016, Lelli2017}, offer insights into galaxy formation processes and even possibly the nature of DM. 
However, the relation between the total mass of the neutral atomic hydrogen (H{\sc i}) and  the total DM mass remains less explored than the aforementioned relations and not fully understood.
Several studies have relied on indirect and model-dependent techniques, such as abundance matching (\citealp{Yang2007, popping2015, Rodriguez-Puebla2015, Chauhan2021, Saraf2024}), galaxy clustering (\citealp{Guo2017, Obuljen2019, Calette2021}), or H{\sc i} spectral stacking (\citealp{Guo2020}), to estimate DM halo masses. While these approaches are powerful for large statistical samples, they can introduce systematic uncertainties, potentially affecting the inferred shape of the scaling relation. A more robust characterisation of the relation between H{\sc i} and DM halo masses requires detailed modelling of individual galaxies, providing a more direct determination of halo properties (e.g., \citealp{Posti2019, Korsaga2023, ManceraPina2025, Ponomareva2026}). 

In this context, \citet{Korsaga2023} analysed the kinematics of individual low-redshift galaxies to investigate the H{\sc i}–to–halo mass relation of disc galaxies. They studied the {\it Spitzer} Photometry and Accurate Rotation Curves (SPARC) sample of 150 disc galaxies from \cite{Lelli2016}, which provides high quality 3.6~$\mu$m photometry and extended H{\sc i} rotation curves, together with the `Local Irregulars That Trace Luminosity Extremes, The H{\sc i} Nearby Galaxy Survey' (LITTLE THINGS) sample of 15 dwarf irregular galaxies \citep{Hunter2012, Oh2015, Iorio2017, Read2017}, and found that rotationally supported disc galaxies across the full mass spectrum, $\rm 7 \leq log (M_{\star}/M_\odot) \leq 11.5$, exhibit an approximately constant H{\sc i}-to-halo mass fraction of $\sim 1.25 \% $ (i.e., $\rm log(M_{\hi}/M_{200}) \ = -1.90$) with a $1\sigma$ scatter of 0.37 dex. This suggests a remarkable self-similarity of disc galaxies and points to mass-independent regulation mechanisms whose physical origins remain unclear, given that numerical simulations such as IllustrisTNG \citep{Nelson2018,Nelson2019} and SIMBA \citep{Dave2019} did not predict it. However, this result may be in tension with other observational studies, such as \citet{Saraf2024}, \citet{Dev2024}, or \citet{Guo2020}, which predict a break in the relation at both the low- and high-mass ends. But these studies rely on indirect and model-dependent methods to estimate DM halo masses, and do not focus solely on disc galaxies, which likely explains their disagreement with \citet{Korsaga2023}, where the halo masses were derived more directly from mass modelling of individual disc galaxies. The present article attempts to gain further insights into these differences.

In the current article, we extend the analysis of \citet{Korsaga2023} by investigating the H{\sc i}-to-halo mass relation in a carefully selected sample of isolated disc galaxies from the Analysis of the interstellar Medium in Isolated GAlaxies (AMIGA) sample \citep{Karachentseva1973,Verdes-Montenegro2005,Sorgho2024}, complemented by galaxies from the Gassendi HAlpha survey of SPirals (GHASP) survey \citep{Epinat2008,Epinat2008b, Korsaga2019b}. 
Indeed, while the SPARC and LITTLE THINGS samples used in \citet{Korsaga2023} exclude strongly interacting systems, their selection did not involve robust criteria to ensure strict isolation. 
In contrast, the AMIGA sample \citep{Verdes-Montenegro2005} is the most carefully constructed sample of isolated galaxies to date, with strict isolation criteria based on local galaxy density and tidal strength \citep{Verley2007, Argudo-Fernandez2013}. Consequently, AMIGA serves as an ideal reference sample  for investigating the H{\sc i}-to-halo mass relation in interaction-free galaxies in the local Universe, where secular evolution has been minimally affected by external processes.
By combining high-resolution H{\sc i} kinematics and/or extended H$\alpha$ and H{\sc i} data with infrared WISE photometry, we construct detailed mass models and derive the DM halo properties. The aim is to examine whether there is any dependence of the H{\sc i}-to-halo mass scaling relation on the environment. Moreover, while the numerical simulations of galaxy formation investigated in \citet{Korsaga2023} provided statistically representative disc galaxy populations in large volumes, their resolutions were not as high as those of zoom-in simulations. Therefore, we also complement the comparison to the predictions  from IllustrisTNG \citep{Nelson2018,Nelson2019} and SIMBA \citep{Dave2019} with an additional comparison to the predictions of the NIHAO (Numerical Investigation of a Hundred Astrophysical Objects; \citealp{Wang2015}) simulations.
The structure of this paper is as follows: 
Section 2 describes the observational data; Section 3 details the simulations used for comparison; Section 4 outlines the mass modelling methodology; Section 5 presents the results and discussion, including comparison with the literature; and finally, Section 6 summarises our conclusions. 

\section{Observations}

\subsection{AMIGA H{\sc i} rotation curves}

To complement the SPARC and LITTLE THINGS (hereafter SPARC+LT) samples studied in \citet{Korsaga2023}, we use the H{\sc i} kinematics data of 23 isolated galaxies from the AMIGA project. 
This project is based on a refinement of the pioneering Catalogue of Isolated Galaxies (CIG; \citealp{Karachentseva1973}). From the CIG sample of 1050 galaxies identified as isolated systems, \citet{Verdes-Montenegro2005} extracted a refined sample of isolated galaxies in the Local Universe. The degree of isolation was re-evaluated using two parameters \citep{Verley2007}: the tidal force exerted on the galaxies by their neighbours (namely $Q < -2$, with $Q$ a dimensionless estimation of the gravitational interaction strength) and the local number density ($n_k$ < 2.4), where $n_k$ can only be determined for galaxies with at least two neighbours. These isolation parameters showed that 950 galaxies, with systemic velocities above $\rm 1500\ km\ s^{-1}$ presented a continuous spectrum of isolation. From this refined sample, \cite{Sorgho2024} further selected 36 isolated galaxies for which high-quality H{\sc i} data were available and for which the H{\sc i} masses were measured \citep{Jones2018}. These H{\sc i} data were obtained either with the Very Large Array (VLA), the Giant Metrewave Radio Synthesis Radio Telescope (GMRT), or the Westerbork Synthesis Radio Telescope (WSRT).

To model the rotation curves of the H{\sc i} data cubes, \cite{Sorgho2024} used the 3D BAROLO (3D-Based Analysis of Rotating Objects via Line Observations) package \citep{DiTeodoro2015}. This package takes as input the H{\sc i} data cube of the galaxy and performs a three-dimensional titled-ring fit in order to derive both kinematic and geometric parameters. In their analysis, \citet{Sorgho2024} provided the initial parameters of the galaxy properties, including the kinematic centre, systemic velocity, inclination, and position angle. For each galaxy, optical parameters were adopted as initial guesses. To improve the robustness of the fits, a 3D mask was constructed for each galaxy using the Source Finding Application's (SoFiA, \citealp{Serra2015}) smooth-and-clip algorithm at a 4$\sigma$ threshold. This mask ensured that only the H{\sc i} emission associated with the target galaxy was included in the modelling. The 3D BAROLO outputs consist of the H{\sc i} rotation curve and the H{\sc i} surface-density profile, computed in concentric annuli, each with its own fitted geometric parameters centred on the galaxy's kinematic centre.
To accurately determine the DM halo parameters through  mass modelling, we require well sampled, high-quality rotation curves.  
Therefore, galaxies exhibiting strongly distorted rotation curves were excluded from the mass modelling. Such distortions can in fact contain valuable information on the dynamical state of the DM halo and on the processes affecting the galaxy; however, when they are associated with interactions or other departures from dynamical equilibrium, they violate the assumptions underlying the axisymmetric mass models adopted here and prevent a physically meaningful decomposition of the rotation curve into its baryonic and DM components (four galaxies). In addition, following \citet{Bosma1978} and the subsequent formalisation by \citet{Begeman1987} galaxies with fewer than five independent rotation-curve bins on either side of the galaxy were excluded because their spatial sampling is insufficient to provide robust constraints on the mass distribution. In total, seven galaxies were excluded on the basis of insufficient independent radial sampling before performing the mass modelling.
Mass models were then constructed for the remaining galaxies. Following the fitting procedure described in Sect. \ref{sect methodology}, six galaxies with best-fitting stellar mass-to-light ratio (M/L) of $\rm M/L< 0.05\ M_{\odot}/L_{\odot}$ and halo masses of $\rm M_{200}<10^{9}\ M_{\odot}$ were also excluded from the scaling relation analysis, as the resulting fits lead to unreliable halo parameter estimates. Applying these selection criteria resulted in a final sample of 23 galaxies used for the scaling relation analysis (see Table \ref{tab:amiga_sample}).

\subsection{GHASP rotation curves}

We further use hybrid H$\alpha$ and H{\sc i} rotation curves of 17 additional galaxies from \cite{Korsaga2019b}. These galaxies were originally selected from the GHASP \citep{Garrido2002, Epinat2008} survey, which comprises 203 nearby spiral and irregular galaxies and provides H$\alpha$ rotation curves with high spectral ($R\sim 10,000$) and spatial ($\sim 2''$) resolutions. These rotation curves were derived using a Fabry-Perot interferometer on the 1.93 m telescope at the Observatoire de Haute Provence and were computed from two-dimensional velocity fields \citep{Epinat2008, Epinat2008b}.
\citet{Korsaga2019b} selected a subsample of 31 galaxies from the initial sample of 203 galaxies for which both H{\sc i} rotation curves and 3.4~$\mu$m photometric data were available. For these galaxies, hybrid rotation curves were constructed by combining H$\alpha$ and H{\sc i} measurements. The use of hybrid rotation curves is particularly advantageous for constraining the distribution of DM in galaxies, as mass model parameters are sensitive not only to the outer regions, which are best traced with H{\sc i} data, but also to the exact velocity gradient in the inner regions. 
These hybrid rotation curves provide strong constraints on the DM halo parameters across both the inner and outer regions of the galaxy rotation curves. For a detailed description of the construction of hybrid rotation curves, we refer the reader to \citet{Korsaga2019b}. 
After removing galaxies that were already included in \citet{Korsaga2023} and excluding galaxies without total H{\sc i} mass measurement, we ended up with a sample of 17 galaxies (see Table \ref{tab:ghasp_sample}). 

\subsection{Mid-infrared photometry}

To trace the stellar component of the 23 AMIGA and 17 additional GHASP galaxies, we use mid-infrared W1 (3.4~$\mu$m) photometric data from the Wide-field Infrared Survey Explorer (WISE) Extended Source Catalog (WXSC; \citealp{Jarrett2013, Jarrett2019}). 
The WISE survey mapped the entire sky in four bands W1 (3.4~$\mu$m), W2 (4.6~$\mu$m), W3 (12~$\mu$m) and W4 (22~$\mu$m) with a field of view of $\rm  47 \times 47 \ arcmin^2$ and an angular resolution of $\sim 6$\arcsec\ in the shortest wavelength band (W1). The W1 band is the most sensitive of the WISE bands and is particularly well suited for probing the stellar population, due to its weaker sensitivities to star formation activities and dust extinction. 
This is especially important, as it enables better constraints on the stellar M/L and consequently, on the DM halo parameters.
The photometric data used in this study were derived following the procedures described in \citet{Parkash2018, Jarrett2013, Jarrett2019, Jarrett2023}. Detailed descriptions of the photometric measurements and data processing for AMIGA and GHASP samples can be found in \citet{Sorgho2024} and \citet{Korsaga2019b} respectively. The W1 surface brightness profile is decomposed into an exponential disc and, when present, a spherical bulge using a S\'ersic function.

\begin{figure*}
  \includegraphics[width=6.3cm]{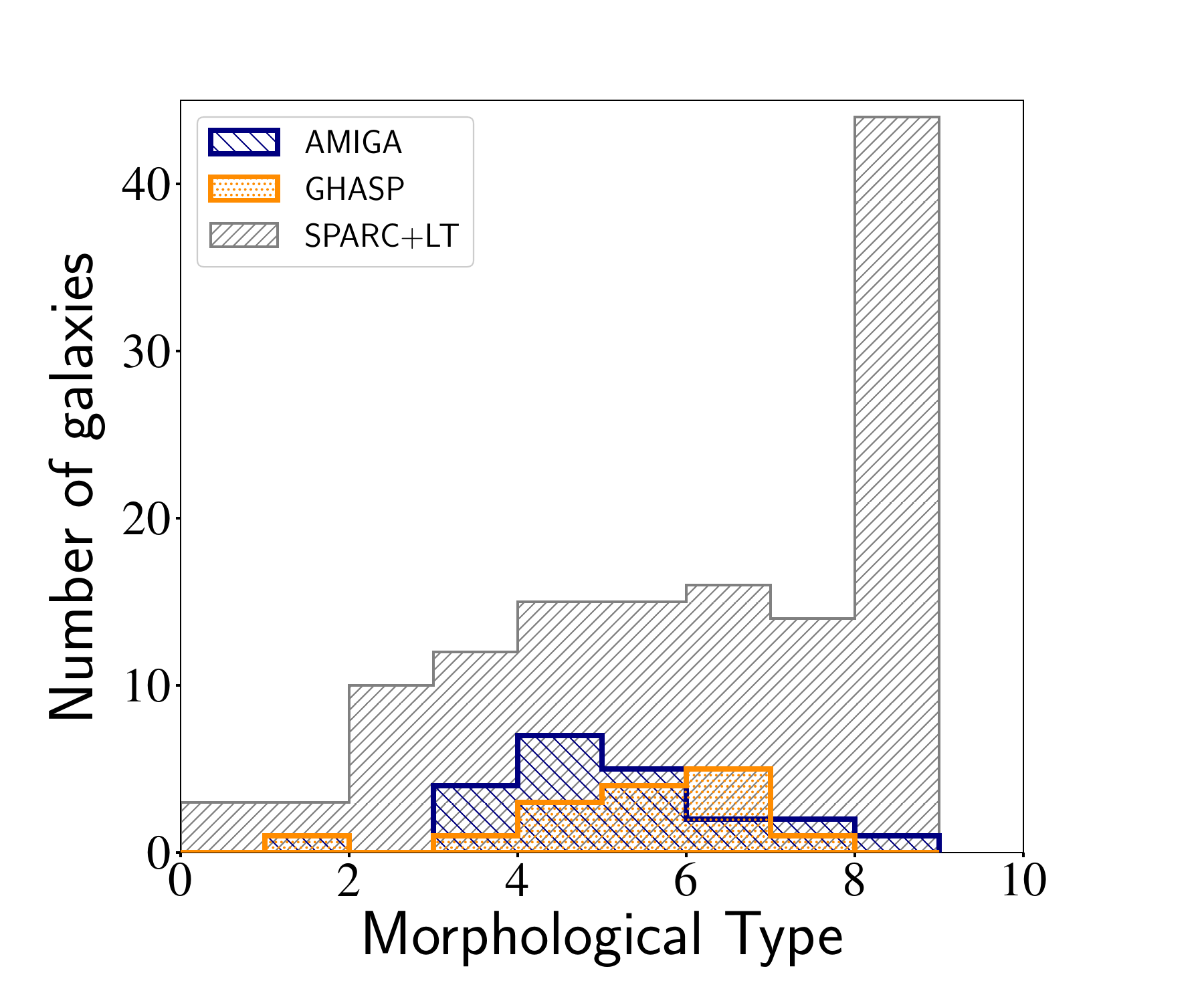}
 \includegraphics[width=6.3cm]{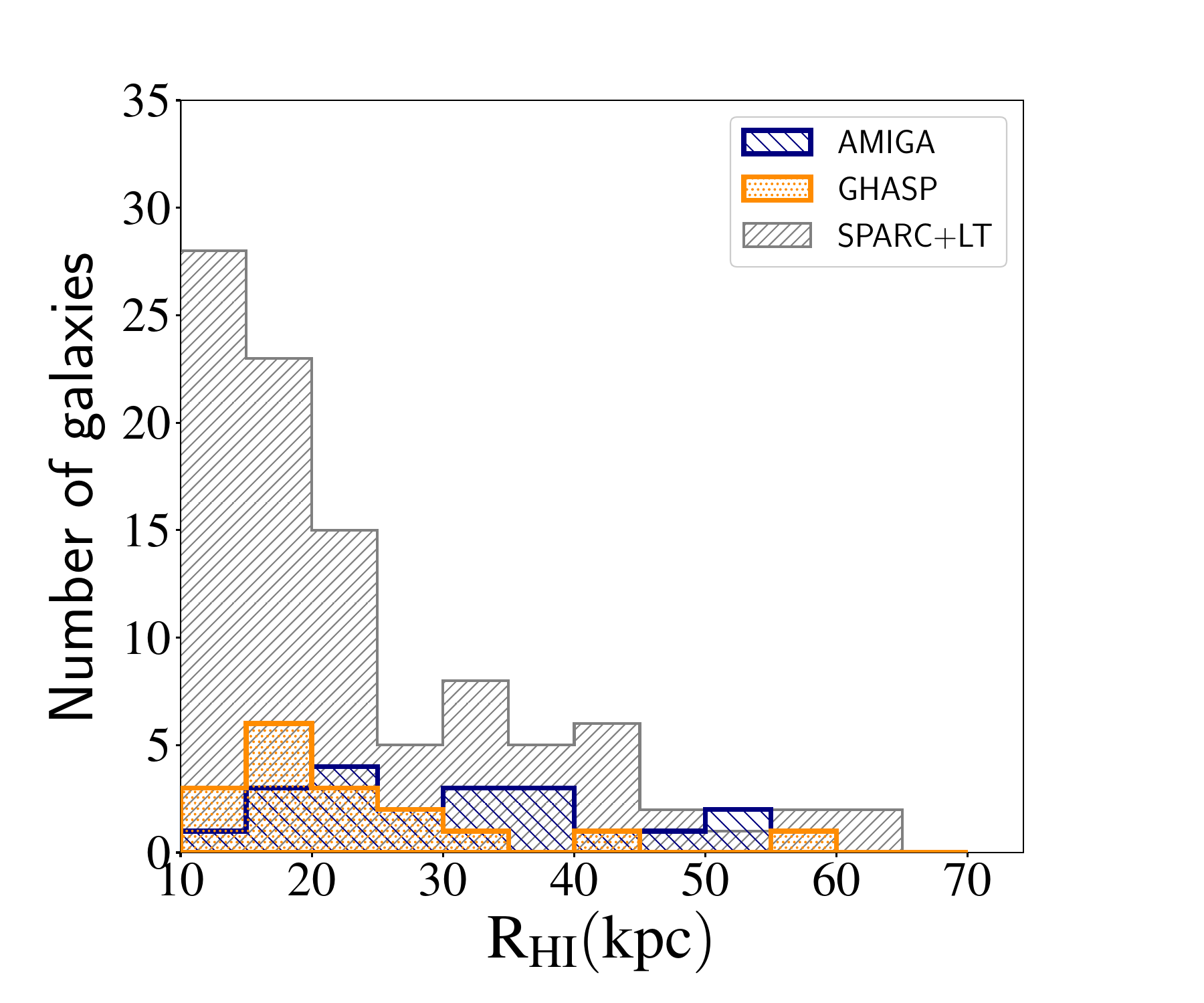}
  \includegraphics[width=6.3cm]{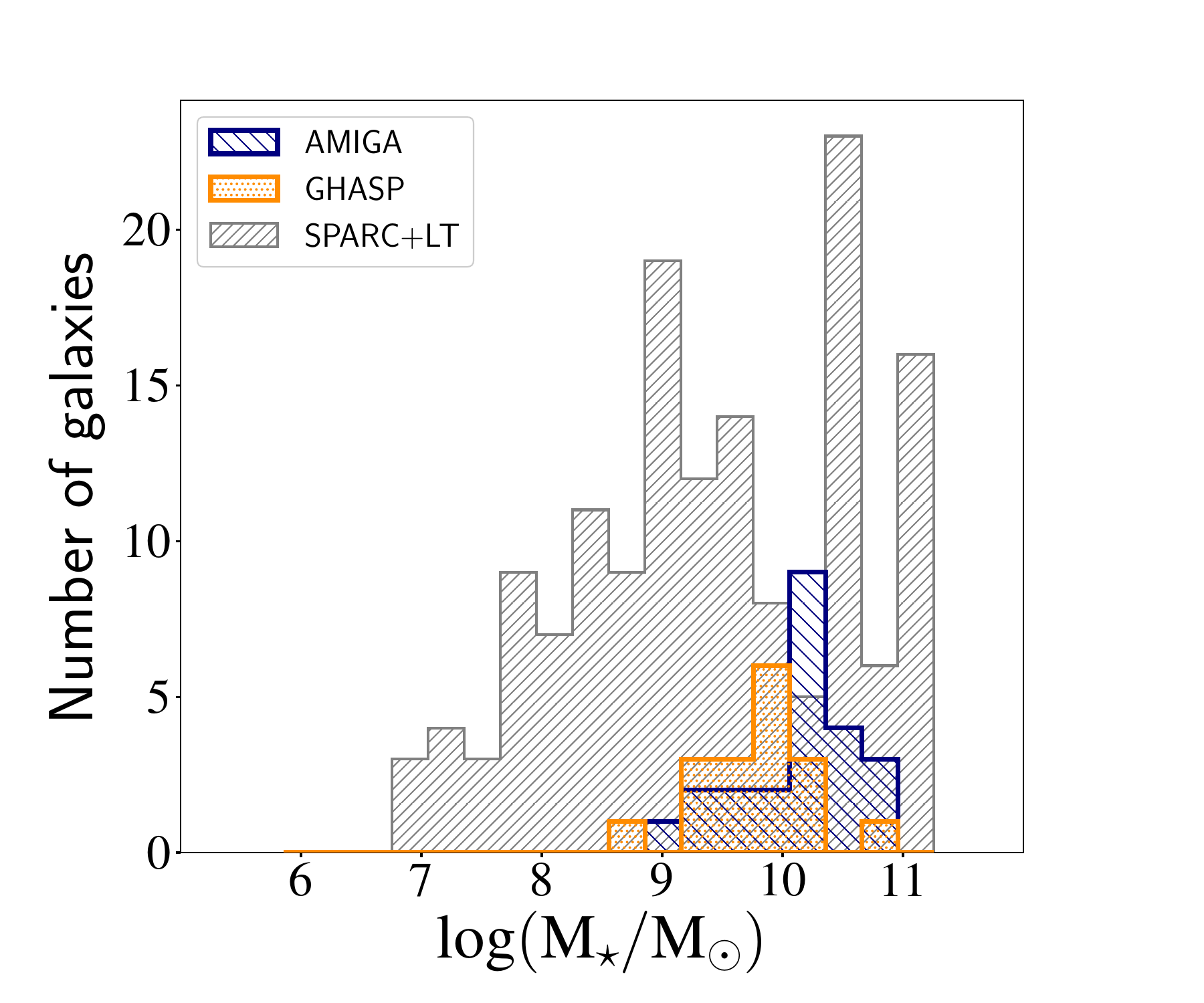}
 \caption{Distributions of morphological type (left, see Sect.~2.3 for description), H{\sc i} radius $\rm R_{HI}$ (centre), and stellar mass (right) for AMIGA, GHASP, and SPARC plus LITTLE THINGS (SPARC+LT) samples.}
 \label{fig:data distribution}
\end{figure*}

The distributions of morphological type, H{\sc i} radius ($\rm R_{HI}$), and stellar mass for the AMIGA, GHASP, and SPARC+LT samples are shown in Fig. \ref{fig:data distribution}. For the AMIGA, GHASP and LITTLE THINGS samples, the morphological classifications are taken from the Third Reference Catalogue of Bright Galaxies \citep[RC3;][]{DeVaucouleurs1991}. For the SPARC sample, we use the numerical Hubble types compiled by \citet{Lelli2016} which are based on the RC3 catalogue \citep{DeVaucouleurs1991}, \citet{Schombert1992}, and the NASA/IPAC Extragalactic Database (NED). The $\rm R_{HI}$ is defined as the radius at which the H{\sc i} surface density reaches $\rm 1M_{\odot}pc^{-2}$. The $\rm R_{HI}$ values are adopted from \citet{Ianjamasimanana2026} for AMIGA, \citet{Naluminsa2021} for GHASP, \citet{Lelli2016} for SPARC, and \citet{Read2017} for LITTLE THINGS. Stellar masses for AMIGA and GHASP samples were estimated from the W1-band luminosity reported by \citet{Jarrett2023} and \citet{Jarrett2013} respectively, and the M/L derived from our rotation curves fits. For SPARC+LT sample, stellar masses were adopted from \citet{Korsaga2023}. Figure \ref{fig:data distribution} shows that the AMIGA and GHASP samples span a similar range of morphological types and H{\sc i} radii as the SPARC+LT sample. However, they are preferentially composed of massive galaxies ($\rm log(M_{\star}/M_{\odot})>9$), whereas SPARC+LT extend to significantly lower stellar masses. Together, these samples provide a broad range in morphology and stellar mass, enabling us to investigate the H{\sc i}-to-halo mass scaling relation over several orders of magnitude in galaxy mass while improving the sampling of the high-mass regime.

\section{Simulations}
To interpret the observational results, we compare them with predictions from state-of-the-art cosmological hydrodynamical simulations. We consider three simulations: IllustrisTNG, SIMBA and NIHAO. The first two have already been used in our previous work \citep{Korsaga2023} but we briefly review them here for completeness. For IllustrisTNG \citep{Nelson2018, Pillepich2019}, we extract galaxies from the highest-resolution realizations TNG50-1 for which the mean baryon and DM particle mass resolutions are respectively $\rm 8.5 \times 10^4 M_{\odot}$ and $\rm 4.5 \times 10^5 M_{\odot}$. We select central subhalos from the group catalogue and cross-match them with the supplementary HI$\rm+ H_2$ catalogue \citep{Diemer2018, Diemer2019}. Stellar masses and DM halo masses ($\rm M_{200}$) are taken from the group catalogue, whereas the atomic gas masses are taken from the HI$\rm+ H_2$ catalogue.
The H{\sc i} masses in TNG50-1 are computed in post-processing following the method described in \citet{Diemer2018}. They are therefore not directly resolved by the simulation and are subject to non-negligible uncertainties. In contrast, for SIMBA \citep{Dave2019}, the H{\sc i} masses are computed self-consistently during the simulation, eliminating the need for a post-processing H{\sc i} assignment. 
For TNG50 simulation, we apply a halo mass cut of $M_{200}\geq10^9 M_\sun$ and obtain a sample of 21,168 central galaxies at redshift z=0. No environmental selection is applied; therefore, the sample contains central galaxies spanning a range of environments, from isolated systems to large-density environment. We have also checked in \citet{Korsaga2023} that disc galaxies in TNG50, most relevant for the comparison with our data, follow the exact same trend as the general population. 
From the SIMBA $(100\ h^{-1} \rm \ Mpc)^3$ volume, we select 3180 central disc galaxies with $M_{200} \geq 10^9 M_\sun$ and a kinematic ratio $v/\sigma \geq 0.8$, selected according to an isolation criterion. The isolation criterion requires that the host halo contains only one galaxy (i.e. the number of centrals in the halo is equal to one). The kinematic ratio is used to separate rotation-dominated from dispersion-dominated systems.  

While IllustrisTNG and SIMBA provide statistically representative galaxy populations in large cosmological volumes, their resolution limits the level of detail accessible in the internal structure of disc galaxies. To overcome this limitation and further strengthen our analysis, we additionally include here a comparison to the NIHAO simulations, which offer a significantly higher resolution. 
The NIHAO project \citep{Wang2015} consists of a sample of 97 hydrodynamical zoom-in simulations, drawn from DM only cosmological simulations with halo masses between $\log(M_{\rm vir}/M_\sun)=9.5$ and $12.3$ \citep{Dutton2014}. The particle mass and the force softening lengths are chosen to resolve the DM mass profile below $0.01 R_{\rm 200}$. The simulations are performed using the Smoothed Particle Hydrodynamics (SPH) method \texttt{gasoline2} \citep{Wadsley2017} on a flat $\Lambda$CDM cosmology \citep{Planck2014}. They include subgrid models describing the turbulent mixing of metals and thermal energy \citep{Wadsley2008}, cooling via hydrogen, helium and other metal lines in a uniform UV ionizing and heating background \citep{Shen2010}, star formation according to the Kennicutt-Schmidt relation when temperature $T<15\ 000\ \rm K$ and density $\rho>10.3~\rm cm^{-3}$ \citep{Stinson2013}, pre-supernova feedback where 13\% of the stellar luminosity is injected into the surrounding interstellar medium \citep{Stinson2013}, supernova feedback where stars of mass between $8-40\ M_\sun$ eject $10^{51}~\rm erg$ and metals 4 Myr after their formation \citep{Stinson2006}, and delayed cooling for 30 Myr after the explosion to prevent the supernova energy to be radiated away. Stellar masses were calculated by summing the masses of all star particles associated with each halo, and range from $5\times 10^4$ to $2\times 10^{11}\ M_\sun$; neutral hydrogen mass was computed using fitting functions obtained from radiative transfer calculations by \citet{Rahmati2013}, as in \citet{Gutcke2017}. The NIHAO simulations produce morphologies, colours and size in line with observations \citep{Wang2015, Stinson2015, Dutton2016a} and yield a variety of DM halo inner slopes from cusps to cores \citep{Tollet2016, Dutton2017, Dekel2017, Freundlich2020a, Freundlich2020b}. We removed the galaxy g1.12e12 ($\rm log(M_{\star}/M_{\odot}) = 10.89$) from the sample. With this exclusion, the high-mass end of the sample does not contain any quenched, dispersion-supported giant elliptical.

\section{Methodology}
\label{sect methodology}
We determined the DM halo parameters by performing individual galaxy mass models using the so-called Dekel-Zhao \citep[DZ;][]{Freundlich2020b} parametrisations of the DM density profile. The DZ profile can be characterized by three free parameters: an inner slope s$_1$, defined here as the absolute value of the logarithmic density slope at 1\% of the virial radius $R_{\rm 200}$, a dimensionless concentration $c$, corresponding to the radius at which the logarithmic slope of the density profile equals $-2$, and the DM halo mass $M_{200}$. A key advantage of the DZ profile is its flexibility, allowing both cored and cuspy density distributions to be modelled within a single parametrisation, in contrast to fixed-slope profiles such as NFW \citep{Navarro1996}. The DZ profile was first applied to observed rotation curves by \citet{Korsaga2023}, who demonstrated that it can reduce the scatter in scaling relations, such as that between H{\sc i} and halo masses, and provide tighter constraints on DM halo parameters than the NFW profile. More recently, \citet{Ciocan2026b,Ciocan2026a} implemented the DZ profile in the three-dimensional dynamical modelling tool GalPaK-3D \citep{Bouche2015} and used it to infer DM density profiles for galaxies at intermediate redshifts ($0.3<z<1.5$). In that lower spatial resolution regime, the DZ profile provided fits comparable to those obtained with the \citet{DiCintio2014} and \citet{Einasto1965} profiles. 

The observed rotation curves are modeled as the quadratic sum of their primary mass components,
\begin{equation}
    \rm V_{circ} = \sqrt{V_{\star}^2 + |V_{gas}|V_{gas} + |V_{halo}|V_{halo}}.
\end{equation}
Each velocity term represents the circular velocity of a test particle in dynamical equilibrium, assuming a gravitational potential free from pressure support or non-circular motions.
$\rm V_{circ}$ is the resulting circular velocity; the stellar contribution is defined as $\rm V_{\star}^2 = (M/L)_{disc} \times |V_{disc}|V_{disc} + (M/L)_{bulge} \times |V_{bulge}|V_{bulge}$, with $\rm (M/L)_{disc}$ and $\rm (M/L)_{bulge}$ the stellar mass-to-light ratios of the disc and bulge, respectively; $\rm V_{halo}$ accounts for the DM halo; and $\rm V_{gas}$ traces atomic hydrogen (H{\sc i}) distribution, scaled by a factor of 1.35 to include helium. In the absence of CO observations for a large fraction of the galaxies, we neglect the molecular gas contribution. For nearby galaxies, molecular gas generally represents a relatively small fraction of the total baryonic mass, with typical molecular-to-stellar mass ratios of a few percent in mass-selected samples \citep[e.g.][]{Saintonge2011, Saintonge2017}, and the total cold-gas reservoir is generally dominated by H{\sc i} \citep{Catinella2018}. Furthermore, the dynamical effect of the molecular gas contribution is partially degenerate with the stellar M/L ratio, particularly where the molecular gas distribution follows that of the stars.
We further assume $\rm (M/L)_{bulge} = 1.4 \times (M/L)_{disc}$ \citep[see, e.g.,][]{Schombert2014}.

In the DZ modelling, the rotation curve is thus described by four free parameters: s$_1$, $c$, $M_{200}$, and the stellar disc mass-to-light ratio $\rm (M/L)_{disc}$. These parameters are constrained using an affine-invariant Markov Chain Monte Carlo (MCMC) approach, implemented with the EMCEE Python package \citep{Foreman-Mackey2013}. We adopt: a flat prior on DM halo mass $\rm M_{200}$ ($\rm 6\leq log\ M_{200} \leq 15$), a Gaussian prior on the halo concentration $c$ that follows the $c-M_{200}$ relation  with a scatter of 0.11 derived from \citet{Freundlich2020b}, a flat prior on s$_1$ ($0\leq s_1 \leq 5$), and a Gaussian prior on the logarithm of $\rm (M/L)_{disc}$ centred on $0.6 M_{\sun}/L_{\sun}$ with a dispersion $\sigma=0.2$ dex. The MCMC method returns full posterior distributions for all parameters with a Bayesian approach. For each parameter, we adopt the median of the marginalised posterior as the best-fitting value, with uncertainties given by the 16th and 84th percentiles. The posterior distributions and mass models of the AMIGA and GHASP sample are shown in Figs. \ref{fig:massmodel} and \ref{fig:ghasp massmodel}, respectively. The mean M/L ratio of the AMIGA galaxies is $M/L_{\rm disc}=0.28\pm 0.17\ M_{\odot}/L_{\odot}$, while the GHASP sample has a mean value of $M/L_{\rm disc}=0.31\pm 0.18\ M_{\odot}/L_{\odot}$. Both values are slightly below the mean value $\rm M/L_{W1}$= $0.35\pm 0.11\ M_{\odot}/L_{\odot}$ reported by \citet{Jarrett2023}, but largely within the scatter. In turn, the value reported by \citet{Jarrett2023} is lower than the typical M/L of $\sim 0.5-0.8\ M_{\odot}/L_{\odot}$ inferred from earlier studies based on 3.6$\mu$m Spitzer photometry  \citep{Meidt2014,Lelli2016}. 

\begin{figure*}
\centering
 \includegraphics[width=15.3cm]{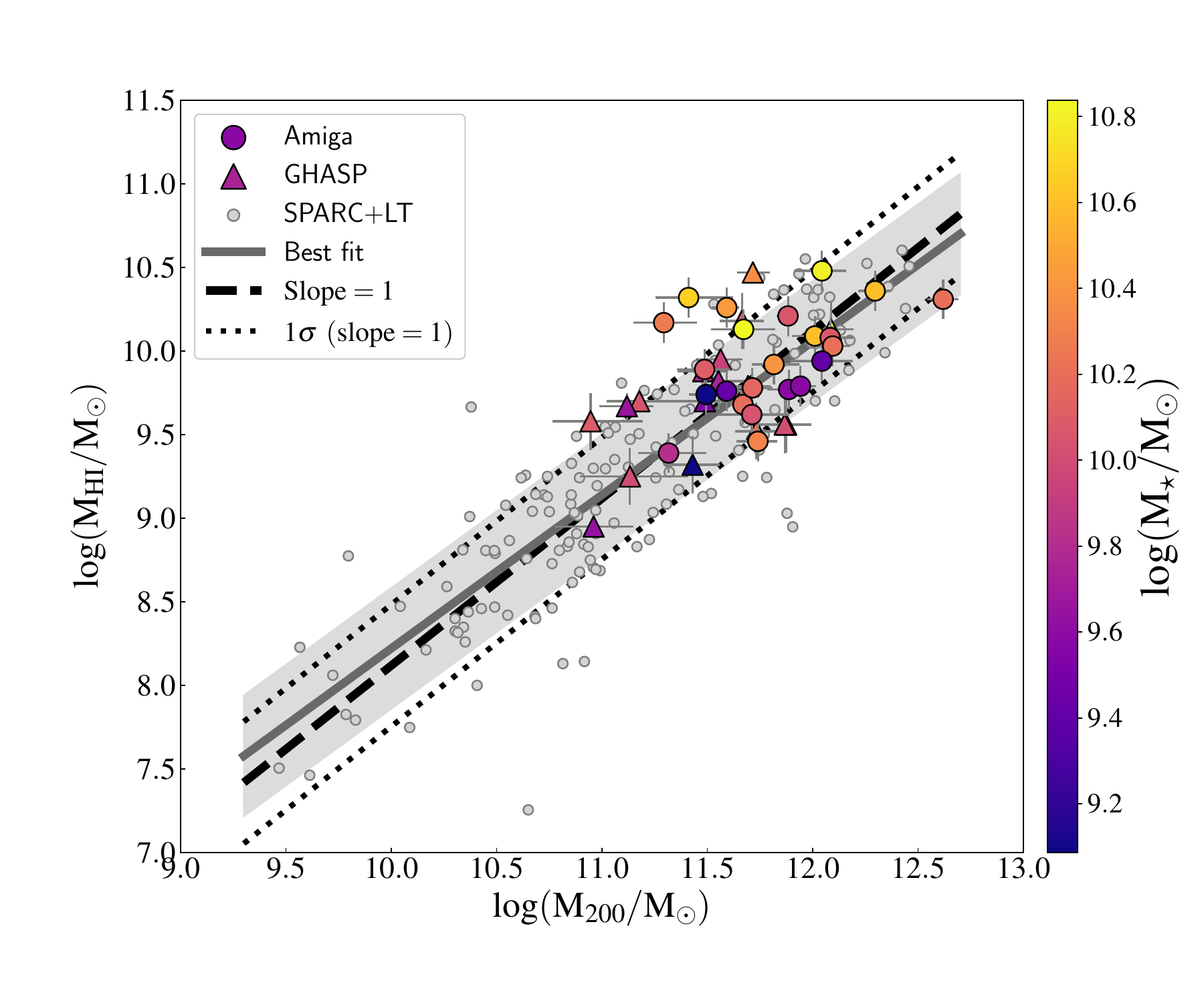}
   \caption{H{\sc i}-mass as a function of halo mass. The isolated AMIGA galaxies are shown as coloured circles, GHASP galaxies as coloured triangles, with colours indicating stellar mass. The SPARC and LITTLE THINGS galaxies from \citet{Korsaga2023} are shown as grey dots. The dashed black line shows the fit with slope fixed to one, whose zero point yields $\rm -1.88\pm 0.37$ for the combined dataset (AMIGA, GHASP, SPARC and LITTLE THINGS), the solid grey line shows the best-fitting relation with free slope $\rm log(M_{\hi}) = (0.92\pm 0.04) \times log(M_{200}) - (0.96\pm 0.46)$. The dotted lines and grey band correspond to the 1$\sigma$ scatters of these two relations (0.37 and 0.36 dex, respectively).}
 \label{fig:MHI-Mhalo}
\end{figure*}

\begin{figure*}
 \includegraphics[width=9.3cm]{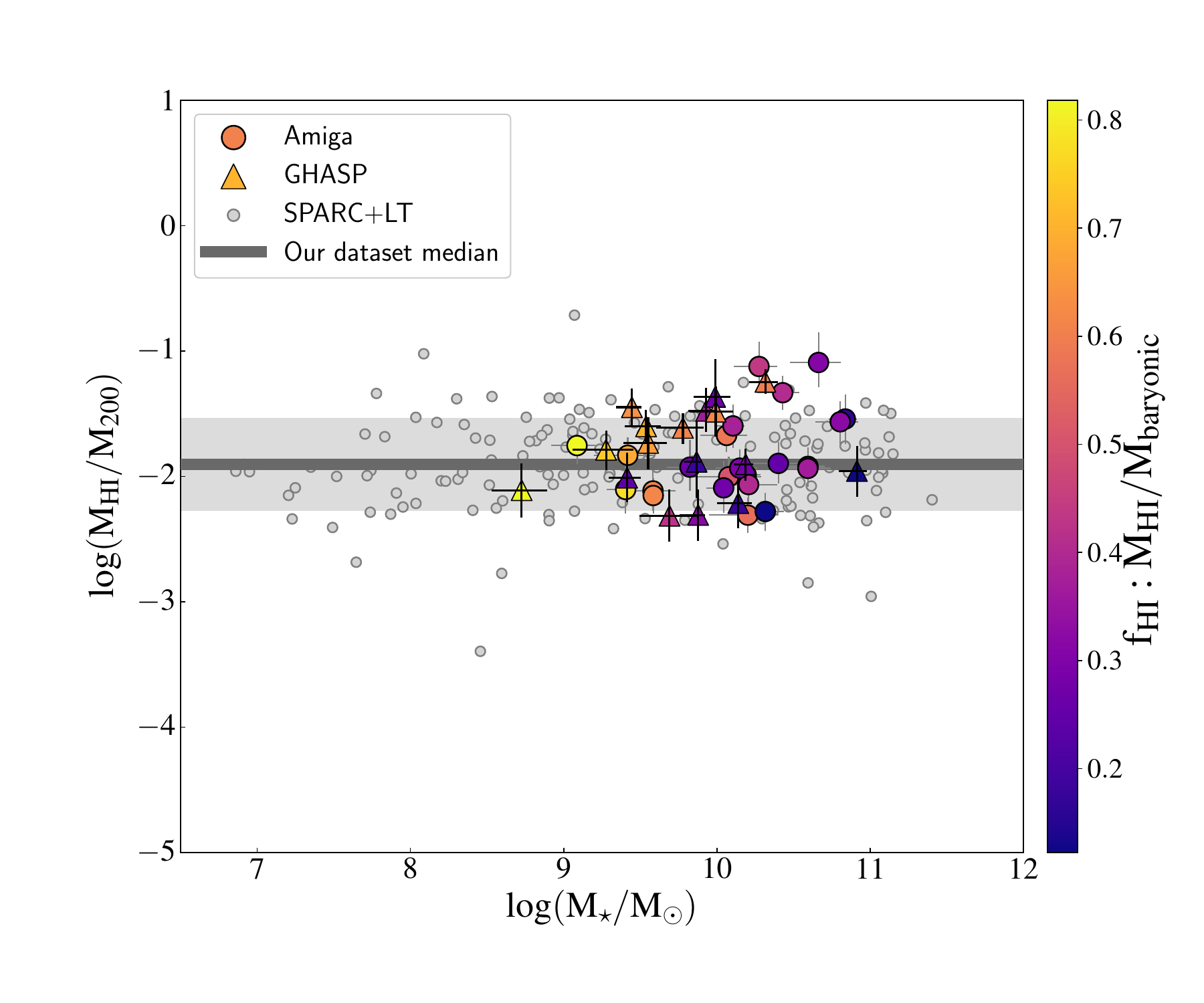}
  \includegraphics[width=9.3cm]{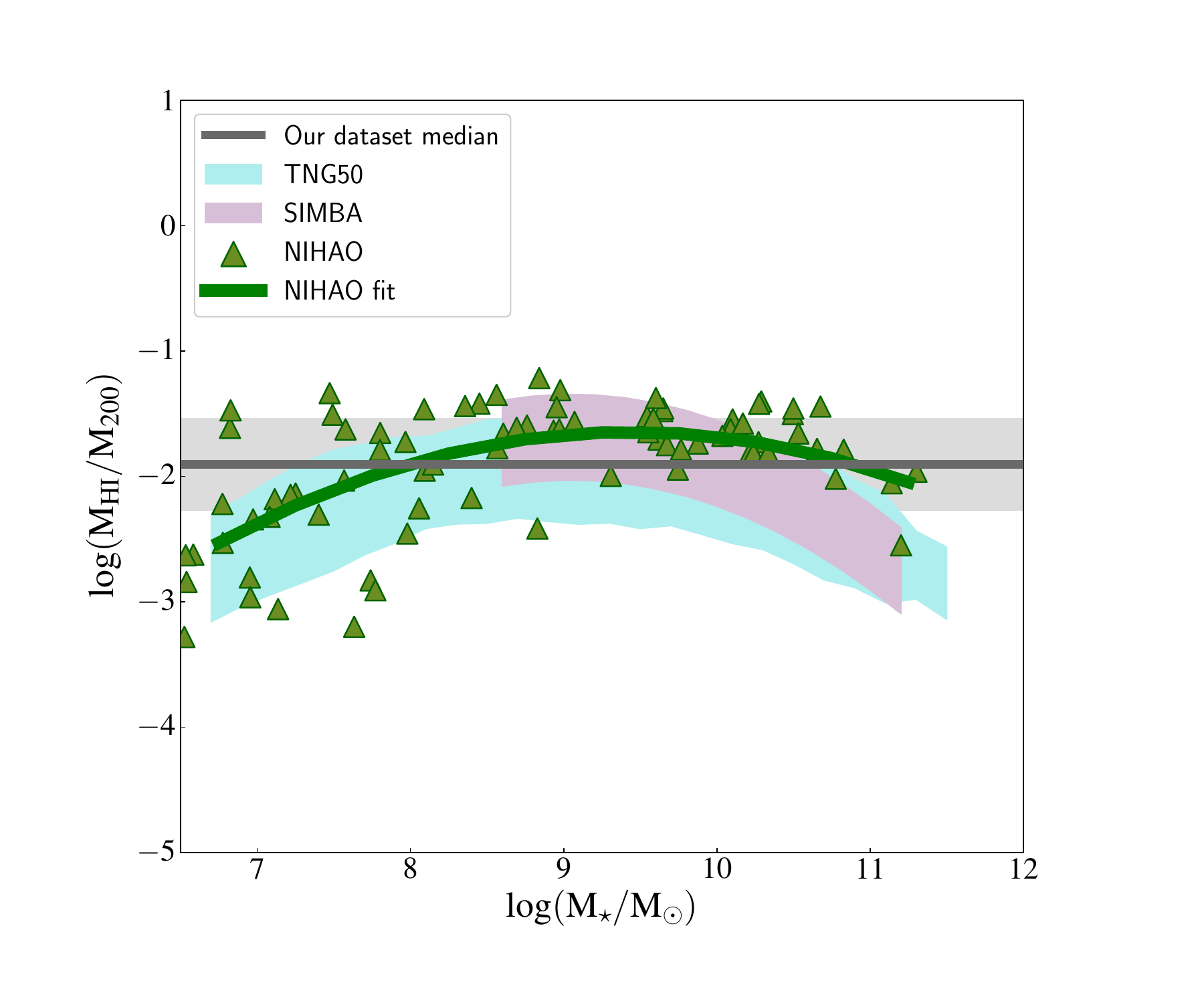}
 \caption{H{\sc i}-to-halo-mass ratio vs. stellar mass. The left panel highlights the observations: isolated AMIGA galaxies are shown as coloured circles, with the colour scale indicating the H{\sc i} fraction ($\rm M_{\hi}/M_{baryonic}$), GHASP galaxies are represented by coloured triangles, SPARC and LITTLE THINGS galaxies are shown as grey dots. The grey horizontal line indicates the median value ($\rm log(M_{\hi}/M_{200}) = -1.90$) of the combined dataset, the grey band shows its $1\sigma$ scatter (0.36 dex). The right panel highlights hydrodynamical simulations : The 1$\sigma$ around the median for SIMBA isolated disc galaxies is shown in magenta band, while the blue band represents the 1$\sigma$ around the median for Illustris TNG-50. The NIHAO zoom-in simulations are plotted in green triangles, with the fit of the binned averages in green line. The median value and 1$\sigma$ scatter of the combined dataset are again indicated by the grey horizontal line and grey band for comparison.}
 \label{fig:MHI/Mh}
\end{figure*}

\section{Results}

\subsection{The H{\sc i}-to-halo mass relation}
\label{section:ratio}

In Fig. \ref{fig:MHI-Mhalo}, we present the H{\sc i} mass as a function of DM halo mass inferred from the rotation curves, including the SPARC+LT sample from \citet{Korsaga2023}.
Both the AMIGA and GHASP galaxies follow the same overall trend as the SPARC+LT sample, indicating that H{\sc i} mass increases monotonously with halo mass over the full range probed, without any sign of a break in the relation at the high-mass end, thus confirming and extending the findings of \citet{Korsaga2023}. 
As shown in the figure, the relation between $\log (\rm M_{\hi}/M_\odot)$ and $\log (\rm M_{200}/M_\odot)$ using the combined dataset (AMIGA, GHASP, and SPARC+LT) is consistent with a linear relation with slope equal to one, with a scatter of 0.37 dex. Accordingly, the Pearson correlation coefficient is equal to 0.86. 
We further perform a fit to the data with a free slope, yielding 
$\rm log(M_{\hi}/M_\odot) = (0.92 \pm 0.04) \times log(M_{200}/M_\odot) - (0.96 \pm 0.46)$, with a 0.36 dex scatter. But we note that the difference between this fit and the one with the slope fixed to 1 is not significant, given that the RMS difference between the two (0.05 dex) is lower than the scatter of each relation. 

The linear relation between H{\sc i} mass and halo mass in log-log space naturally implies an approximately constant H{\sc i}-mass-to-halo-mass ratio (hereafter $\rm M_{\hi}/M_{200}$), as shown on the left panel of Fig. \ref{fig:MHI/Mh}.
The median value of the ratio for the combined dataset (AMIGA, GHASP, and SPARC+LT) is $\rm log(M_{\hi}/M_{200}) = -1.90$ with a scatter of 0.36 dex. Just for the AMIGA sample, we find -1.93 with a scatter of 0.33 dex; for the GHASP sample, -1.79 and 0.32 dex. This overall result fits well with our previous result for the SPARC+LT sample, namely $\rm log(M_{\hi}/M_{200}) = -1.90$ and a $1\sigma$ scatter of 0.37 dex. Two-sample Kolmogorov-Smirnov \citep[KS;][]{Kolmogorov1933, Smirnov1936} tests further show no statistically significant differences among the residual distributions of the three samples: AMIGA versus GHASP ($D$ = 0.22, $p$-value = 0.61), AMIGA versus SPARC+LT ($D$ = 0.17, $p$-value = 0.56), and GHASP versus SPARC+LT ($D$ = 0.19, $p$-value = 0.58.) The inclusion of AMIGA galaxies, which are largely unaffected by external processes such as tidal interactions or ram-pressure stripping, suggests that strong environmental perturbations are not required to establish the observed constancy of the $\rm M_{\hi}/M_{200}$. However, environmental processes may still modify this relation, particularly in very dense environments such as groups and central regions of galaxy clusters, where gas stripping is expected to play an important role. Our results support the idea that the cold gas reservoir remains fundamentally connected to the gravitational potential and the DM halo, making H{\sc i} a particularly sensitive tracer of the galaxy-halo connection. 

\subsection{Comparison to simulations}
\label{section:ratiosim}

The right panel of Fig. \ref{fig:MHI/Mh} compares our observational results with predictions from the state-of-the-art cosmological hydrodynamical simulations, namely SIMBA, Illustris TNG-50, and NIHAO. Despite differences in the simulation techniques, subgrid models, and the definition of H{\sc i} (e.g. on-the-fly or post-processed), all simulations show a dependence of the H{\sc i}-to-halo mass ratio on stellar mass. In particular, the $\rm M_{\hi}/M_{200}$ ratio decreases at the high-mass end, in contrast to the observations. 
Interestingly, the break in the $\rm M_{\hi}/M_{200}$ ratio occurs roughly at the same mass ($M_* \gtrsim 10^{10} M_\odot$) across all three simulations. Since NIHAO does not include the AGN feedback implemented in SIMBA and IllustrisTNG but still exhibits this break, the divergence cannot be solely driven by AGN activity. This stellar mass corresponds approximately to the critical halo mass scale ($M_{200} \sim 10^{12} M_\odot$) above which two fundamental transitions are expected to occur: supernova feedback becoming inefficient at expelling gas \citep{Dekel1986}, leading to gas consumption via star formation within the deep potential well, and a stable virial shock forming, cutting off cold-stream replenishment \citep{Birnboim2003, Dekel2006, Dekel2009}. Consequently, the drop in H{\sc i} across the simulations appears to reflect these concomitant mass-dependent physical thresholds (see also \citealt{Dekel2019}) and their interplay with subgrid or post-processed H{\sc i} partitioning rather than AGN physics alone. But the flat observational relation indicates that disc galaxies may actually circumvent these theoretical bottlenecks to regulate their atomic gas reservoirs.
We however note that the interpretation for NIHAO remains limited by small-number statistics at the high-mass end, as only a few galaxies with $\rm M_{\star} > 10^{10.5}$~M$_\sun$ are available. 

To provide a quantitative assessment of the discrepancy between observational and simulation datasets, we perform a two-dimensional KS test.
We focus on galaxies with $\rm M_{\star}> 10^{10}~M_{\sun}$, the domain where theoretical models often predict deviations from a constant $\rm M_{\hi}/M_{200}$ ratio. 
The KS test is a non-parametric test used to compare probability distributions and determine whether two samples are drawn from the same distribution. Using the public code \texttt{ndtest}\footnote{Written by Zhaozhou Li, \url{https://github.com/syrte/ndtest}}, we apply the 2D KS test in the ($\rm  M_\star$, $\rm M_{\hi}/M_{200}$) plane. We use the SPARC+LT sample as reference and renormalize the stellar mass by its mean and standard deviation to account for first-order selection effects. For the  AMIGA and GHASP observational samples, we obtain $p$-values of $0.23$ and $0.51$, respectively, indicating statistical consistency with the reference sample. In contrast, {\it all} simulation datasets yield $p$-values near zero ($<0.006$), highlighting significant deviations from the observed distributions. Using the whole observational dataset as reference, we similarly obtain near-zero $p$-values for the simulation datasets ($<0.006$). 
Finally, to further compare the trend of $\rm M_{\hi}/M_{200}$ as a function of $\rm  M_\star$ in observations and simulations, we carry out linear fits to the data for $\rm M_{\star}> 10^{10}~M_{\sun}$ and find that while the slopes for the observational samples are compatible with a constant ratio ($-0.11\pm 0.17$ for SPARC+LT, $0.60\pm 0.40$ for AMIGA, and $0.00\pm 0.58$ for GHASP), there is on the contrary a clear decreasing trend at more than $2\sigma$ for all the simulated samples (with slopes $-0.50\pm 0.04$ for TNG50, $-0.71\pm 0.06$ for SIMBA, and $-0.53\pm 0.24$ for NIHAO).

\begin{figure*}
 \includegraphics[width=9.8cm]{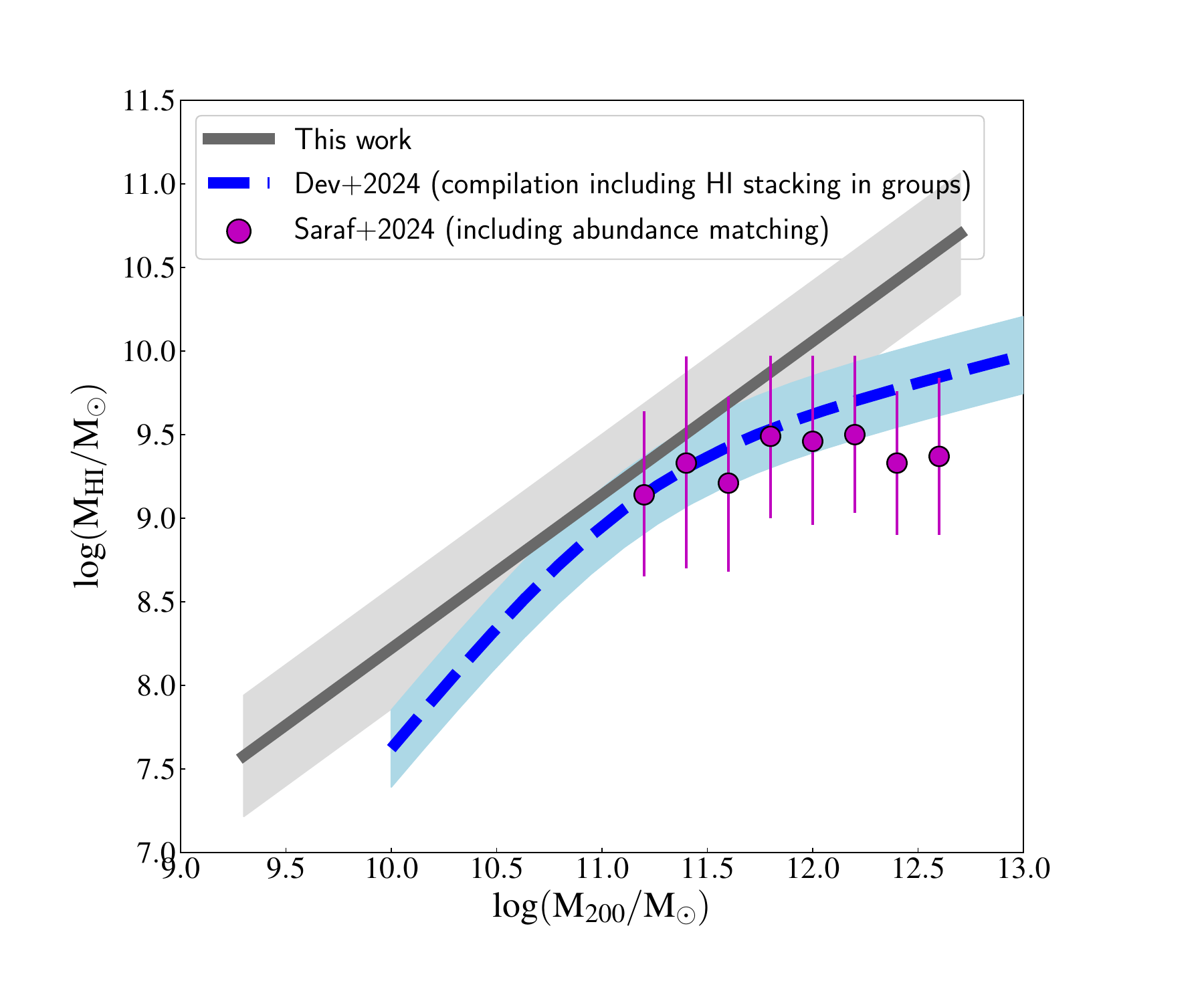}
  \includegraphics[width=9.8cm]{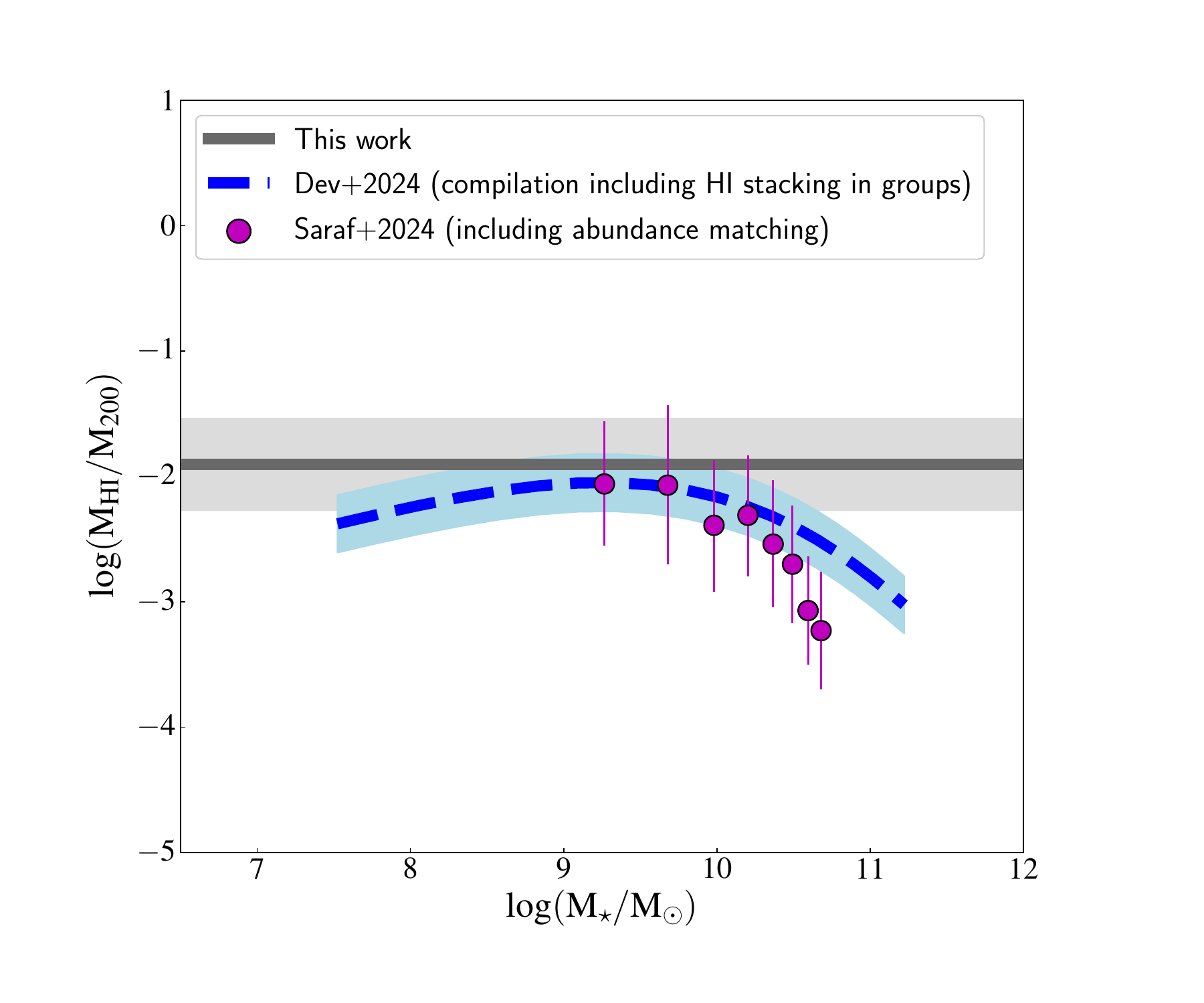}\\
   \includegraphics[width=9.8cm]{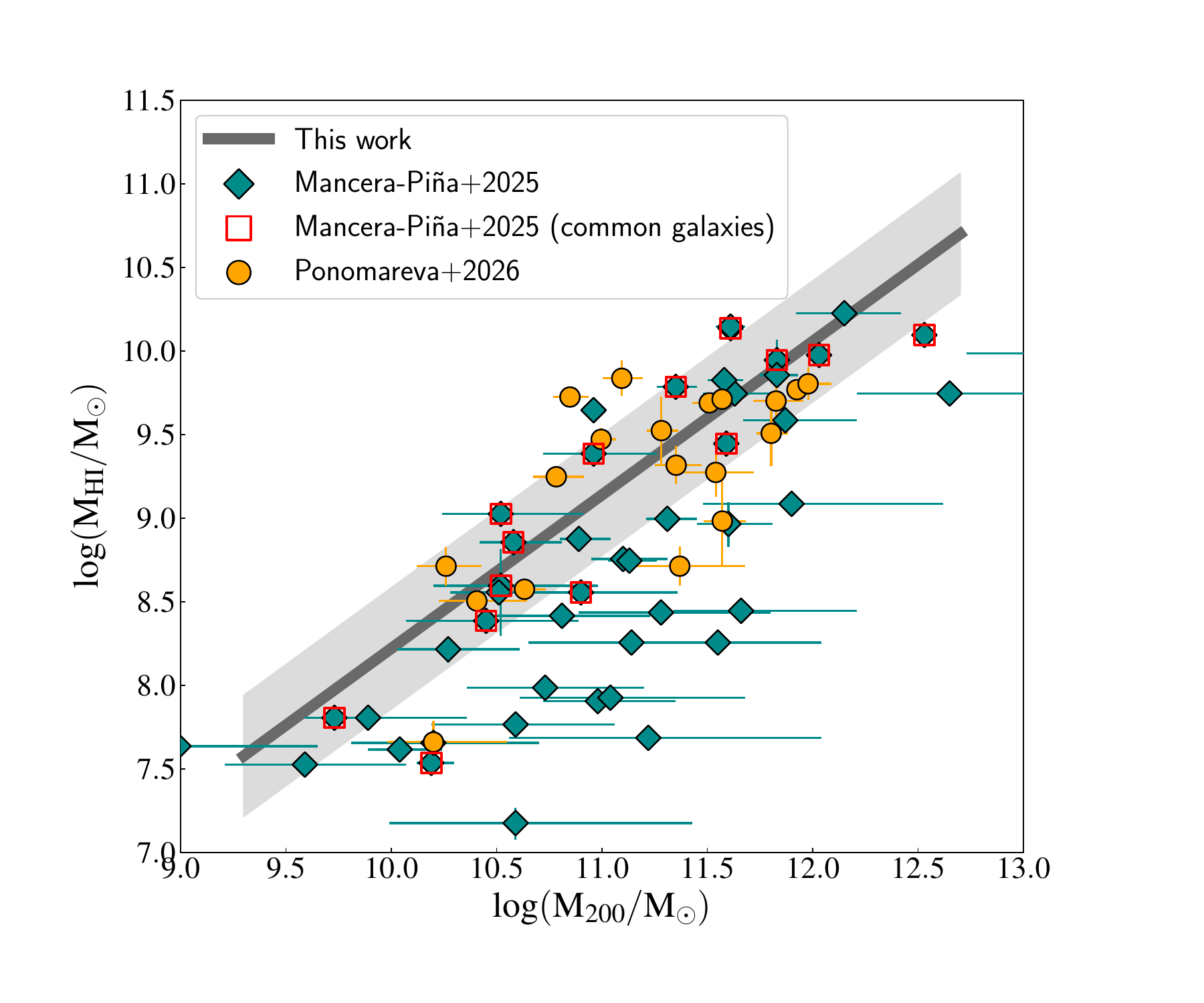}
  \includegraphics[width=9.8cm]{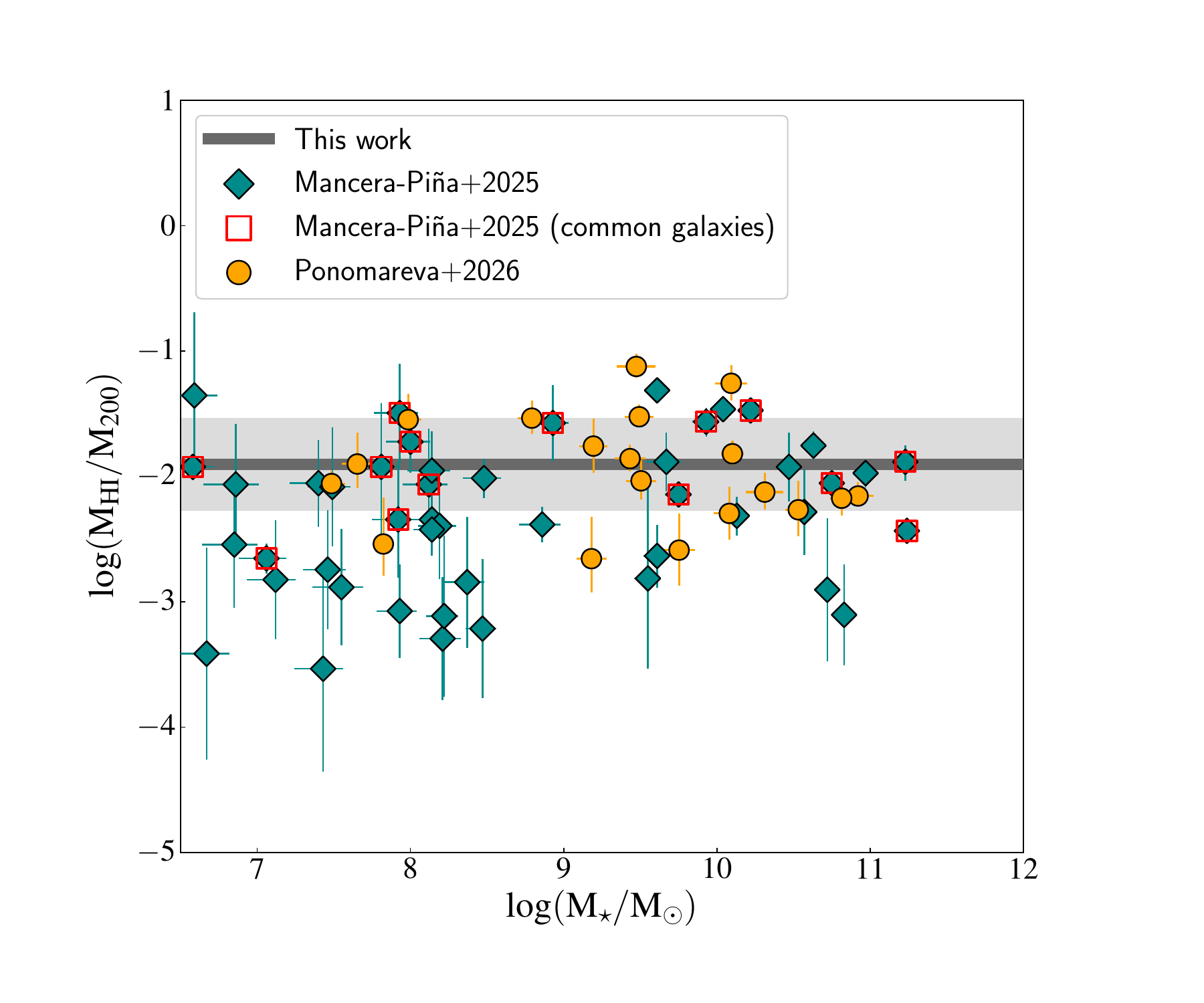}
 \caption{Comparison of the H{\sc i}-to-halo-mass relation (left panel) and the H{\sc i}-to-halo-mass ratio (right panel) obtained in this work with other observational measurements. The magenta dots show the data from \citet{Saraf2024}, the blue dashed line represents the relation from \citet{Dev2024}, the orange circles are from \citet{Ponomareva2026}, and the cyan diamonds from \citet{ManceraPina2025}. Galaxies common to the \citet{ManceraPina2025} sample and our dataset are highlighted by red squares.}
 \label{fig:comparison}
\end{figure*}

\subsection{Comparison to other observational studies}
\label{comparison with litterature}

Figure \ref{fig:comparison} compares the H{\sc i}-to-halo mass relations of Figs.~\ref{fig:MHI-Mhalo} and \ref{fig:MHI/Mh} with other observational studies. These studies rely on different galaxy samples and methods for estimating the DM halo masses, providing a useful context for our results.

\citet{Dev2024} compiled different measurements from the literature and proposed fitting functions for stellar-to-halo-mass and H{\sc i}-to-halo mass relations, from which it it also possible to derive a relation for the H{\sc i}-mass-to-halo-mass ratio as a function of stellar mass. The compilation includes (i) H{\sc i} stacking measurements in groups \citep{Guo2020, Dev2023, Rhee2023}, whose halo masses $M_{200}>10^{11}M_\odot$ were determined through halo occupation models or group velocity dispersions; (ii) SPARC isolated galaxies \citep{Lelli2016} between $10^{10}M_\odot<M_{200}<10^{12} M_\odot$, with the halo masses obtained by \citet{Posti2019} from individual H{\sc i} rotation curves with an NFW parametrisation; and (iii) 11 additional galaxies with individual H{\sc i} measurements and rotation curves obtained with the Giant Metrewave Radio Telescope (GMRT) analysed by \citet{Biswas2022, Biswas2023} with the NFW parametrisation, yielding $4 \times 10^{10}M_\odot<M_{200}<6\times 10^{11} M_\odot$.
The resulting H{\sc i}-to-halo mass relation is described by a double power-law with a break at $ \rm M_{200}\sim 10^{11} M_\odot$, but we note that \citet{Dev2024} essentially use only one data point below that mass to yield the break such that a single power law is equally satisfactory. Furthermore, the SPARC data points they use between $10^{10}M_\odot<M_{200}<10^{12} M_\odot$ are well fitted by our single power-law, as already shown in \citet{Korsaga2023}. 
Why then, does their relation depart from ours at $ \rm M_{200}> 10^{11} M_\odot$? This trend is actually driven by the stacks in galaxy groups, where the dense environment is expected to deplete galaxies of their gas due to gravitational and hydrodynamical perturbations (e.g. \citealt{Gunn1972, Larson1980, Abadi1999, Quilis2000, Boselli2022}), where different observations have shown that there is indeed atomic and molecular gas depletion (e.g. \citealt{Catinella2013}, \citealt{Denes2016}, \citealt{For2021}, \citealt{Kleiner2021} for the atomic gas; \citealt{Lee2022}, \citealt{Toni2026}, \citealt{Freundlich2026} for the molecular gas), and where we can speculate that gas depletion becomes even more pronounced as the group mass increases. Moreover, the techniques to obtain these masses are both less precise and less direct than for rotationally-supported galaxies in the field : it has always been clear that our results here and in \citet{Korsaga2023} applied only to {\it disc} galaxies, which is not the case for these high-stellar mass galaxies in groups.

In broad agreement with \citet{Dev2024} at $ \rm M_{200}> 10^{11} M_\odot$, \citet{Saraf2024} found a nearly constant median H{\sc i} mass of $\rm log(M_{\hi}/M_{\sun}) \sim 9.40$ in a sample of central group galaxies from the extended GALEX Arecibo SDSS Survey (xGASS; \citealp{Catinella2018}). Again, these are not all {\it disc} galaxies, and we further note that halo masses are estimated using abundance matching for massive haloes ($\rm M_{200} \geq 10^{11.5} M_\odot$) and conditional stellar mass function at lower mass. Such methods rely on cosmological simulations \citep[cf. e.g. ][]{Wechsler2018} and a mismatch between observations and abundance matching estimates at $ \rm M_{200}\gtrsim 10^{11} M_\odot$ was also reported by \citet{Posti2019} in the case of the stellar-to-halo-mass relation (SHMR). In this case, the discrepancy is notably driven by the fact that while the galaxy population at $ \rm M_{200}> 10^{11} M_\odot$ is dominated by passive elliptical galaxies for which the SHMR declines, there is still a population of disc galaxies for which the SHMR increases monotonically \citep{Posti2021}. A similar trend is probably at play for the H{\sc i}-to-halo mass relation since we target only rotationally-supported disc galaxies.

More recently, and with a method much closer to ours, \citet{ManceraPina2025} analysed 49 nearby gas-rich dwarf and massive disc galaxies coming from different surveys, including some in common with us \citep{Iorio2017, Koribalski2018, Walter2008, Heald2011, DiTeodoro2021}, spanning six orders of magnitude in stellar mass.
Figure~\ref{fig:comparison} shows their $\rm M_{\hi}$ and $\rm M_{200}$ measurements, which yield significant uncertainties and scatter ($\rm 1\sigma \sim 0.58 dex$), particularly at the low-mass end, but appear broadly consistent with both our single power-law relation (linear in log-log space) and with the double power-law relation of \citet{Dev2024}. Interestingly, galaxies with small error bars appear well in line with our constant ratio, while those that have a smaller gas-to-halo mass ratio tend to have much larger individual uncertainties. To facilitate the comparison, we report the typical uncertainty on the halo mass $\rm M_{200}$ in our analysis: the median uncertainty in our combined dataset is $\sigma_-$ = -0.14 dex and $\sigma_+=$+0.14 dex, compared to $\sigma_-$ = -0.24 dex and $\sigma_+$ = +0.34 dex for the full \citet{ManceraPina2025} sample. Interestingly, the galaxies common to both studies (red squares in Fig.~\ref{fig:comparison}), which closely follow our best-fitting relation, indeed have substantially smaller halo mass uncertainties in the \citet{ManceraPina2025} analysis (median $\sigma_-$ = -0.09 dex, $\sigma_+$ = +0.11 dex) than the remainder of their sample. Note that \citet{ManceraPina2025} derived halo masses from individual rotation curve decompositions assuming a core-NFW halo profile \citep{Read2016b,Read2016a}, which means that their individual halo masses are slightly different than ours, but for galaxies in common, the global relation is preserved despite individual differences. Even galaxies not in common with our study but having small uncertainties mostly lie on our relation. On the other hand, the largest apparent deviations from our relation are predominantly associated with galaxies having the least constrained halo masses. 

Finally, \citet{Ponomareva2026} further presented individual mass models for a sample of 20 disc galaxies at $0 < z < 0.08$ selected from the MeerKAT International GigaHertz Tiered Extragalactic Exploration (MIGHTEE) survey \citep{Heywood2024}. Their halo masses $\rm M_{200}$ were derived from mass modelling combining resolved H{\sc i} kinematics, Spitzer photometry, and resolved stellar mass surface density, assuming a standard NFW halo density profile. Figure~\ref{fig:comparison} shows their $\rm M_{\hi}$ and $\rm M_{200}$ measurements: although the scatter is significant, we find that these measurements agree well with our results. 

More quantitatively, 2D KS tests for galaxies with $\rm M_\star>10^{10} M_\odot$ using the SPARC+LT sample as reference (cf. Sect.~\ref{section:ratiosim}) yields p-values of 0.47 and 0.13 for the \cite{ManceraPina2025} and \cite{Ponomareva2026} samples, respectively, indicating that they are compatible with being drawn from the same distribution as SPARC+LT. Furthermore, linear fits to the data at $\rm M_\star>10^{10} M_\odot$ yield slopes compatible with no trend in both cases ($-0.30\pm 0.32$ for the sample of \citealt{ManceraPina2025}, $-0.34\pm 0.53$ for the sample of \citealt{Ponomareva2026}).
Taken together, individual galaxy measurements based on resolved kinematics in rotationally-supported galaxies therefore provide additional observational support for a monotonous H{\sc i}-to-halo mass relation, reinforcing the picture that isolated, rotationally-supported discs follow a remarkably simple scaling between their H{\sc i} content and DM halo mass. The main discrepancies with our results arise from studies based on galaxy groups or from halo masses derived using indirect, model-dependent techniques such as abundance matching. The discrepancy with simulations, on the other hand, is very real.

\begin{table*}
\caption{AMIGA sample. The columns list (1) the distance in Mpc; (2) the morphological type; (3) the inclination; (4) the logarithm of the atomic gas mass in $\rm M_{\sun}$ from \citet{Sorgho2024}; (5) the logarithm of the stellar mass in $\rm M_{\sun}$; (6)-(9) the logarithm of the DM halo mass in $\rm M_{\sun}$, the inner slope s1, the logarithmic of the concentration $c$ and the M/L derived from the Dekel-Zhao profile.}
\label{tab:amiga_sample}
\centering
\begin{tabular}{lcccccccccc}
\hline
CIG ID & Dist (Mpc) & Type & $i(^\circ)$ & $\log(M_{\rm{H\,I}})$ & $\log(M_\star)$ & $\log(M_{{200}})$ & $s_1$ & $\log(c)$ & M/L\\
 & (1) & (2) & (3) & (4) & (5) & (6) & (7) & (8) & (9)\\
\hline
1000 & 76.2 & 6 & 75.8 & $9.77\pm 0.12$ & $9.58_{-0.19}^{+0.15}$& $11.89_{-0.10}^{+0.10}$ & $0.14_{-0.10}^{+0.18}$ & $1.04_{-0.05}^{+0.05}$ & $0.20_{-0.05}^{+0.06}$ \\
1004 & 32.5 & 3 & 46.6 & $10.09\pm 0.12$ & $10.59_{-0.16}^{0.11}$ & $12.01_{-0.10}^{+0.11}$ & $1.04_{-0.62}^{+0.51}$ & $1.23_{-0.08}^{+0.08}$ & $0.26_{-0.07}^{+0.07}$ \\
1019 & 47.8 & 5 & 54.9 & $10.21\pm 0.12$ & $10.06_{-0.17}^{+0.11}$ & $11.88_{-0.05}^{+0.05}$ & $0.43_{-0.30}^{+0.47}$ & $1.09_{-0.06}^{+0.06}$ & $0.17_{-0.04}^{+0.04}$ \\
102  & 66.6 & 4 & 34.8 & $10.31\pm 0.12$ & $10.20_{0.25}^{+0.20}$ & $12.62_{-0.07}^{+0.08}$ & $0.07_{-0.05}^{+0.10}$ & $1.18_{-0.03}^{+0.03}$ & $0.09_{-0.02}^{+0.02}$ \\
103  & 20.5 & 5 & 59.9 & $9.39\pm 0.12$ & $9.82_{-0.15}^{+0.11}$ & $11.32_{-0.18}^{+0.15}$ & $1.58_{-0.25}^{+0.17}$ & $1.16_{-0.09}^{+0.09}$ & $0.26_{-0.06}^{+0.07}$ \\
123  & 78.0 & 3 & 51.6 & $10.32\pm 0.12$ & $10.66_{-0.18}^{+0.14}$ & $11.41_{-0.21}^{+0.16}$ & $1.34_{-0.65}^{+0.42}$ & $1.14_{-0.09}^{+0.09}$ & $0.20_{-0.05}^{+0.06}$ \\
134  & 68.4 & 4 & 63.3 & $9.68\pm 0.12$ & $10.19_{-0.15}^{+0.09}$ & $11.67_{-0.05}^{+0.06}$ & $0.61_{-0.42}^{+0.49}$ & $1.06_{-0.06}^{+0.05}$ & $0.24_{-0.05}^{+0.05}$ \\
147  & 35.9 & 4 & 48.3 & $10.13\pm 0.12$ & $10.84_{-0.11}^{+0.07}$ & $11.67_{-0.15}^{+0.16}$ & $1.05_{-0.68}^{+0.65}$ & $1.22_{-0.10}^{+0.10}$ & $0.39_{-0.09}^{+0.10}$ \\
159  & 57.1 & 5 & 76.8 & $10.08\pm 0.12$ & $10.08_{-0.24}^{+0.21}$ & $12.08_{-0.11}^{+0.10}$ & $0.11_{-0.08}^{+0.16}$ & $1.14_{-0.06}^{+0.05}$ & $0.09_{-0.02}^{+0.02}$ \\
292  & 23.4 & 3 & 77.2 & $9.89\pm 0.12$ & $10.10_{-0.11}^{0.07}$ & $11.49_{-0.12}^{+0.13}$ & $0.80_{-0.53}^{+0.59}$ & $1.18_{-0.08}^{+0.08}$ & $0.30_{-0.06}^{+0.06}$ \\
314  & 34.0 & 5 & 39.0 & $10.17\pm 0.12$ & $10.27_{-0.16}^{+0.12}$ & $11.29_{-0.16}^{+0.15}$ & $0.71_{-0.49}^{+0.56}$ & $1.06_{-0.09}^{+0.09}$ & $0.19_{-0.04}^{+0.04}$ \\
361  & 91.5 & 5 & 63.8 & $10.48\pm 0.12$ & $10.80_{-0.16}^{+0.12}$ & $12.04_{-0.11}^{+0.13}$ & $1.22_{-0.72}^{+0.52}$ & $1.27_{-0.09}^{+0.10}$ & $0.23_{-0.05}^{+0.06}$ \\
421  & 89.7 & 3 & 39.1 & $10.26\pm 0.12$ & $10.43_{-0.17}^{+0.11}$ & $11.59_{-0.06}^{+0.07}$ & $0.88_{-0.59}^{+0.60}$ & $1.02_{-0.08}^{+0.08}$ & $0.24_{-0.06}^{+0.06}$ \\
463  & 28.3 & 8 & 68.8 & $9.76\pm 0.12$ & $9.42_{-0.22}^{+0.18}$ & $11.59_{-0.08}^{+0.08}$ & $1.05_{-0.19}^{+0.12}$ & $0.93_{-0.06}^{+0.06}$ & $0.21_{-0.06}^{+0.09}$ \\
512  & 20.2 & 6 & 32.1 & $9.62\pm 0.12$ & $10.04_{-0.11}^{+0.02}$ & $11.71_{-0.18}^{+0.16}$ & $1.23_{-0.34}^{+0.25}$ & $1.30_{-0.08}^{+0.07}$ & $0.70_{-0.27}^{+0.49}$ \\
551  & 31.0 & 7 & 68.7 & $9.74\pm 0.12$ & $9.09_{-0.17}^{+0.12}$ & $11.49_{-0.08}^{+0.08}$ & $0.20_{-0.15}^{+0.25}$ & $0.96_{-0.05}^{+0.05}$ & $0.27_{-0.07}^{+0.08}$ \\
581  & 100.8 & 4 & 75.0 & -- & $10.59_{-0.19}^{+0.15}$ & $12.30_{-0.08}^{+0.08}$ & $0.23_{-0.16}^{+0.23}$ & $1.19_{-0.05}^{+0.05}$ & $0.20_{-0.05}^{+0.06}$ \\
604  & 22.9 & 1 & 65.2 & $9.46\pm 0.12$ & $10.32_{-0.17}^{+0.12}$ & $11.74_{-0.09}^{+0.10}$ & $0.76_{-0.35}^{+0.38}$ & $1.31_{-0.07}^{+0.07}$ & $0.23_{-0.06}^{+0.06}$ \\
676  & 78.2 & 4 & 75.0 & $10.03\pm 0.12$ & $10.21_{-0.22}^{+0.19}$ & $12.10_{-0.08}^{+0.08}$ & $0.17_{-0.12}^{+0.19}$ & $1.21_{-0.05}^{+0.04}$ & $0.13_{-0.03}^{+0.04}$ \\
736  & 19.0 & 4 & 70.3 & $9.78\pm 0.12$ & $10.15_{-0.15}^{+0.12}$ & $11.71_{-0.09}^{+0.10}$ & $0.39_{-0.27}^{+0.38}$ & $1.15_{-0.07}^{+0.06}$ & $0.21_{-0.04}^{+0.05}$ \\
744  & 35.3 & 7 & 33.0 & $9.79\pm 0.12$ & $9.58_{-0.07}^{+0.02}$ & $11.94_{-0.07}^{+0.08}$ & $0.48_{-0.33}^{+0.36}$ & $1.11_{-0.05}^{+0.05}$ & $0.78_{-0.25}^{+0.24}$ \\
85   & 35.9 & 10 & 15.9 & $9.94\pm 0.12$ & $9.40_{-0.12}^{+0.04}$ & $12.04_{-0.15}^{+0.15}$ & $1.53_{-0.41}^{+0.28}$ & $1.37_{-0.09}^{+0.08}$ & $0.63_{-0.24}^{+0.42}$ \\
983  & 66.1 & 4 & 62.0 & $9.92\pm 0.12$ & $10.40_{-0.18}^{+0.14}$ & $11.82_{-0.12}^{+0.11}$ & $0.21_{-0.15}^{+0.26}$ & $1.17_{-0.07}^{+0.06}$ & $0.23_{-0.06}^{+0.07}$ \\
\hline
\end{tabular}
\end{table*}

\begin{table*}
\caption{GHASP galaxies. (1) the distance in Mpc, (2) the morphological type, (3) the inclination, (4) the logarithm of the atomic gas mass in $\rm M_{\sun}$, (5) the logarithm of the stellar mass in $\rm M_{\sun}$, (6)-(9) the logarithm of the DM halo mass in $\rm M_{\sun}$, the inner slope s1, the logarithmic of the concentration $c$ and the M/L derived from the Dekel-Zhao profile.}
\label{tab:ghasp_sample}
\centering
\begin{tabular}{lccccccccc}
\hline
Name & Dist(Mpc) &Type & $i(^\circ)$ & $\log(M_{\rm{H\,I}})$ & $\log(M_\star)$ & $\log(M_{\rm{200}})$ & $s_1$ & $\log(c_2)$ & M/L \\
 & (1) & (2) & (3) & (4) & (5) & (6) & (7) & (8) & (9)\\
\hline
UGC10359 & 16.0 & 6 &$44\pm 1 $& $9.88\pm0.06$ & $9.54_{-0.14}^{+0.09}$ & $11.48_{-0.12}^{+0.12}$ & $0.24_{-0.18}^{+0.28}$ & $1.13_{-0.06}^{+0.06}$ & $0.30_{-0.07}^{+0.08}$ \\
UGC10470 & 21.2 & 4 &$34\pm 9$& $9.95\pm0.06$ & $9.78_{-0.17}^{+0.14}$ & $11.56_{-0.10}^{+0.11}$ & $0.73_{-0.30}^{+0.31}$ & $1.28_{-0.06}^{+0.06}$ & $0.14_{-0.03}^{+0.03}$ \\
UGC11852 & 80.0 & 1 &$ 47\pm 7 $& $10.47\pm0.06$ & $10.32_{-0.11}^{+0.08}$ & $11.71_{-0.08}^{+0.08}$ & $0.43_{-0.24}^{+0.25}$ & $1.27_{-0.07}^{+0.07}$ & $0.26_{-0.05}^{+0.04}$ \\
UGC12754 & 8.9 & 6 &$53\pm 5$ & $9.25\pm0.17$ & $9.87_{-0.08}^{+0.04}$ & $11.14_{-0.27}^{+0.24}$ & $0.80_{-0.45}^{+0.40}$ & $1.21_{-0.09}^{+0.08}$ & $0.66_{-0.20}^{+0.22}$ \\
UGC1913  & 9.3 & 7 &$48\pm 9$ & $9.70\pm0.06$ & $9.28_{-0.22}^{+0.18}$ & $11.49_{-0.15}^{+0.15}$ & $0.11_{-0.08}^{+0.15}$ & $1.06_{-0.05}^{+0.05}$ & $0.13_{-0.03}^{+0.03}$ \\
UGC2080  & 13.7 & 6 & $25\pm 9$ & $9.70\pm0.06$ & $9.93_{-0.12}^{+0.07}$ & $11.18_{-0.18}^{+0.16}$ & $0.87_{-0.56}^{+0.51}$ & $1.13_{-0.09}^{+0.09}$ & $0.43_{-0.11}^{+0.12}$\\
UGC2800  & 20.6 & 10 & $52\pm 1 $& $9.32\pm0.17$ & $8.73_{-0.19}^{+0.17}$ & $11.43_{-0.13}^{+0.13}$ & $0.20_{-0.14}^{+0.24}$ & $1.04_{-0.06}^{+0.06}$ & $0.09_{-0.02}^{+0.02}$  \\
UGC2855  & 17.5 & 5 & $68\pm 2 $& $9.56\pm0.17$ & $9.69_{-0.19}^{+0.18}$ & $11.88_{-0.12}^{+0.12}$ & $0.29_{-0.18}^{+0.22}$ & $1.24_{-0.05}^{+0.05}$ & $0.05_{-0.01}^{+0.01}$ \\
UGC3574  & 21.8 & 6 &$19\pm 1$ & $9.56\pm0.17$ & $9.88_{-0.12}^{+0.04}$ & $11.87_{-0.11}^{+0.12}$ & $1.12_{-0.43}^{+0.25}$ & $1.29_{-0.07}^{+0.06}$ & $0.65_{-0.25}^{+0.37}$ \\
UGC3734  & 15.9 & 5 &$43\pm 7$ & $8.95\pm0.06$ & $9.41_{-0.12}^{+0.09}$ & $10.96_{-0.19}^{+0.18}$ & $0.49_{-0.35}^{+0.49}$ & $1.09_{-0.09}^{+0.09}$ & $0.19_{-0.03}^{+0.03}$ \\
UGC4284  & 9.8 & 6 &$59\pm 9 $& $9.67\pm0.08$ & $9.45_{-0.10}^{+0.06}$ & $11.12_{-0.13}^{+0.13}$ & $0.39_{-0.27}^{+0.36}$ & $1.16_{-0.07}^{+0.06}$ & $0.50_{-0.13}^{+0.15}$\\
UGC5251  & 21.5 & 4 &$73\pm 6$ & $9.82\pm0.17$ & $9.55_{-0.16}^{+0.12}$ & $11.55_{-0.12}^{+0.13}$ & $0.20_{-0.14}^{+0.23}$ & $1.04_{-0.06}^{+0.06}$ & $0.18_{-0.04}^{+0.04}$ \\
UGC6537  & 18.0 & 5 &$47\pm 5$ & $9.52\pm0.17$ & $10.14_{-0.14}^{+0.09}$ & $11.73_{-0.10}^{+0.10}$ & $0.43_{-0.23}^{+0.27}$ & $1.25_{-0.06}^{+0.05}$ & $0.36_{-0.10}^{+0.10}$ \\
UGC6778  & 18.0 & 5 &$49\pm 4$ & $9.79\pm0.06$ & $10.19_{-0.08}^{+0.05}$ & $11.69_{-0.11}^{+0.12}$ & $0.70_{-0.29}^{+0.38}$ & $1.32_{-0.06}^{+0.06}$ & $0.35_{-0.05}^{+0.05}$ \\
UGC7766  & 9.0 & 6 &$69\pm 3$ & $9.58\pm0.17$ & $9.99_{-0.14}^{+0.10}$ & $10.94_{-0.25}^{+0.18}$ & $0.91_{-0.59}^{+0.49}$ & $1.06_{-0.09}^{+0.09}$ & $0.27_{-0.06}^{+0.07}$  \\
UGC9858  & 38.2 & 4 & $75\pm 2$ & $10.18\pm0.17$ & $9.99_{-0.17}^{+0.11}$ & $11.66_{-0.11}^{+0.11}$ & $0.74_{-0.48}^{+0.42}$ & $1.12_{-0.07}^{+0.07}$ & $0.23_{-0.06}^{+0.06}$ \\
UGC9969  & 39.7 & 3 & $61\pm 1$ & $10.13\pm0.17$& $10.91_{-0.12}^{+0.06}$ & $12.09_{-0.11}^{+0.12}$ & $1.25_{-0.31}^{+0.30}$ & $1.45_{-0.08}^{+0.06}$ & $0.53_{-0.16}^{+0.20}$ \\
\hline
\end{tabular}
\end{table*}

\section{Conclusions}
In this work, we have analysed the relation between H{\sc i} mass and DM halo mass using a sample of strictly isolated AMIGA disc galaxies together with GHASP, SPARC and LITTLE-THINGS disc galaxies in the local Universe. We modeled galaxy rotation curves using the Dekel-Zhao (DZ) parametrisation of the density profile to derive DM halo parameters and investigate the scaling relations between neutral hydrogen H{\sc i} mass and the DM halo mass. Our results were compared with predictions from the cosmological simulations IllustrisTNG, SIMBA and NIHAO, as well as with other observational studies. The main results are summarised as follows:
\begin{itemize}
\item We find a linear relation between H{\sc i} mass and DM halo masses in log-log space over the full range of halo masses, with a 1$\sigma$ scatter of 0.36 dex for the combined dataset (see Fig.~\ref{fig:MHI-Mhalo}).
\item Accordingly, the H{\sc i}-mass-to-halo-mass ratio remains constant over five orders of magnitude in stellar mass ($\rm 7 \leq log (M_{\star}/M_\odot) \leq 11.5$), with  $\rm log(M_{\hi}/M_{200}) = -1.90$ and a scatter of 0.36 dex (left panel of Fig.~\ref{fig:MHI/Mh}), indicating a scale-invariant behaviour for discs.
\item The presence of the same scaling behaviour in the strictly isolated AMIGA sample shows that major environmental perturbations are not required to establish the relation.
\item This conclusion is further reinforced by GHASP, where the large spatial coverage and high spectral resolution of well-behaved galaxies enable a much finer kinematic analysis.
\item In contrast to observations, the cosmological simulations SIMBA, TNG50, and NIHAO, predict a break in the H{\sc i}-to-halo mass ratio at high masses ($\rm M_{\star} > 10^{10}~M_\sun$).
\item Other observational studies in the literature based on actual resolved kinematics in rotationally-supported galaxies provide additional observational support for a constant  H{\sc i}-mass-to-halo-mass ratio in discs.
\end{itemize}

The constant  H{\sc i}-to-halo mass ratio across different galaxy populations suggests that the physical processes governing the coupling between neutral hydrogen and DM haloes are self-similar in disc galaxies. Whilst the present scaling relation is a global one, its possible connection to earlier radially resolved studies \citep[e.g.,][]{Carignan1990, Jobin1990} finding an approximately constant {\it local} $\Sigma_{\rm HI}/\Sigma_{\rm halo}$ ratio within individual galaxies, should be investigated in future works. A systematic study across a larger sample could clarify this connection and its implications for the H{\sc i}-to-halo mass relation. 

Extending our analysis to higher redshift will also be crucial for understanding the evolution of this relation over cosmic time. Preliminary results appear to extend the relation out to at least $z= 0.08$. Radio telescopes such as the upcoming Square Kilometre Array (SKA) and its precursor MeerKAT will provide the sensitivity and resolution required to probe the H{\sc i} content of galaxies at even greater lookback times and thus  offer a more complete picture of the mechanisms influencing the H{\sc i}-to-halo mass relation.

\section*{Acknowledgments}
MK thanks Gauri Sharma for useful discussions.
This work used the Spanish Prototype of an SRC service and support funded by the Spanish Ministry of Science and Innovation (MCIN), by the Regional Government of Andalusia and by the European Regional Development Fund (ERDF). MK, LVM, AS, PK, BN, RI acknowledge financial support from the grant CEX2021-001131-S funded by MICIU/AEI/10.13039/501100011033, and from the grants PID2021-123930OB-C21 and PID2024-155817OB-I00 funded by MICIU/AEI/10.13039/501100011033 and by ERDF/EU. MK acknowledges that this project has received funding through the SAFE – "Supporting At-Risk Researchers with Fellowships in Europe" project, funded by the European Union under Grant Agreement No. 101148426. FB and AVM acknowledge support by Tamkeen under the NYU Abu Dhabi Research Institute grant CASS. 
NIHAO simulations were carried out on the high-performance computing resources at New York University Abu Dhabi.

\bibliographystyle{aa}
\bibliography{MyCollection} 



\begin{appendix}
\section{Rotation curve fits for AMIGA galaxies}  
\label{app:AMIGA massmodel}

Figure \ref{fig:massmodel} shows the model rotation curve fits (posterior distribution and rotation curve with its decomposition in different components) for galaxies from AMIGA sample using the Dekel-Zhao (DZ) parametrisation.

\begin{figure*}
 \centering
  \includegraphics[width=6cm]{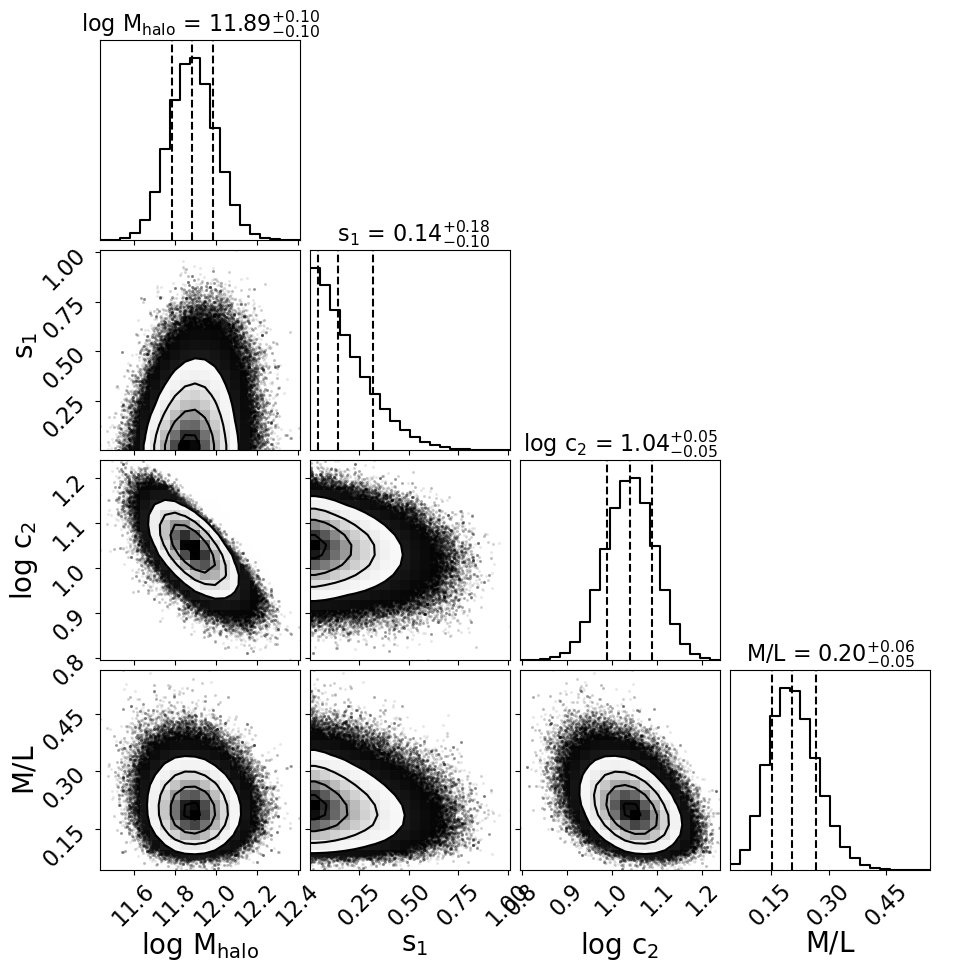}
  \includegraphics[width=7.cm]{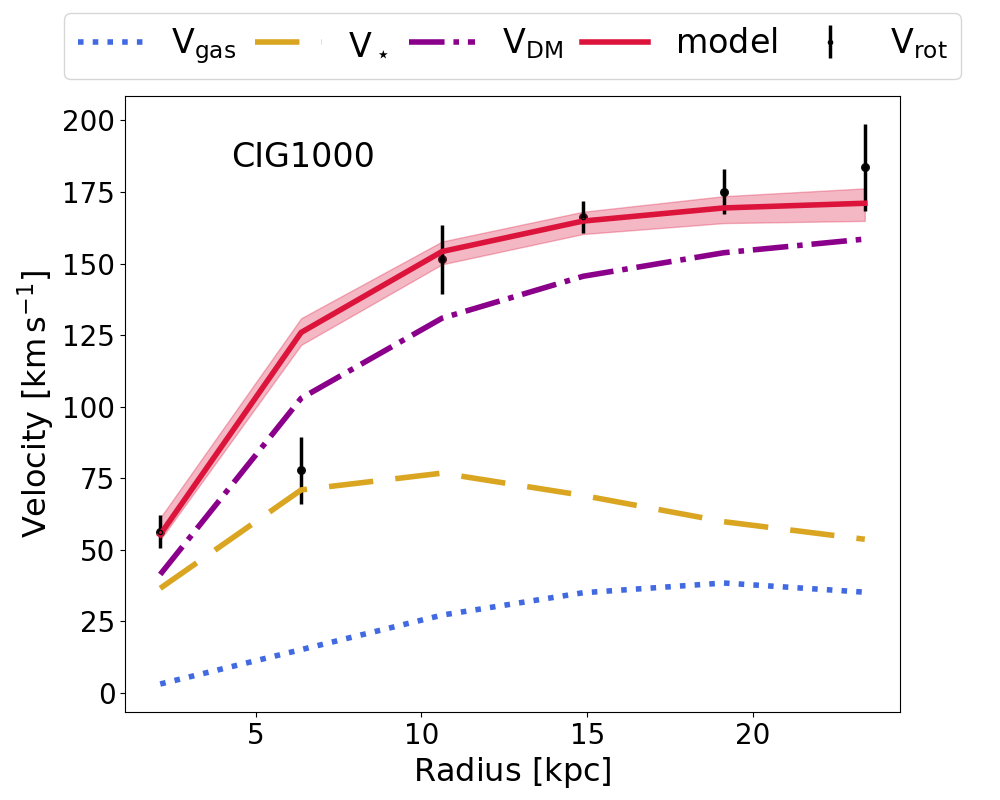}\\
 \includegraphics[width=6.cm]{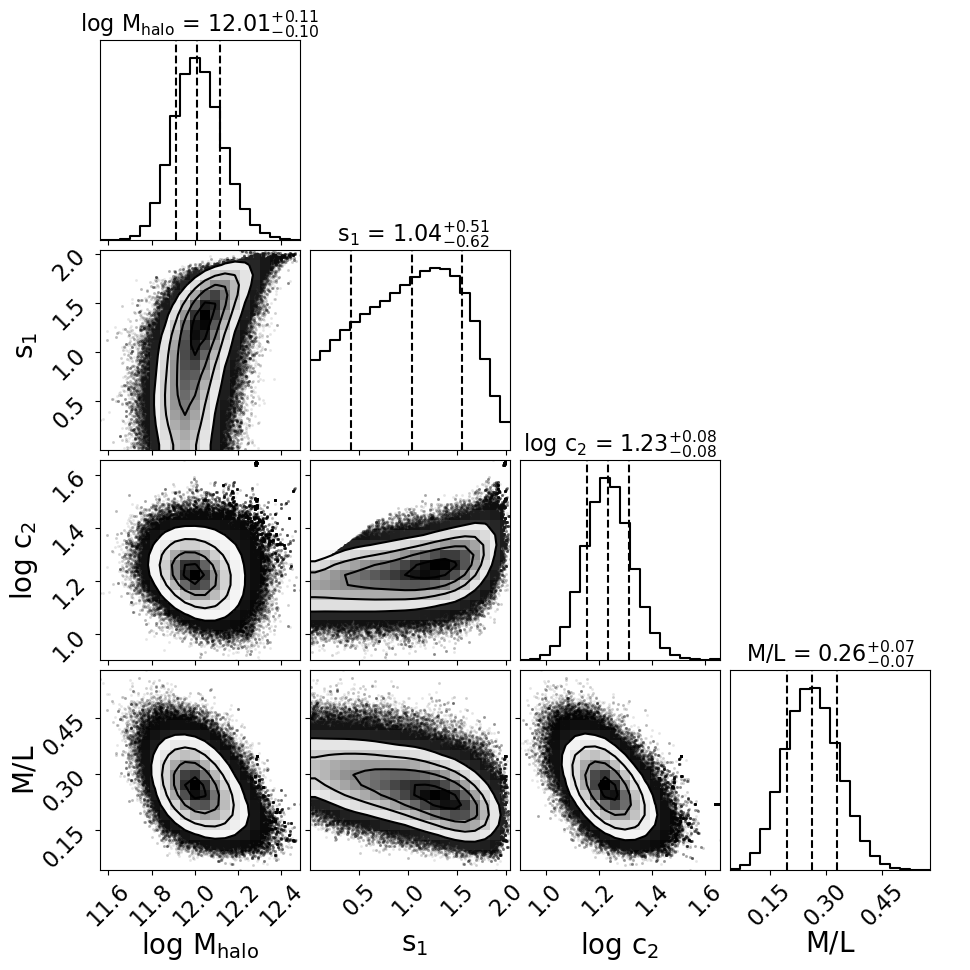}
 \includegraphics[width=7.cm]{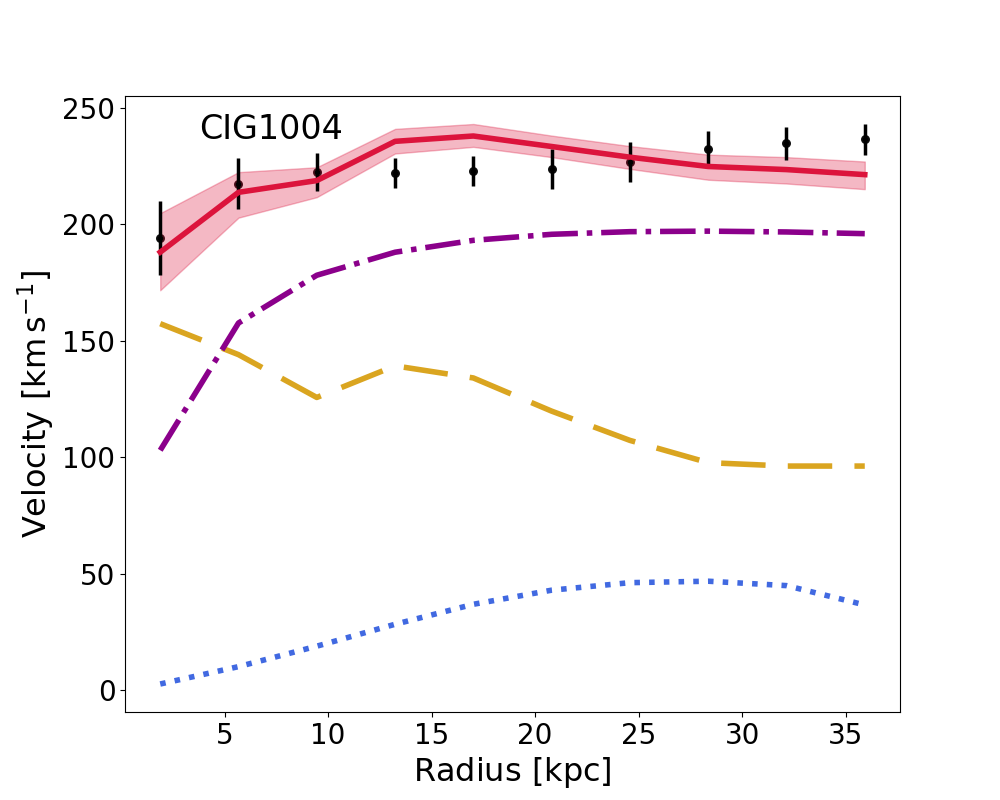}\\
    \includegraphics[width=6.cm]{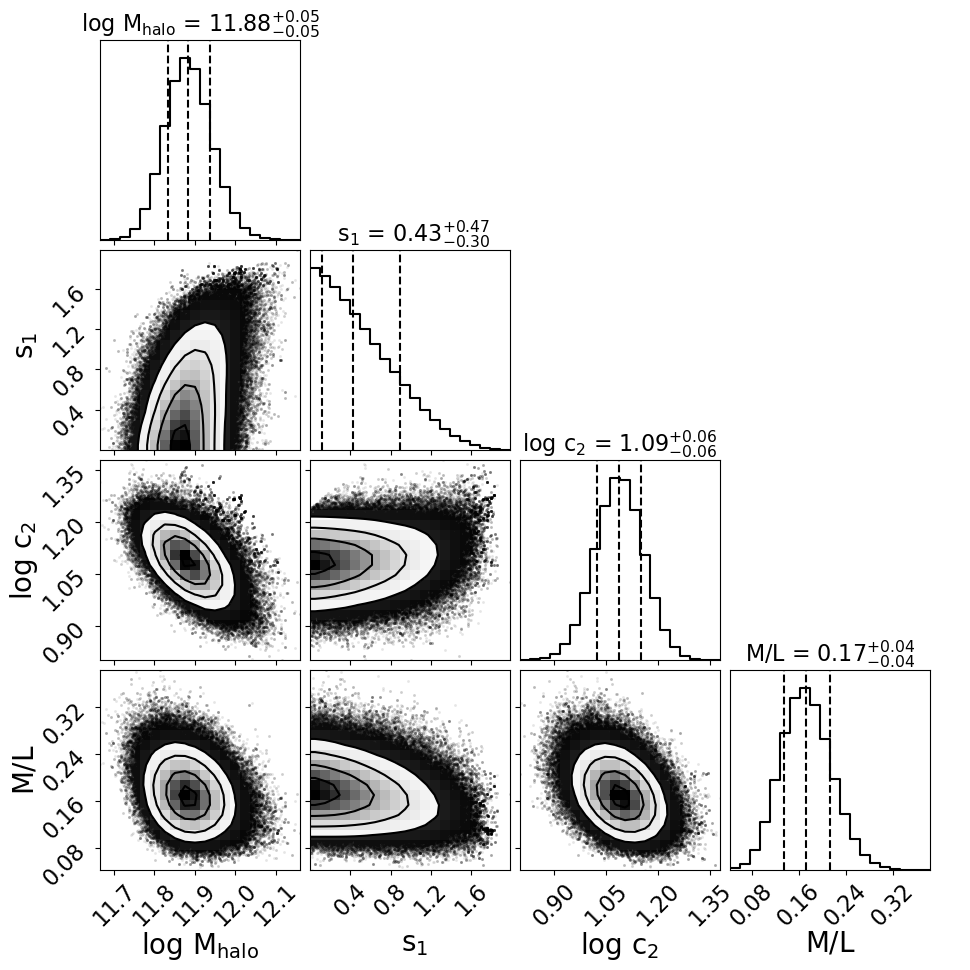}
\includegraphics[width=7.cm]{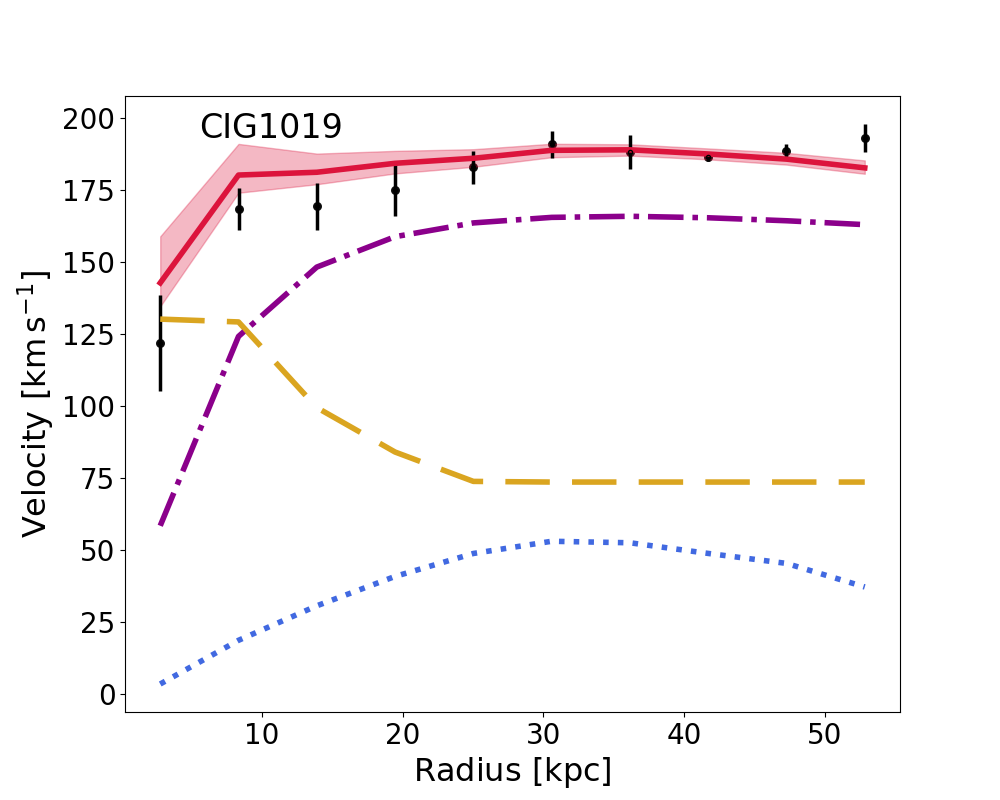}\\
   \includegraphics[width=6.cm]{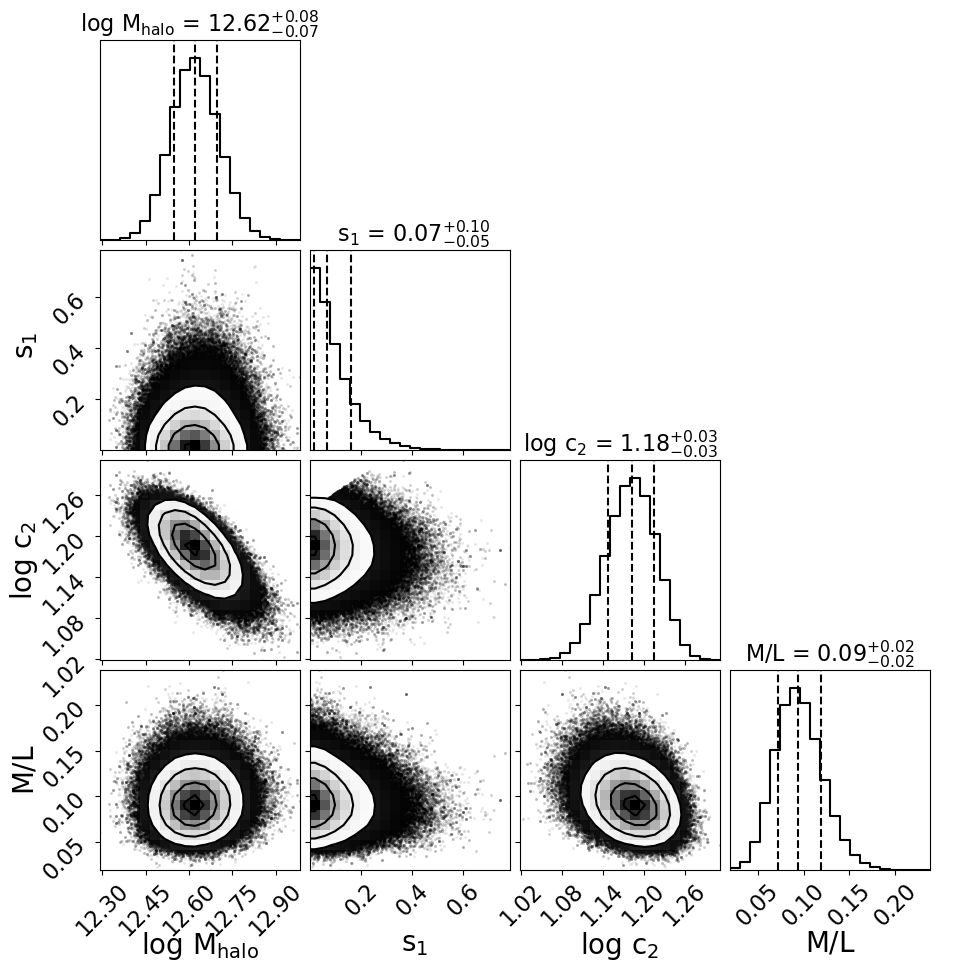}
 \includegraphics[width=7.cm]{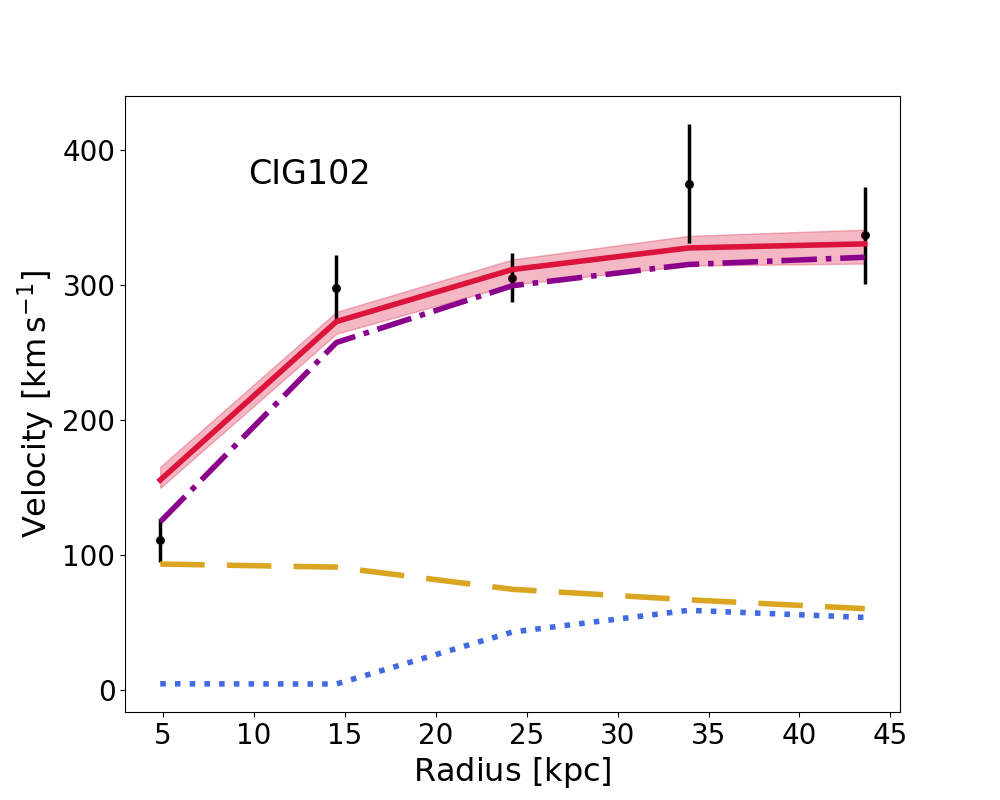}\\
     \caption{Galaxies from the AMIGA sample. The left panels show the posterior distributions of the model parameters, while the right panels present the corresponding mass models obtained using the DZ parametrisation.}
 \label{fig:massmodel}
    \end{figure*}

\addtocounter{figure}{-1}
\begin{figure*}
    \centering
    \includegraphics[width=6.cm]{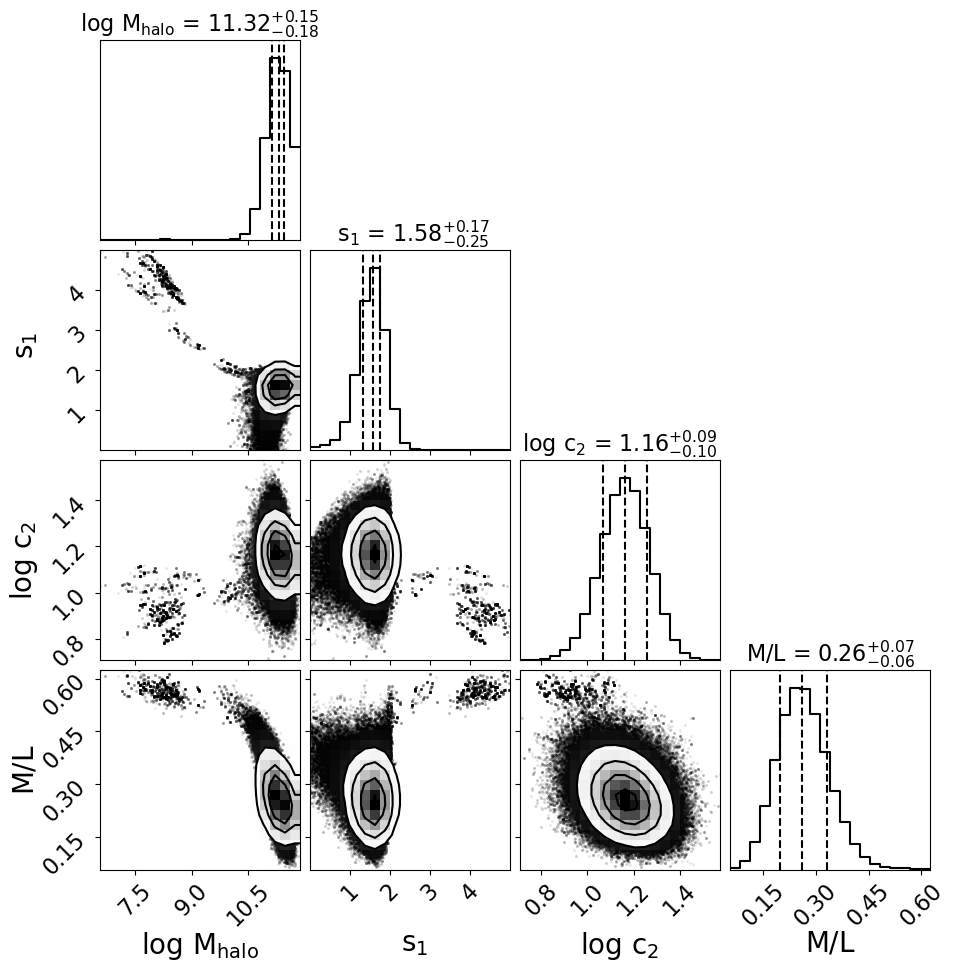}
\includegraphics[width=7.cm]{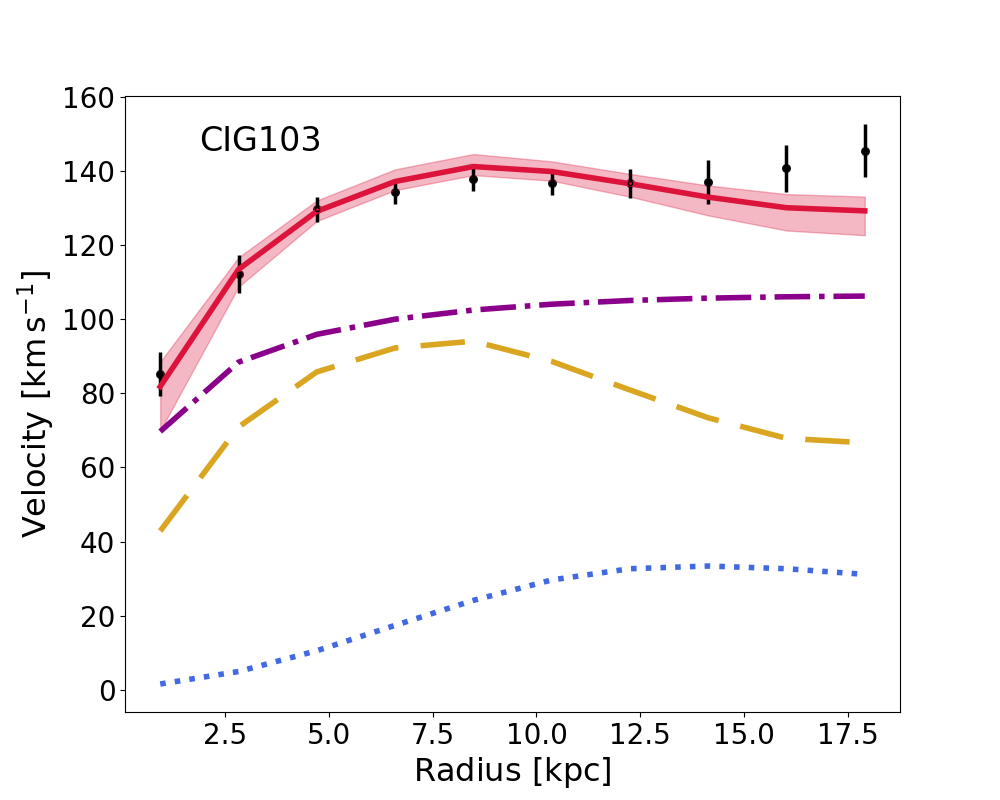}\\
    \includegraphics[width=6.cm]{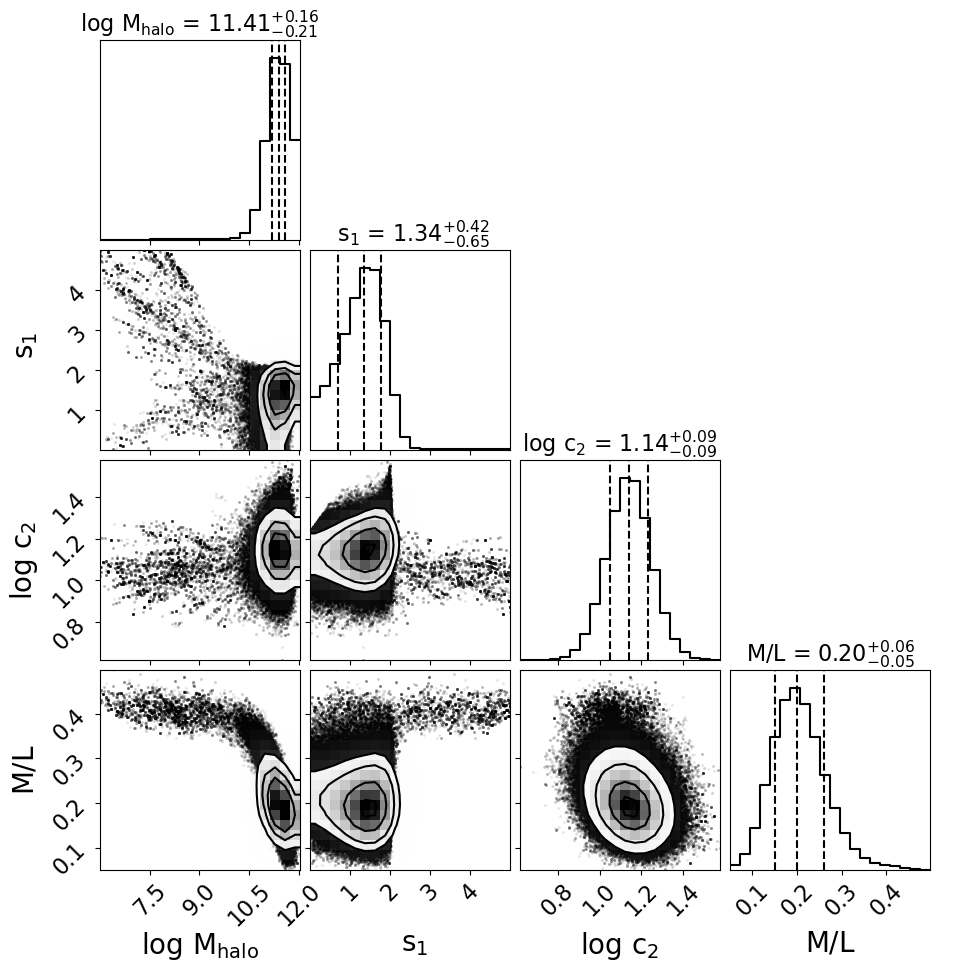}
\includegraphics[width=7.cm]{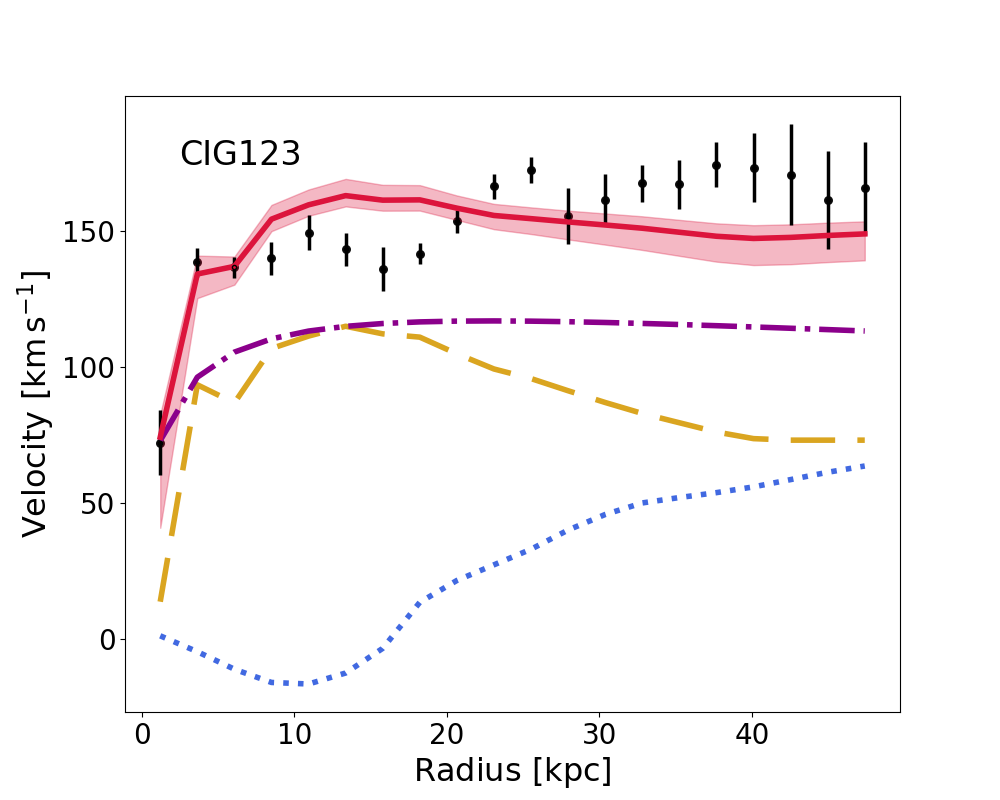}\\
  \includegraphics[width=6.cm]{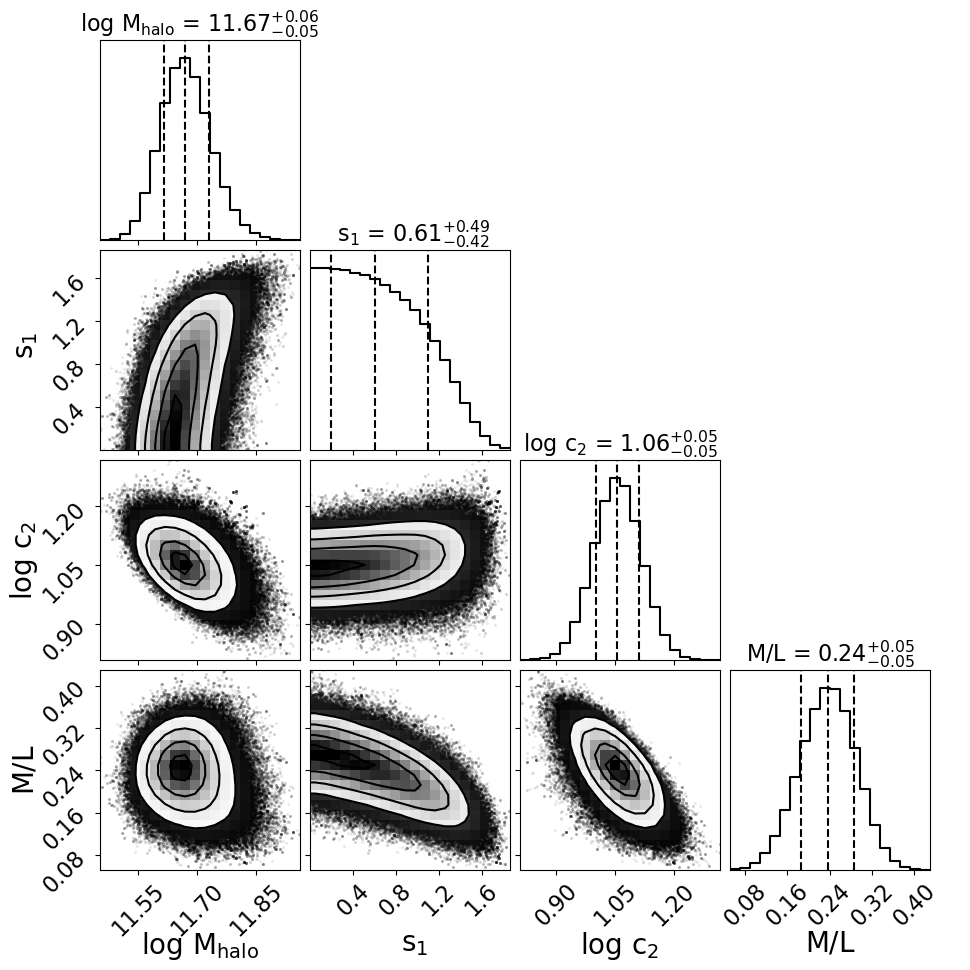}
  \includegraphics[width=7.cm]{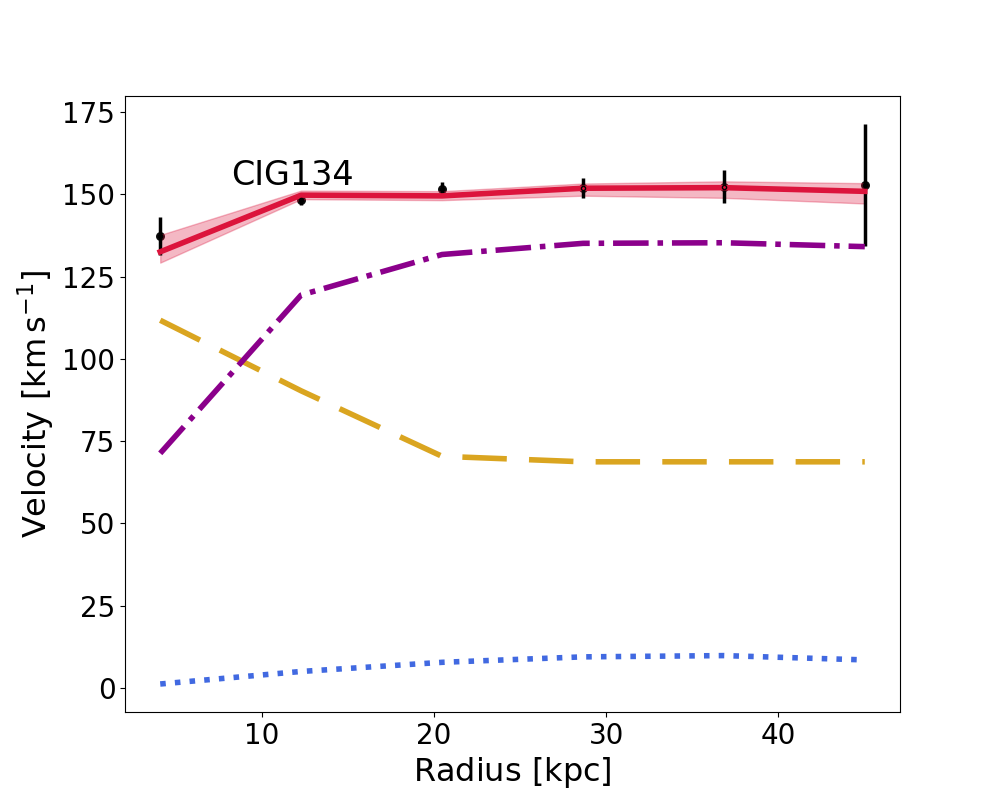}\\
  \includegraphics[width=6.cm]{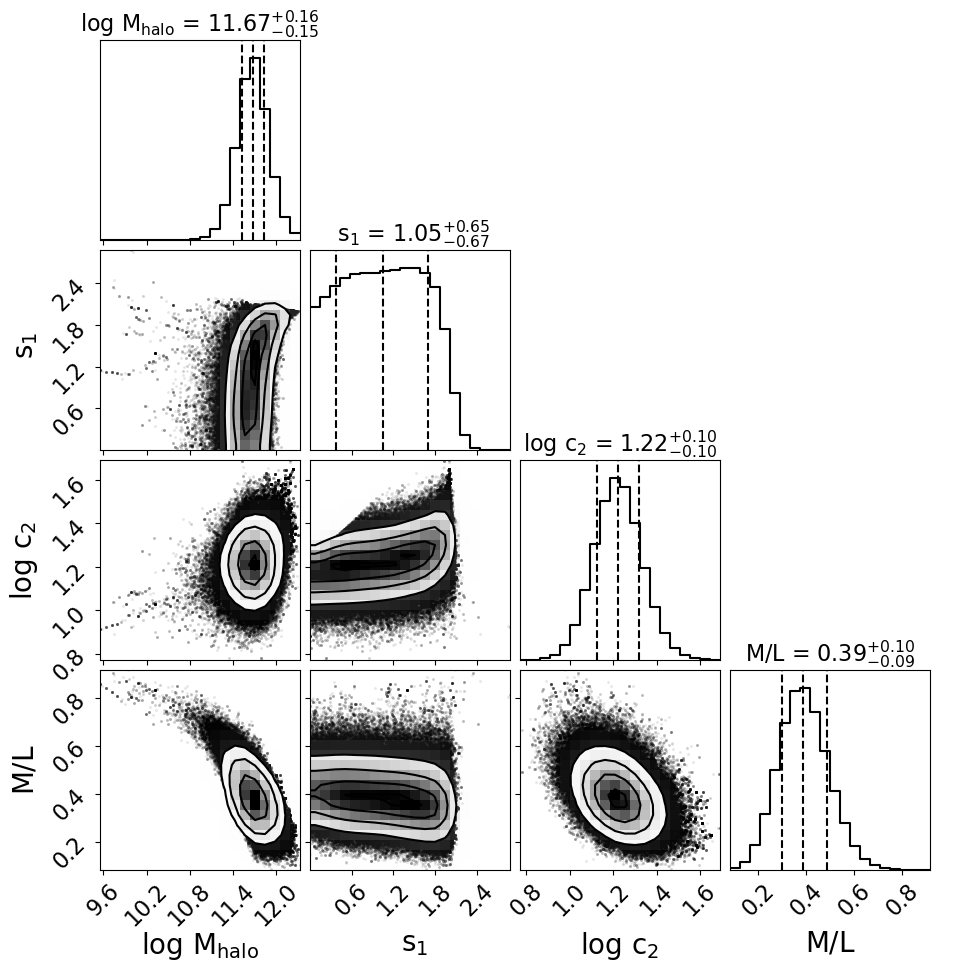}
  \includegraphics[width=7.cm]{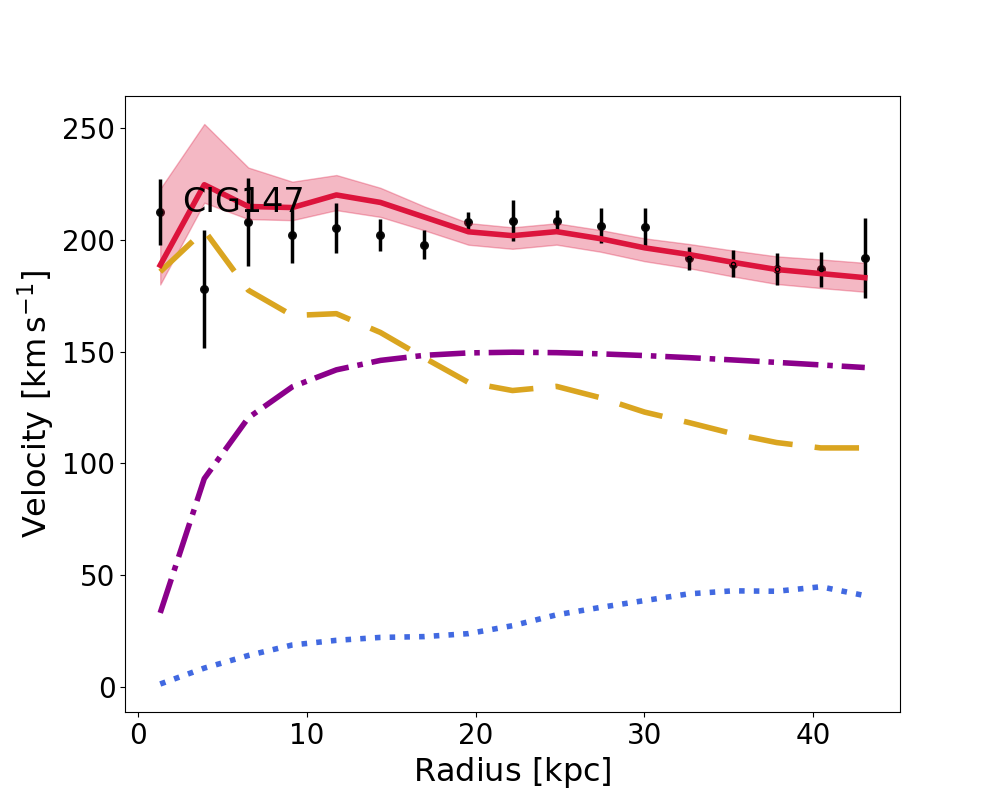}\\
     \caption[]{Continued}
\end{figure*}

 \addtocounter{figure}{-1}
\begin{figure*}
    \centering  
  \includegraphics[width=6.cm]{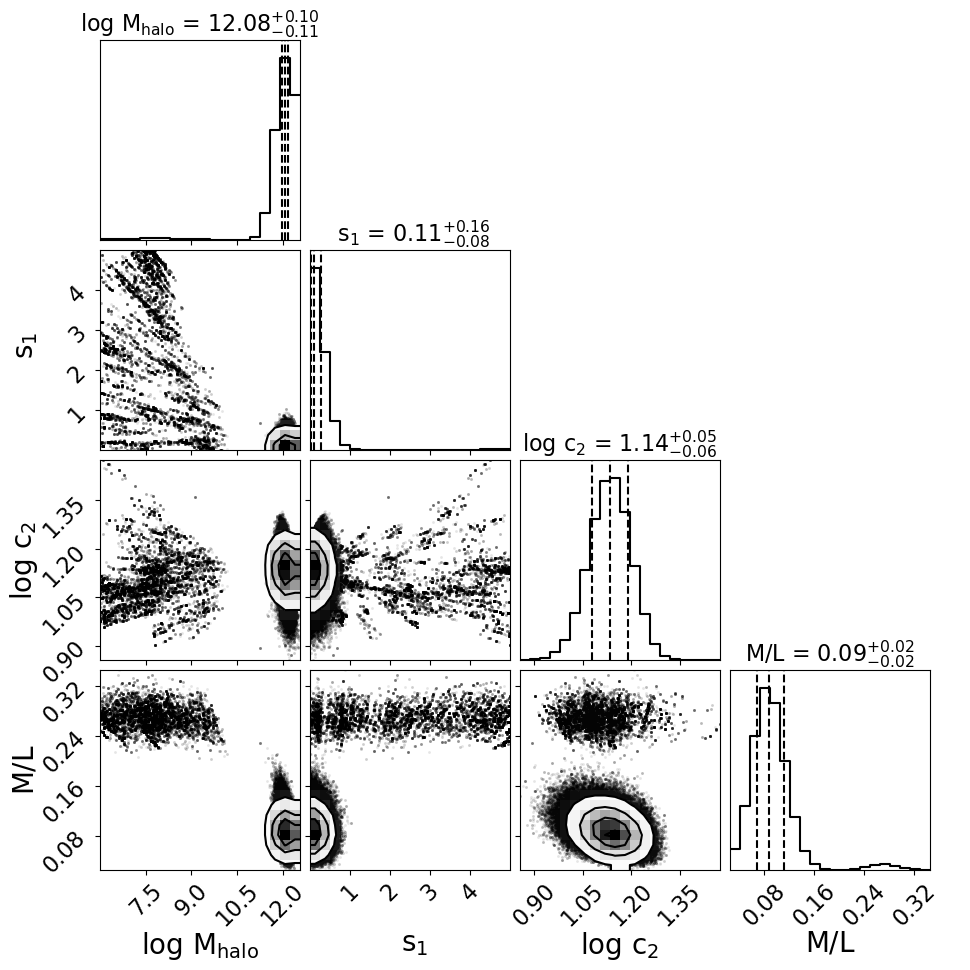}
  \includegraphics[width=7.cm]{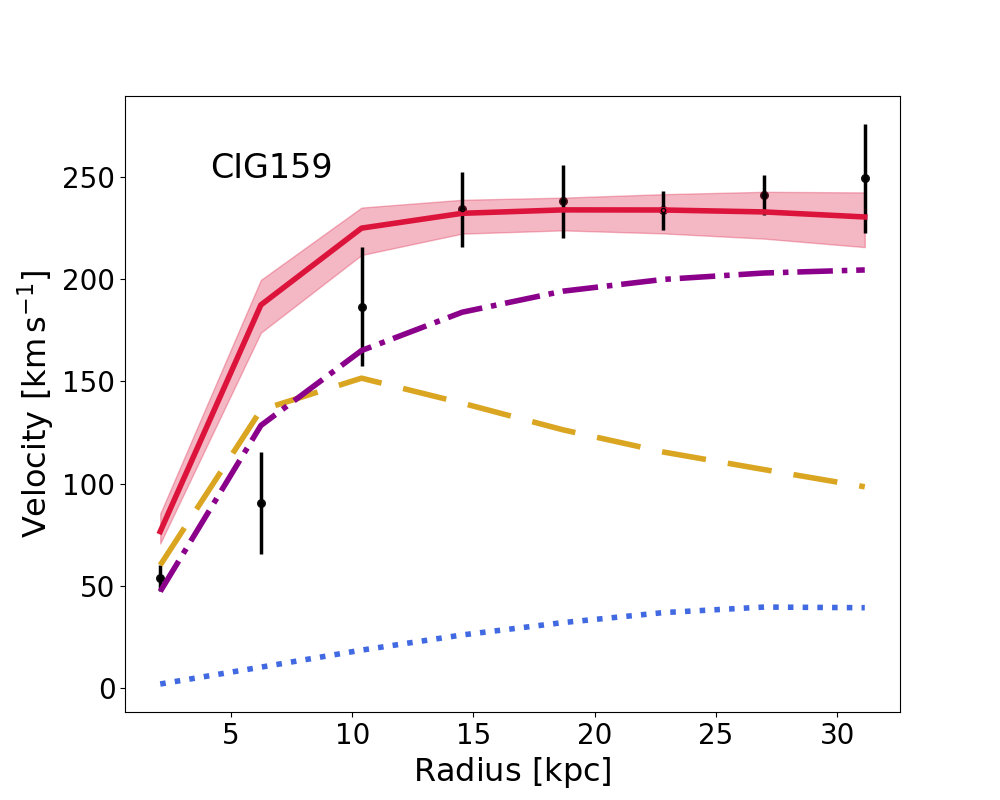}\\
       \includegraphics[width=6.cm]{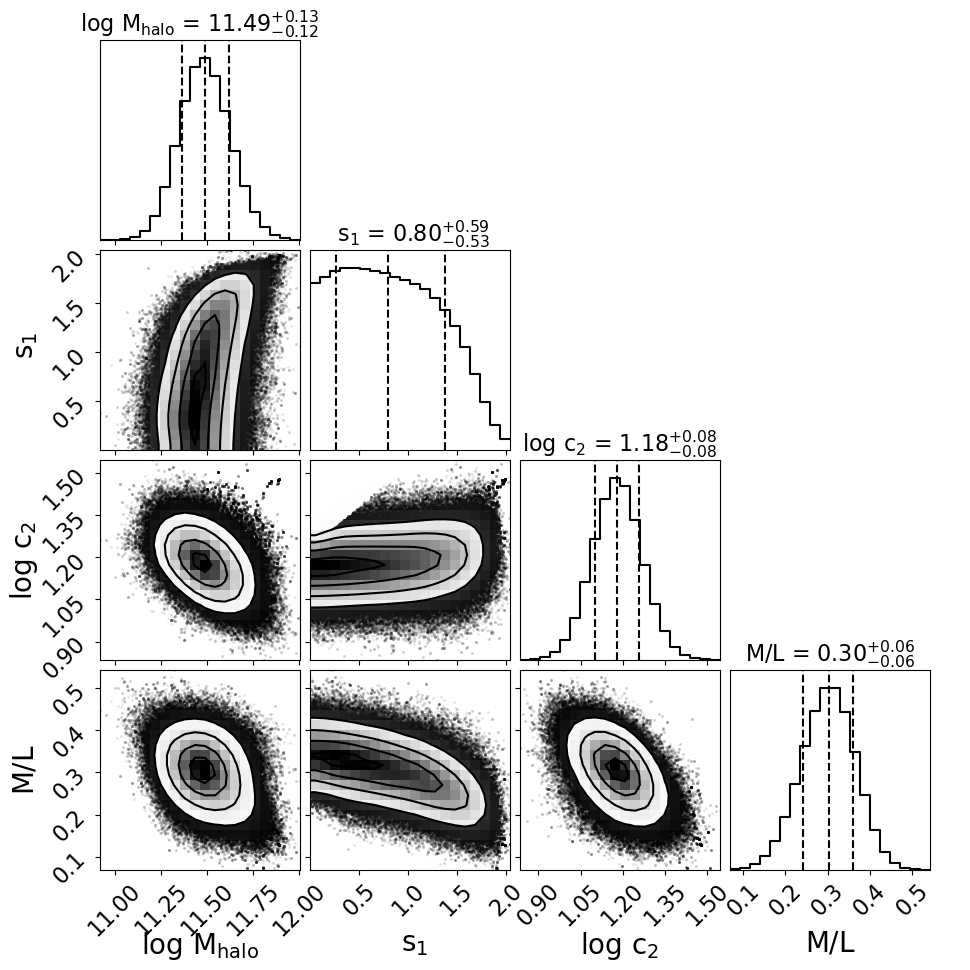}
 \includegraphics[width=7.cm]{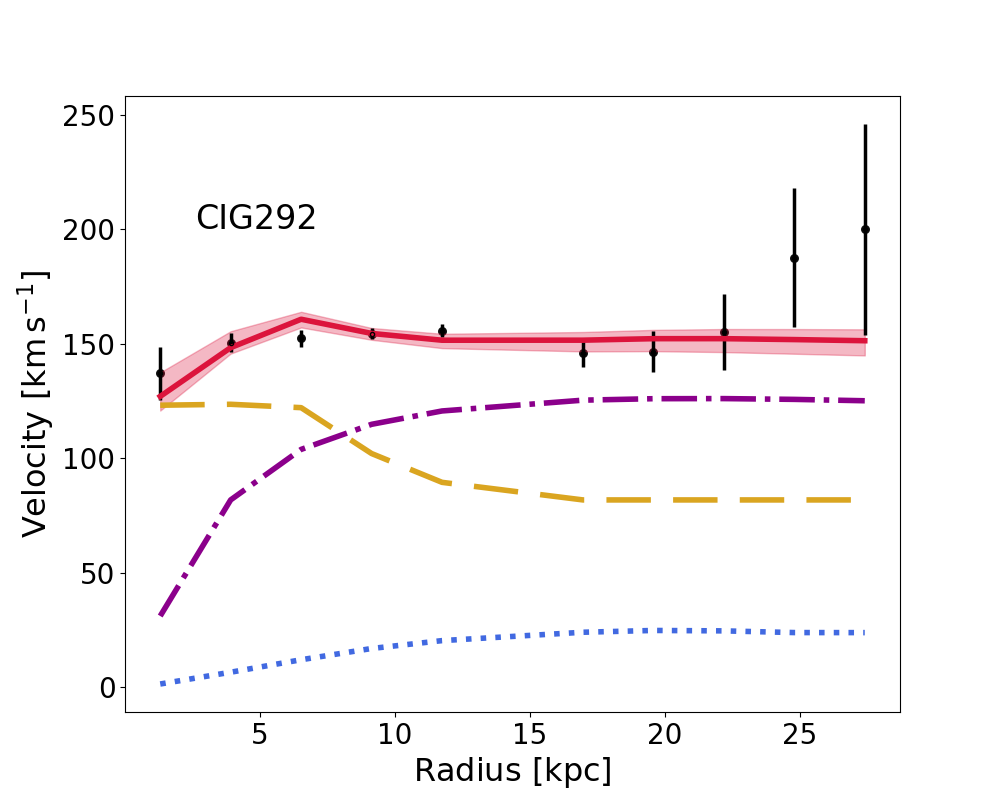}\\
   \includegraphics[width=6.cm]{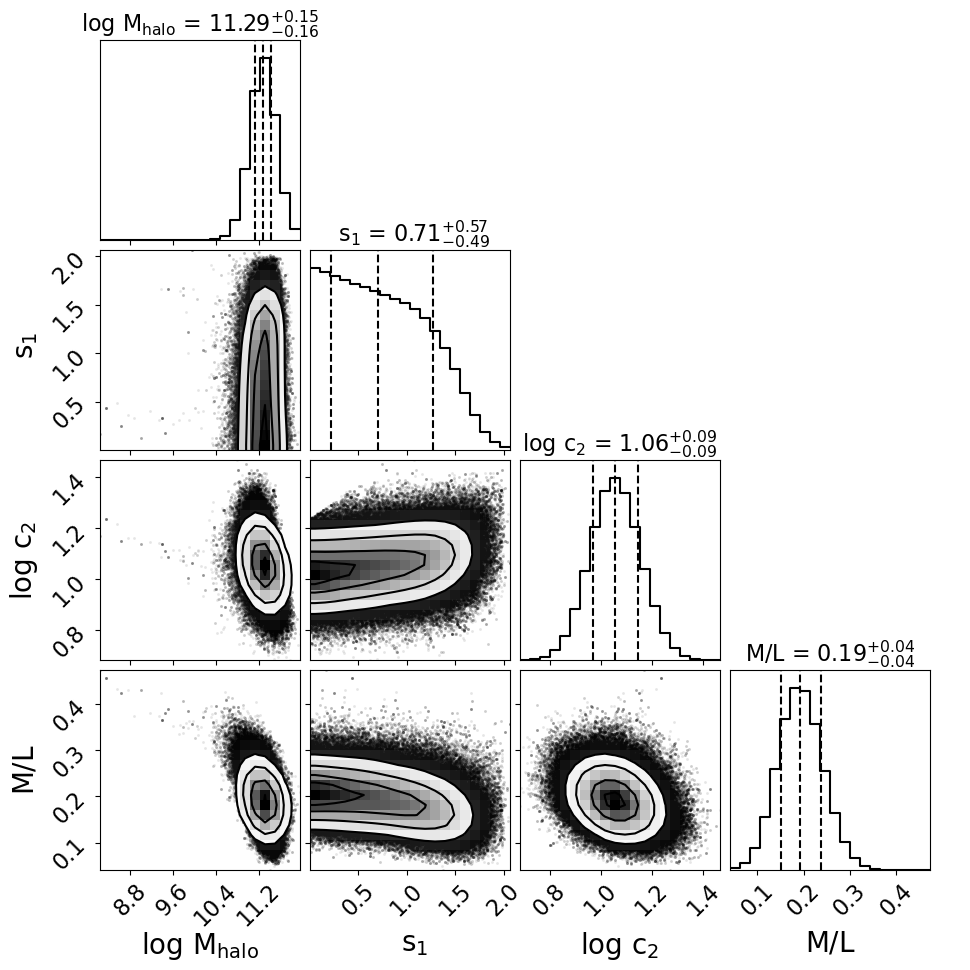}
 \includegraphics[width=7.cm]{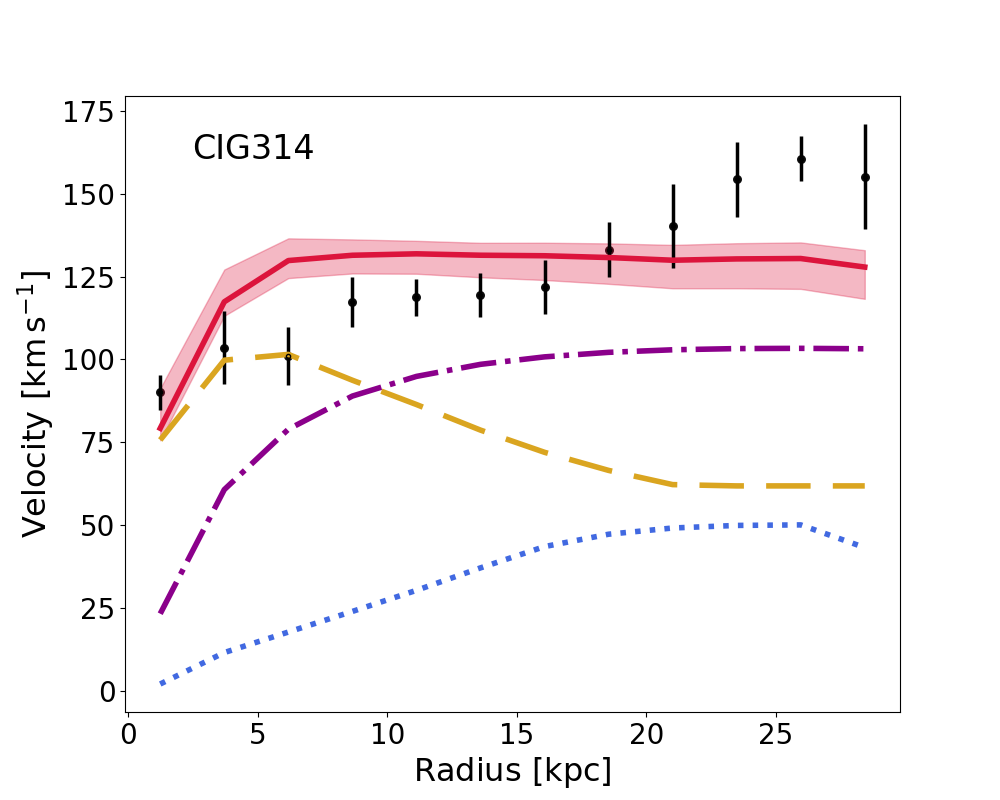}\\
   \includegraphics[width=6.cm]{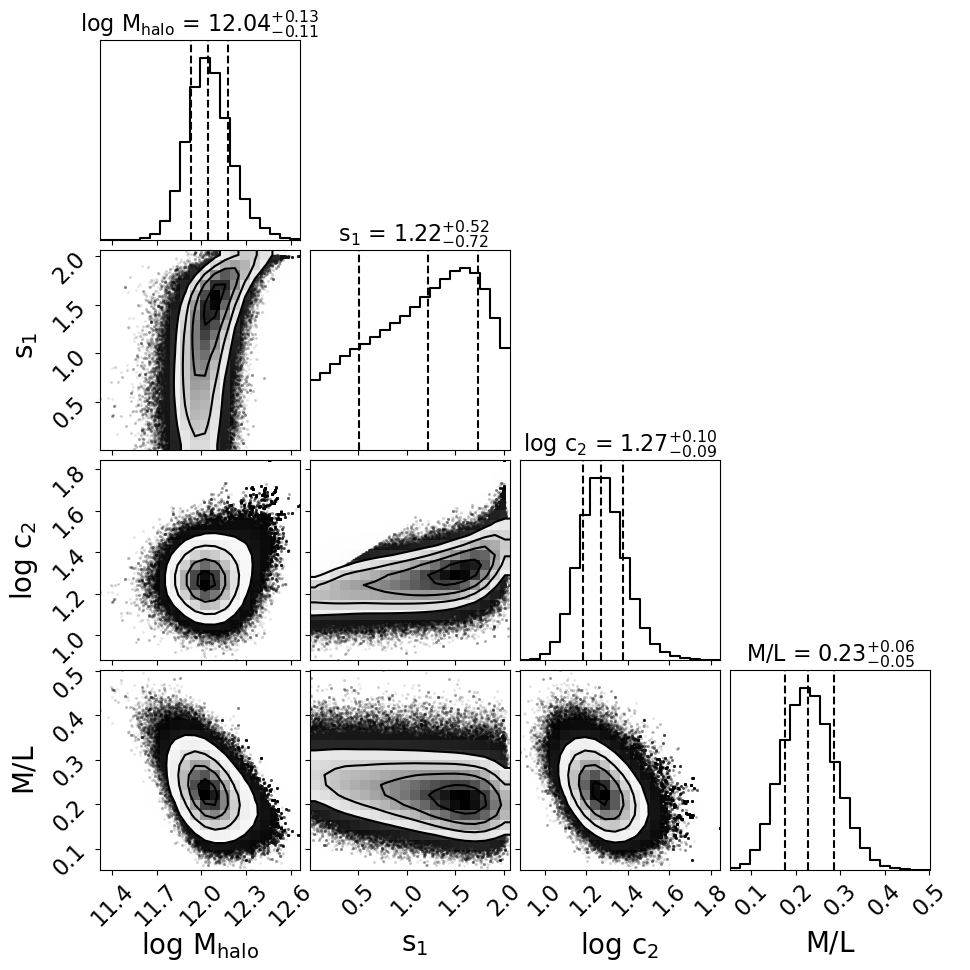}
 \includegraphics[width=7.cm]{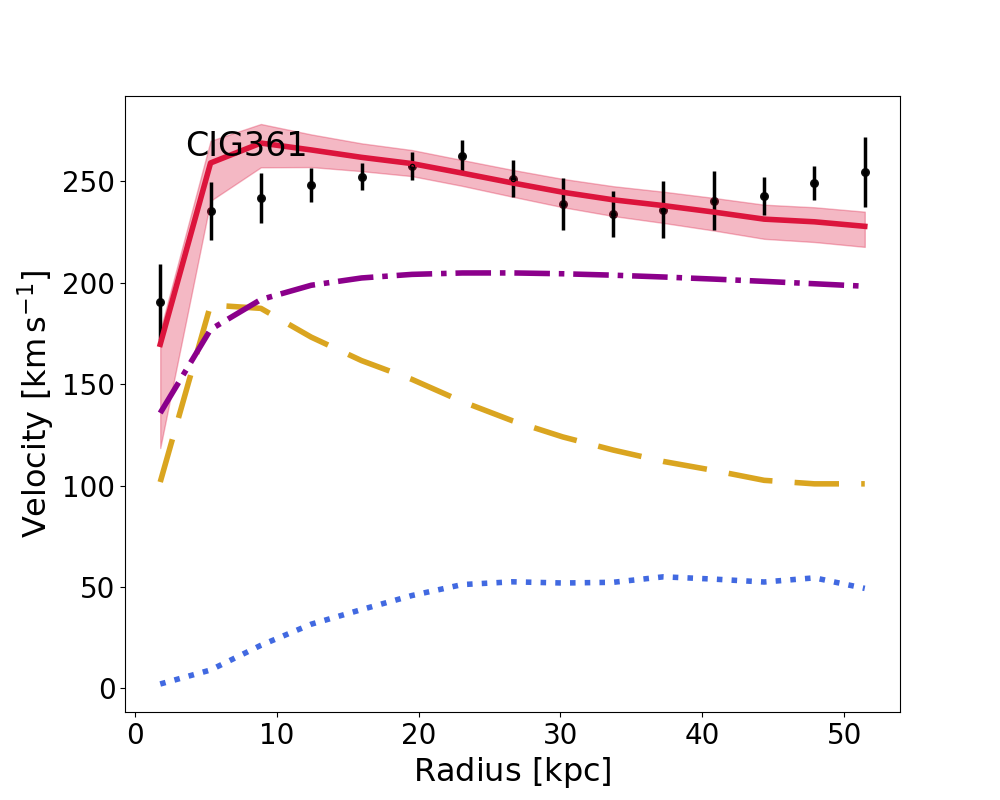}\\
      \caption[]{Continued}
\end{figure*}

 \addtocounter{figure}{-1}
\begin{figure*}
    \centering  
  \includegraphics[width=6.cm]{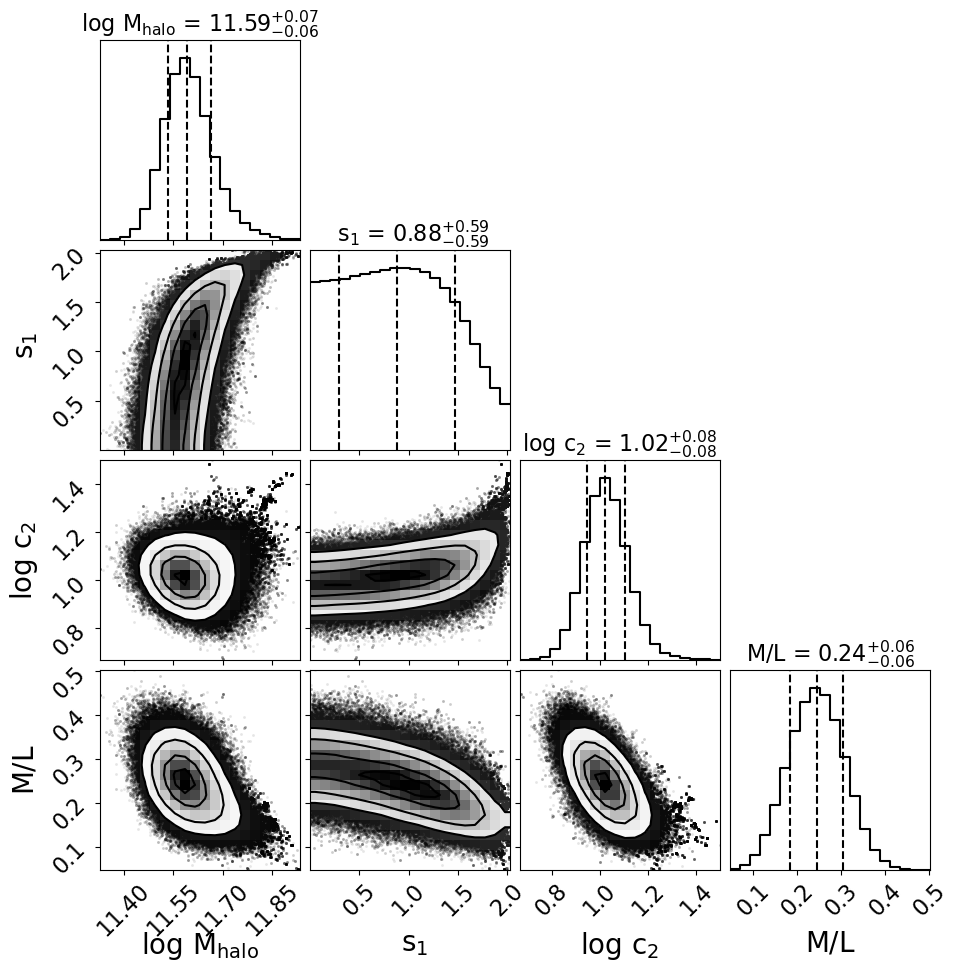}
  \includegraphics[width=7.cm]{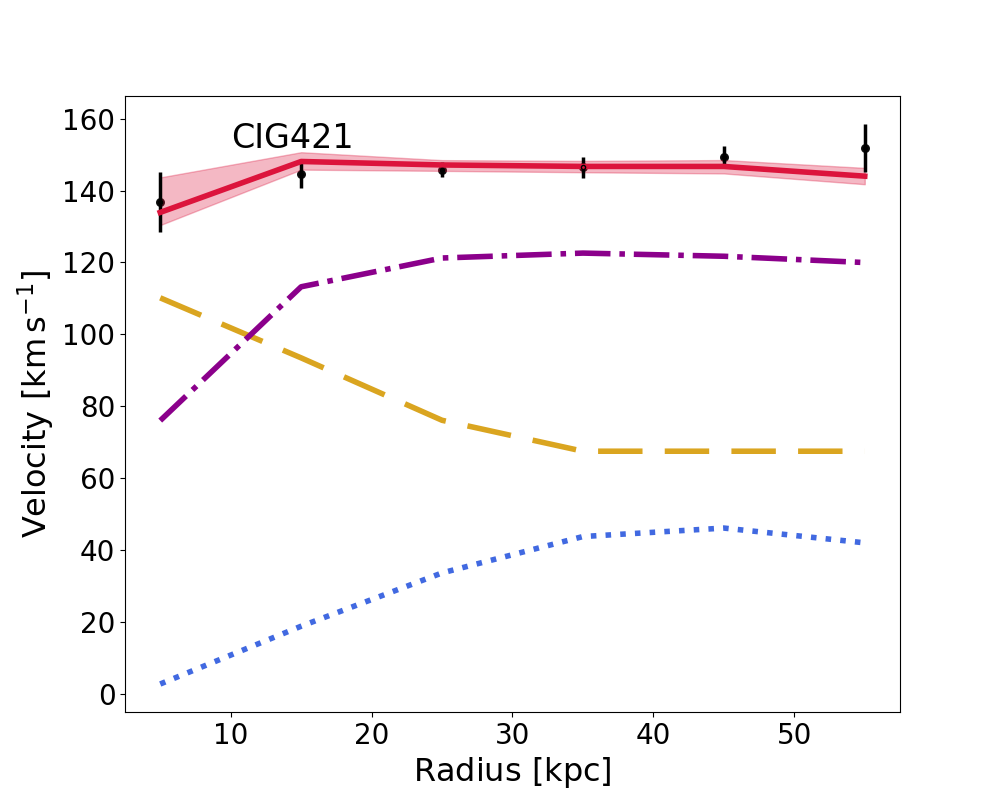}\\ \includegraphics[width=6.cm]{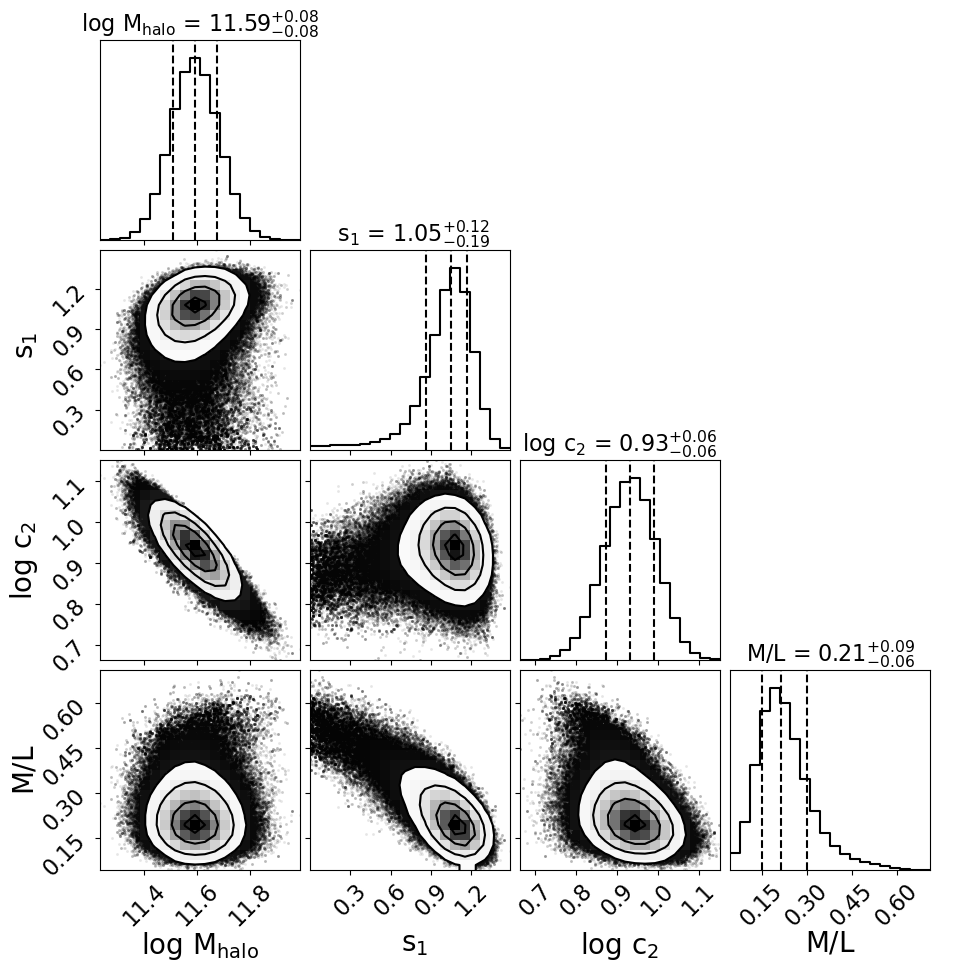}
 \includegraphics[width=7.cm]{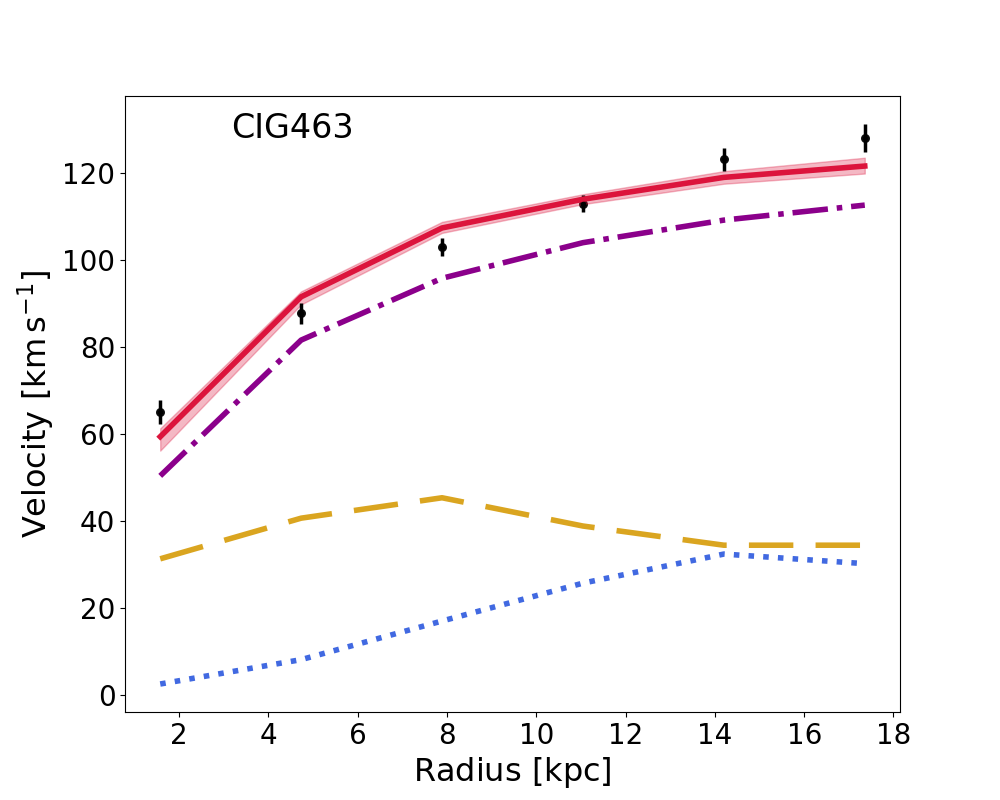}\\
   \includegraphics[width=6.cm]{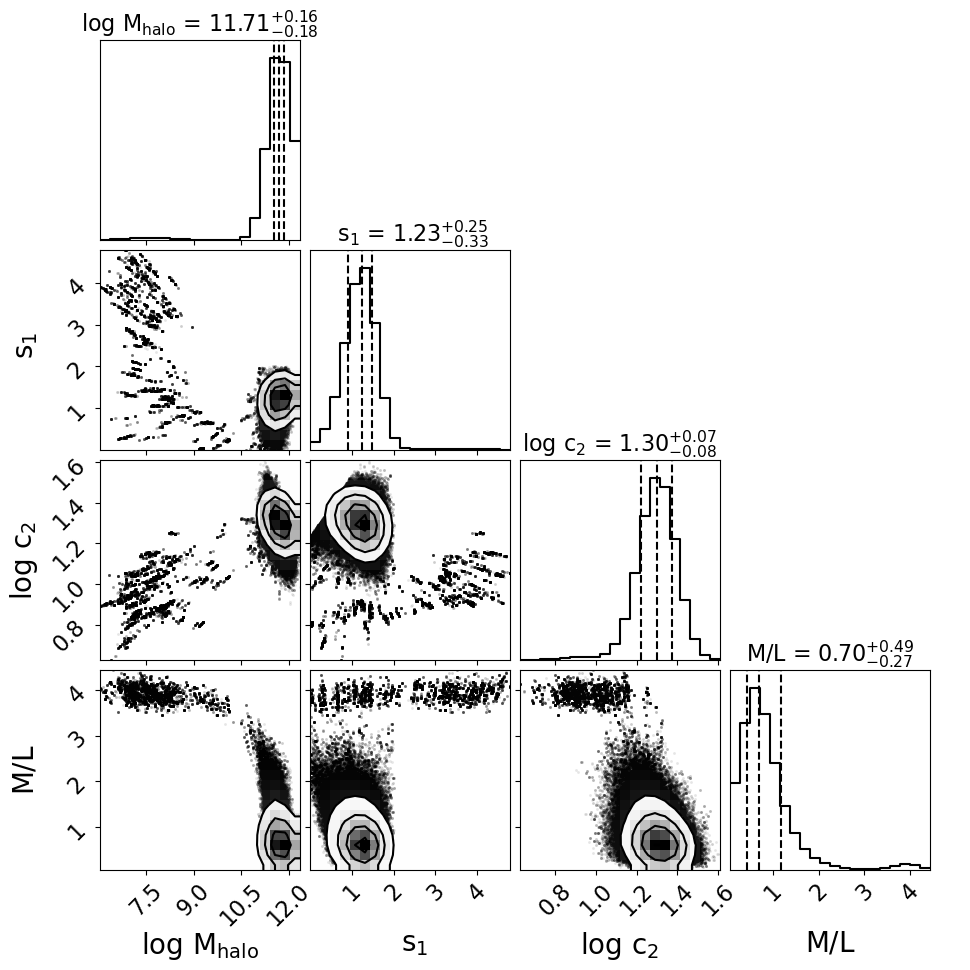}
 \includegraphics[width=7.cm]{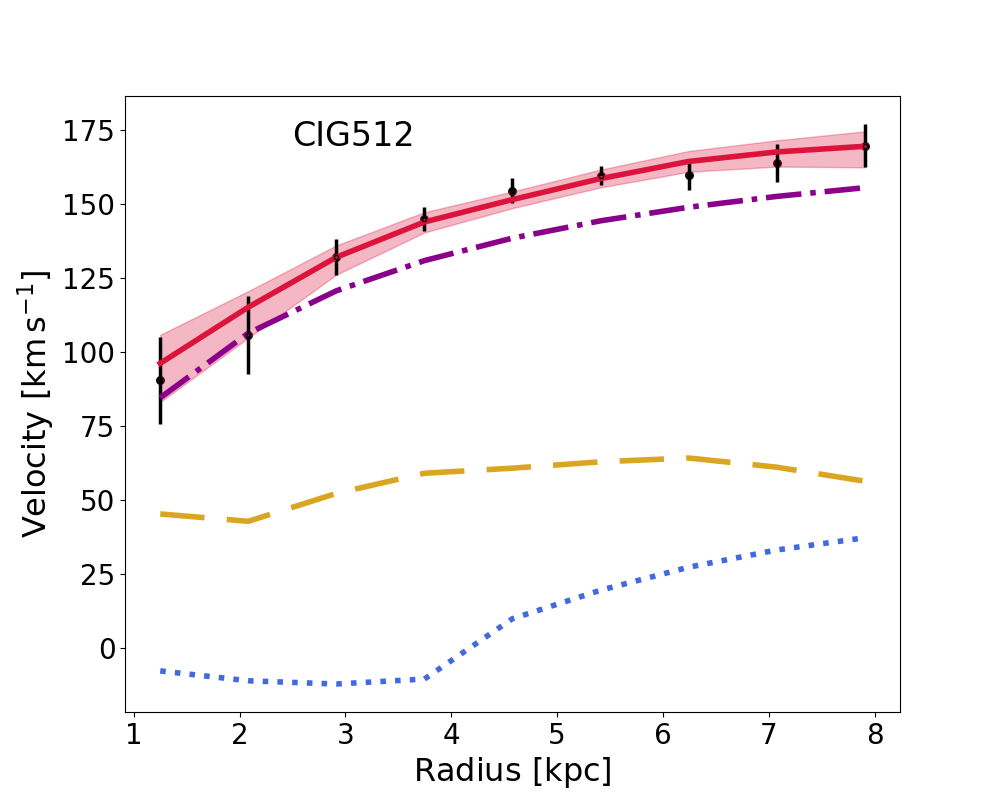}\\
  \includegraphics[width=6.cm]{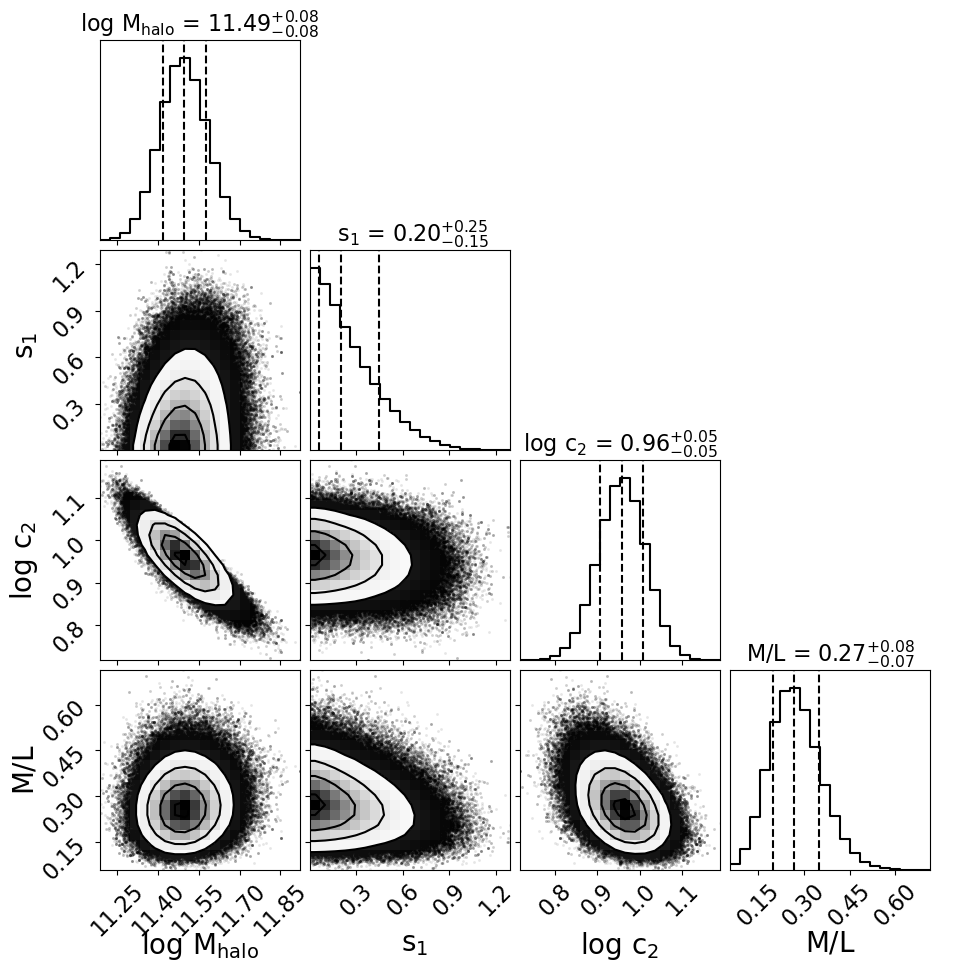}
  \includegraphics[width=7.cm]{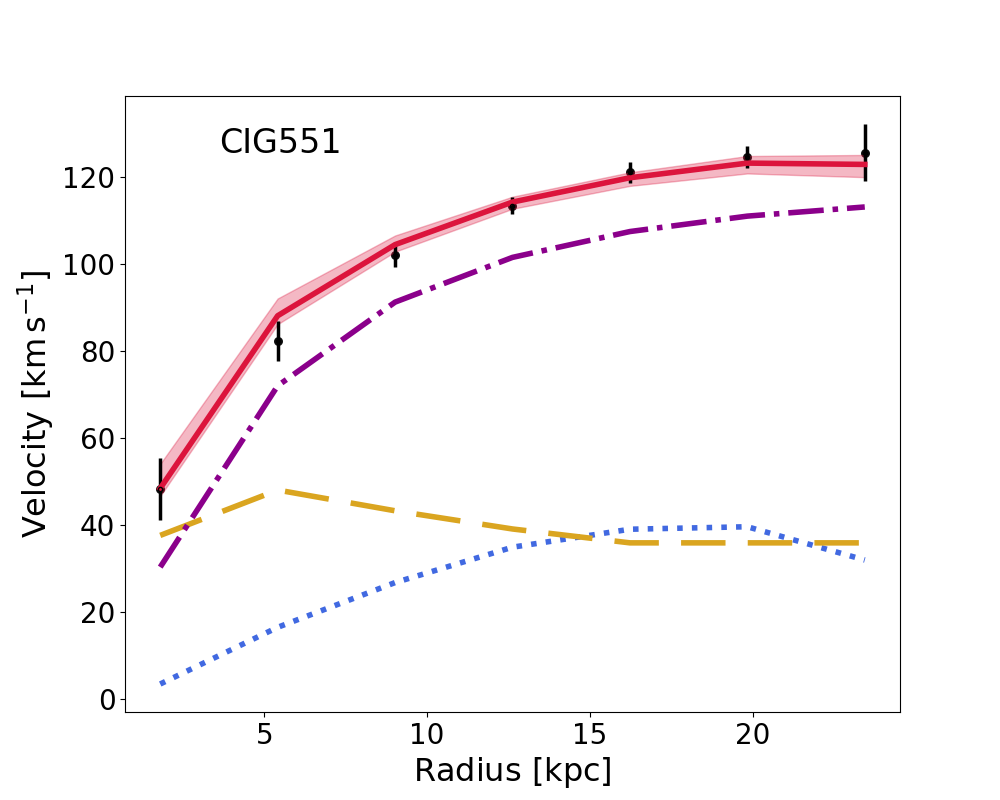}\\
      \caption[]{Continued}
\end{figure*}

 \addtocounter{figure}{-1}
\begin{figure*}
    \centering  
   \includegraphics[width=6.cm]{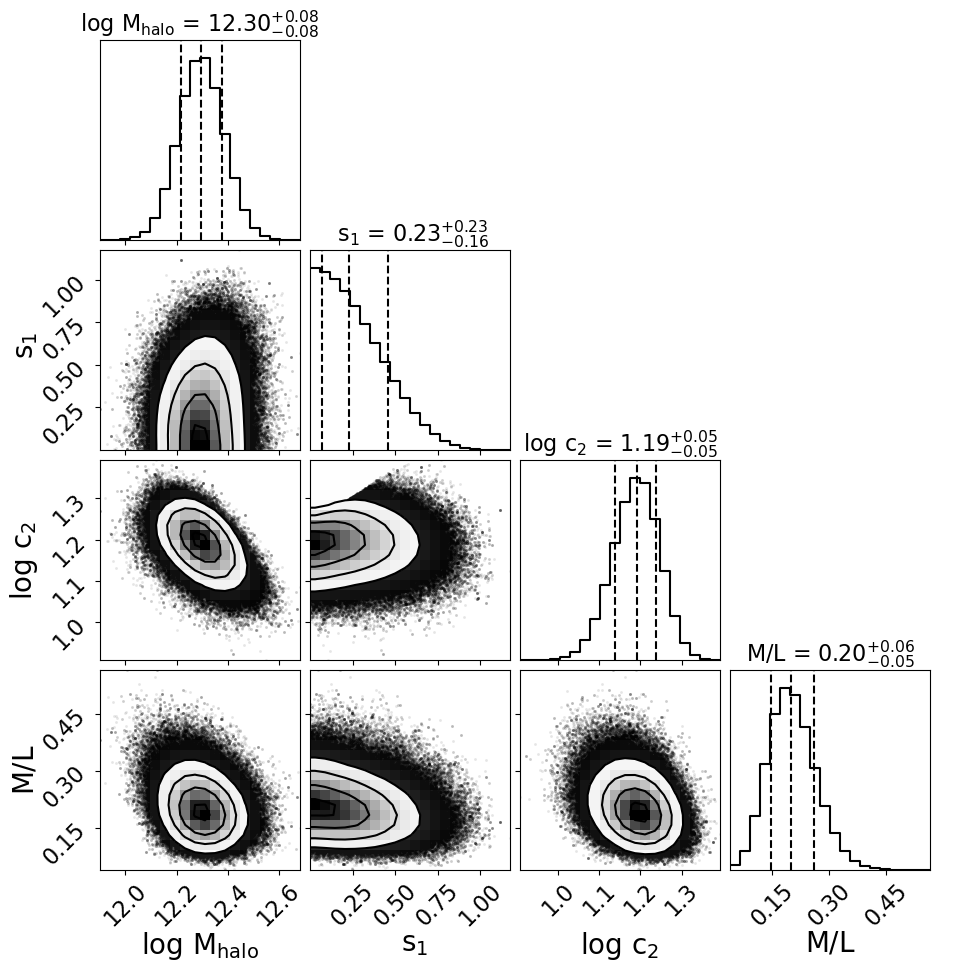}
 \includegraphics[width=7.cm]{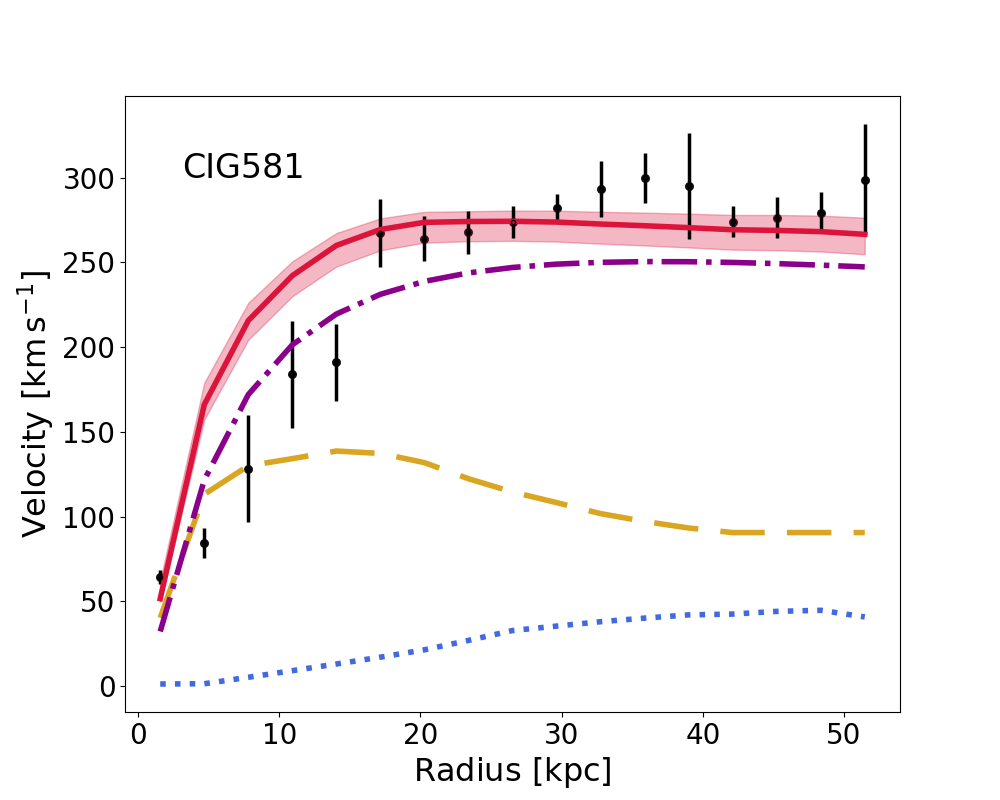}\\
  \includegraphics[width=6.cm]{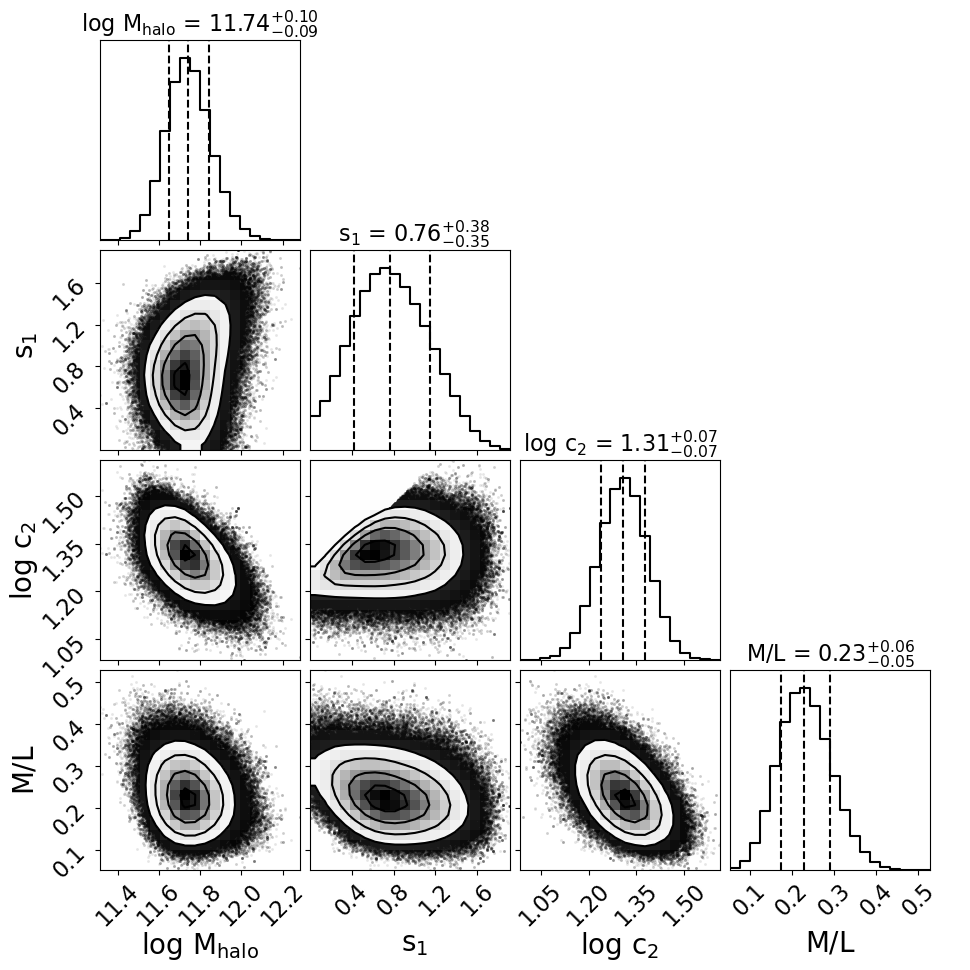}
  \includegraphics[width=7.cm]{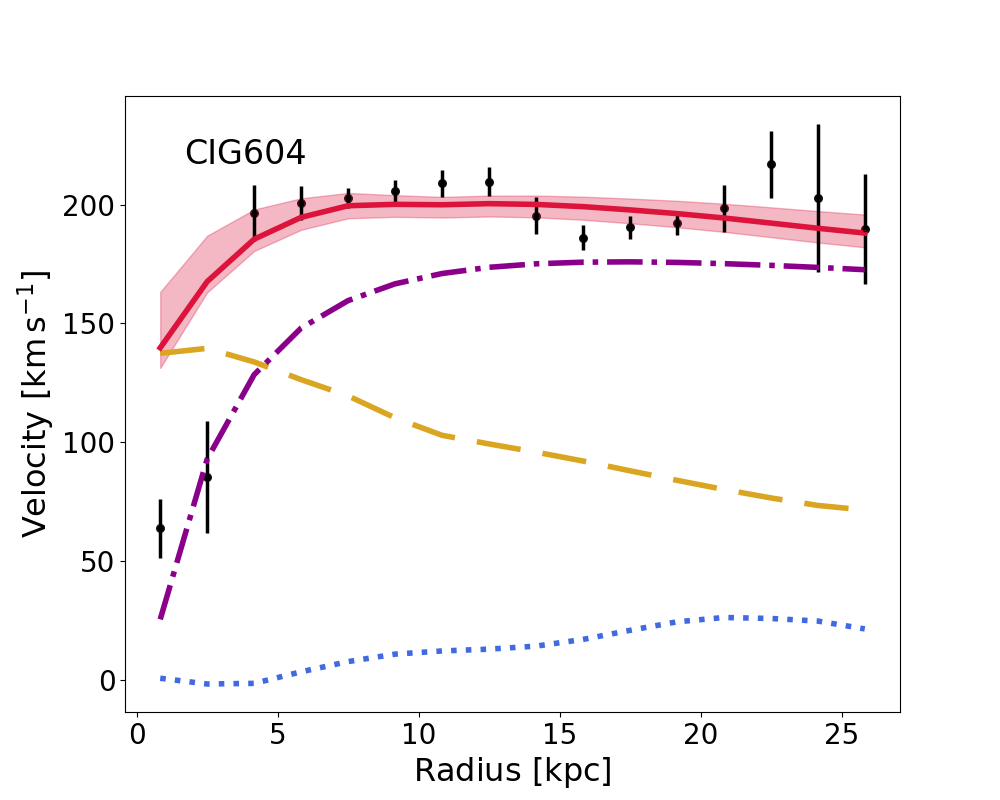}\\
  \includegraphics[width=6.cm]{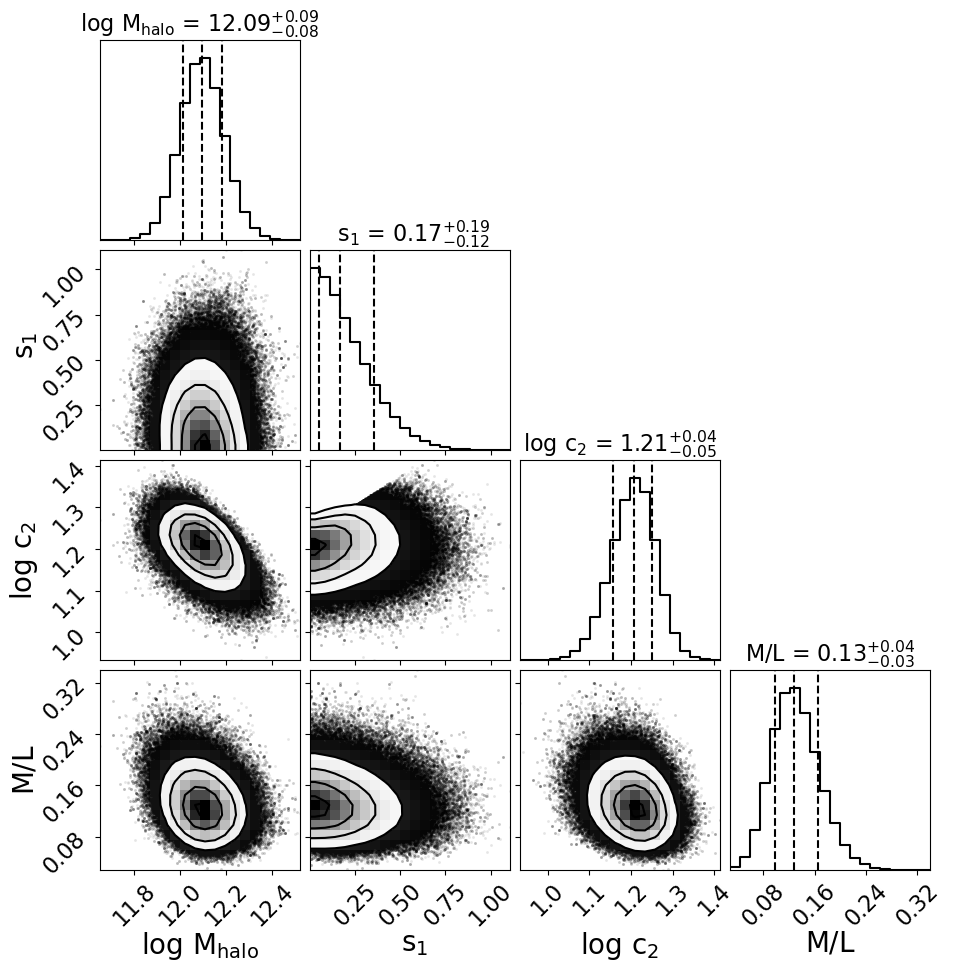}
  \includegraphics[width=7.cm]{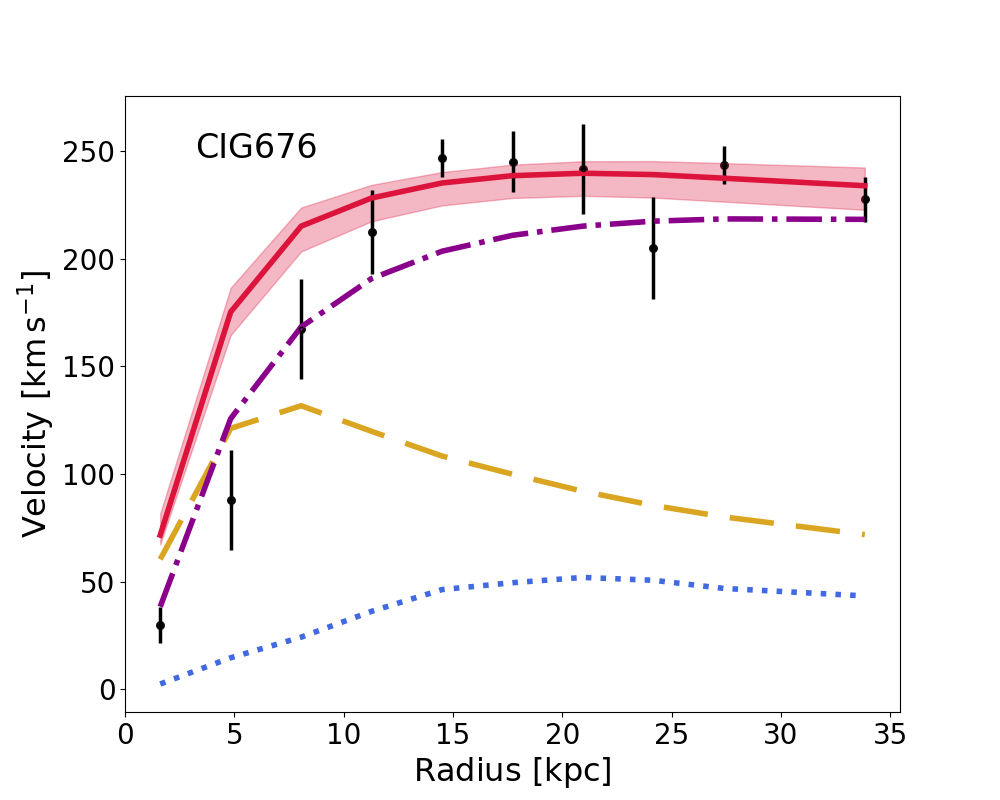}\\
  \includegraphics[width=6.cm]{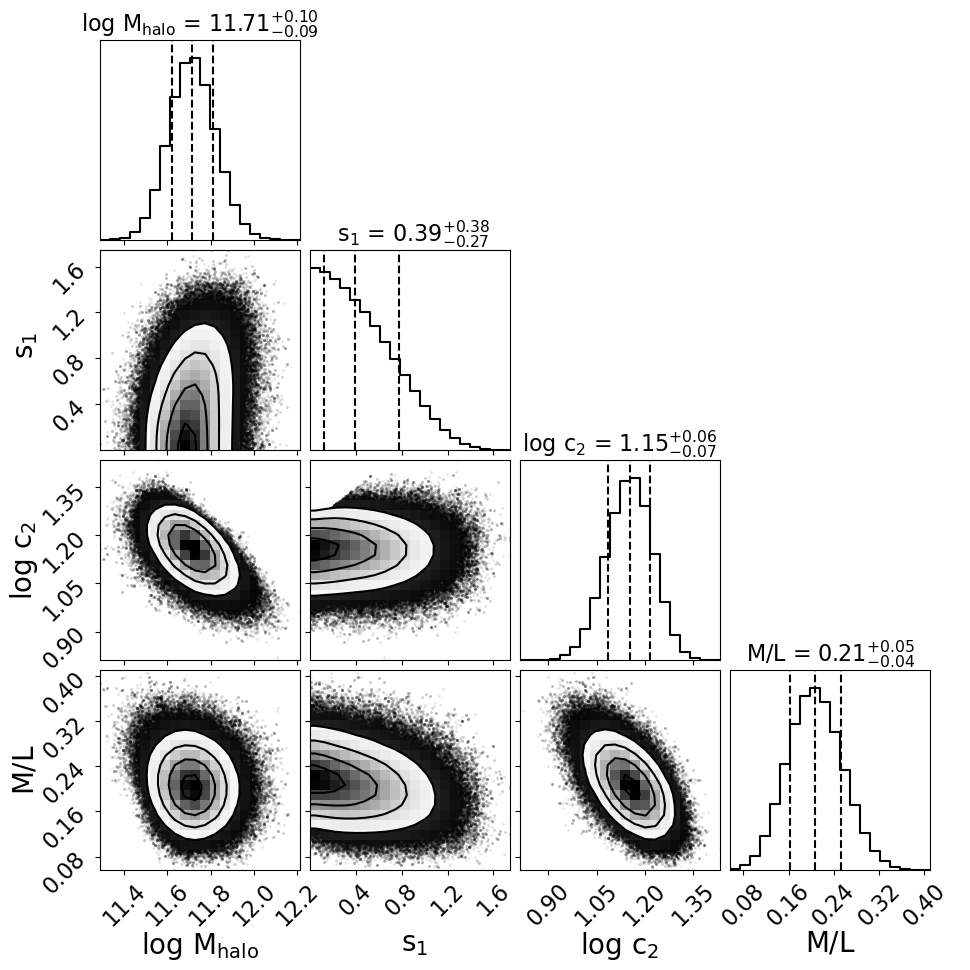}
  \includegraphics[width=7.cm]{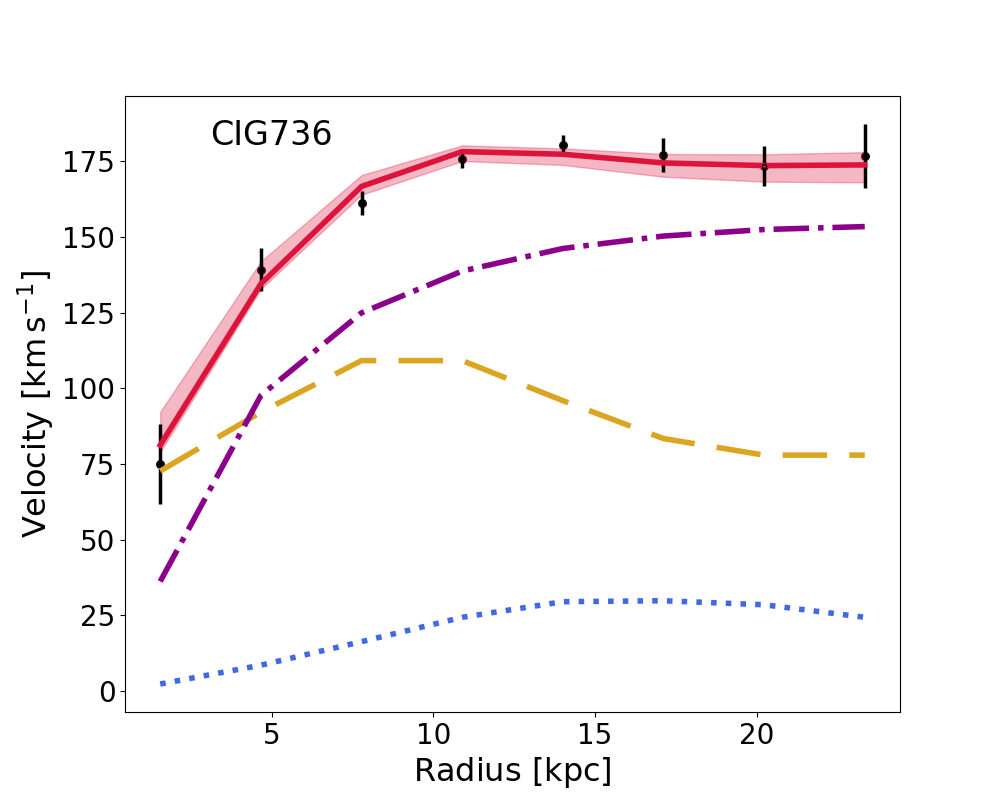}\\
      \caption[]{Continued}
\end{figure*}
  
 \addtocounter{figure}{-1}
\begin{figure*}
    \centering  
  \includegraphics[width=6.cm]{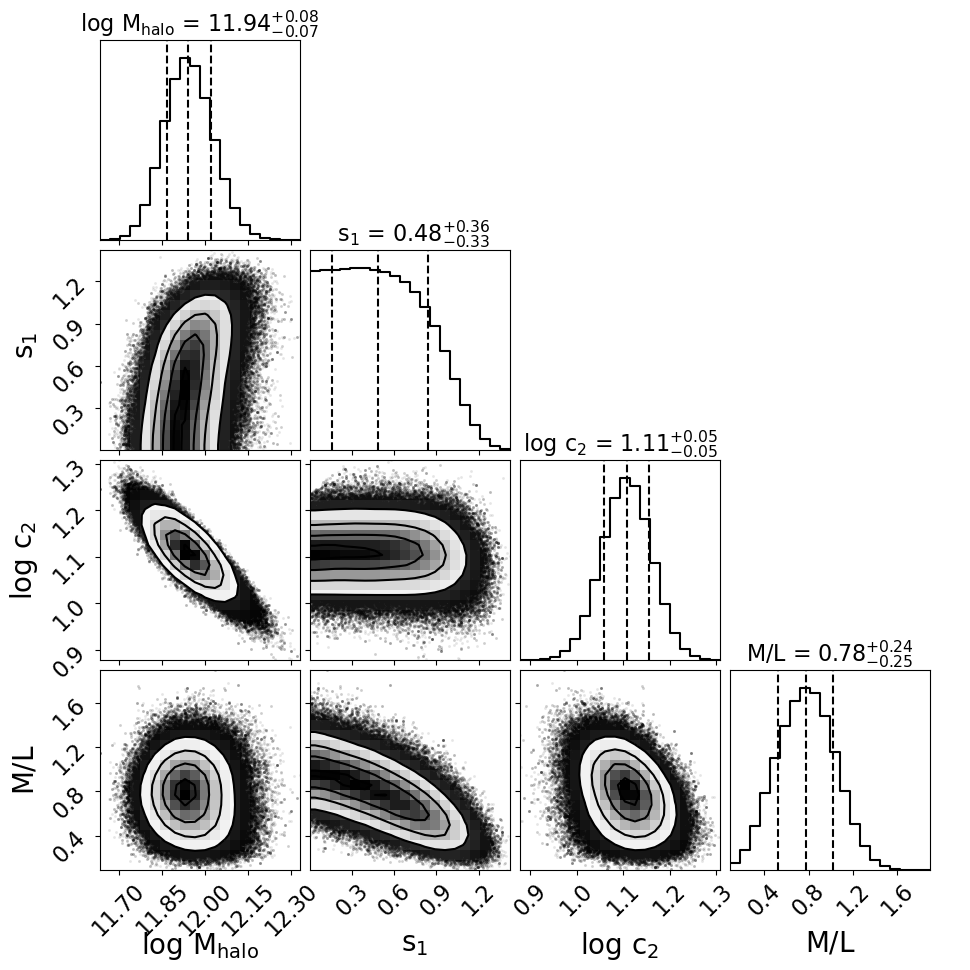}
  \includegraphics[width=7.cm]{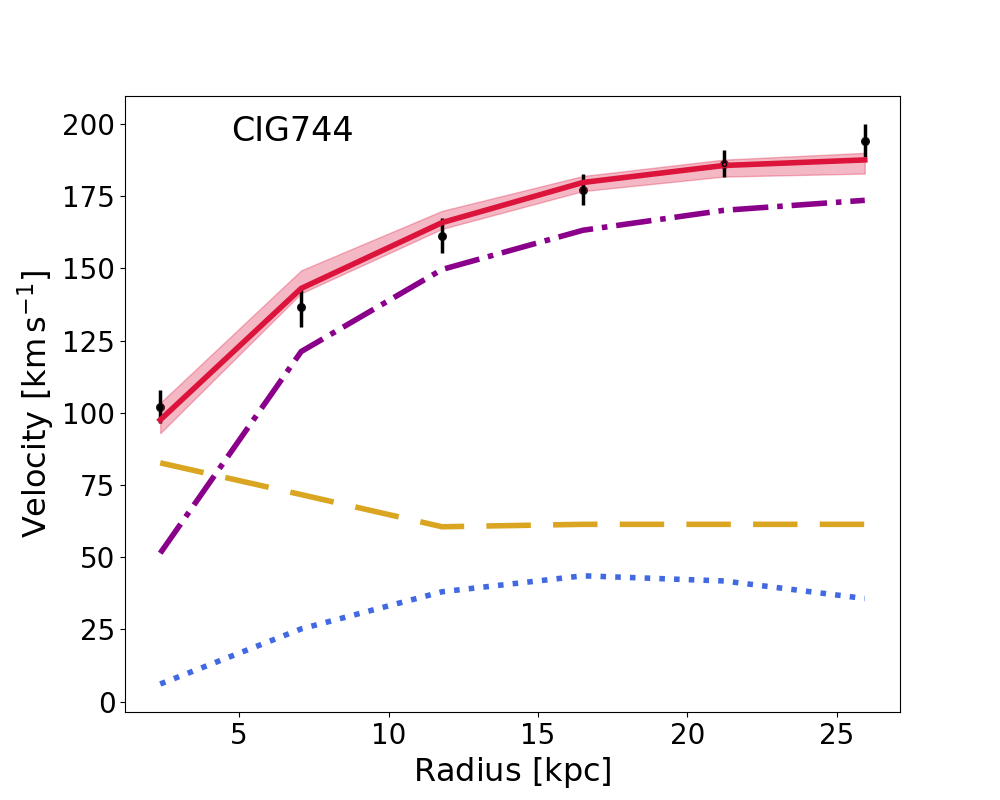}\\
  \includegraphics[width=6.cm]{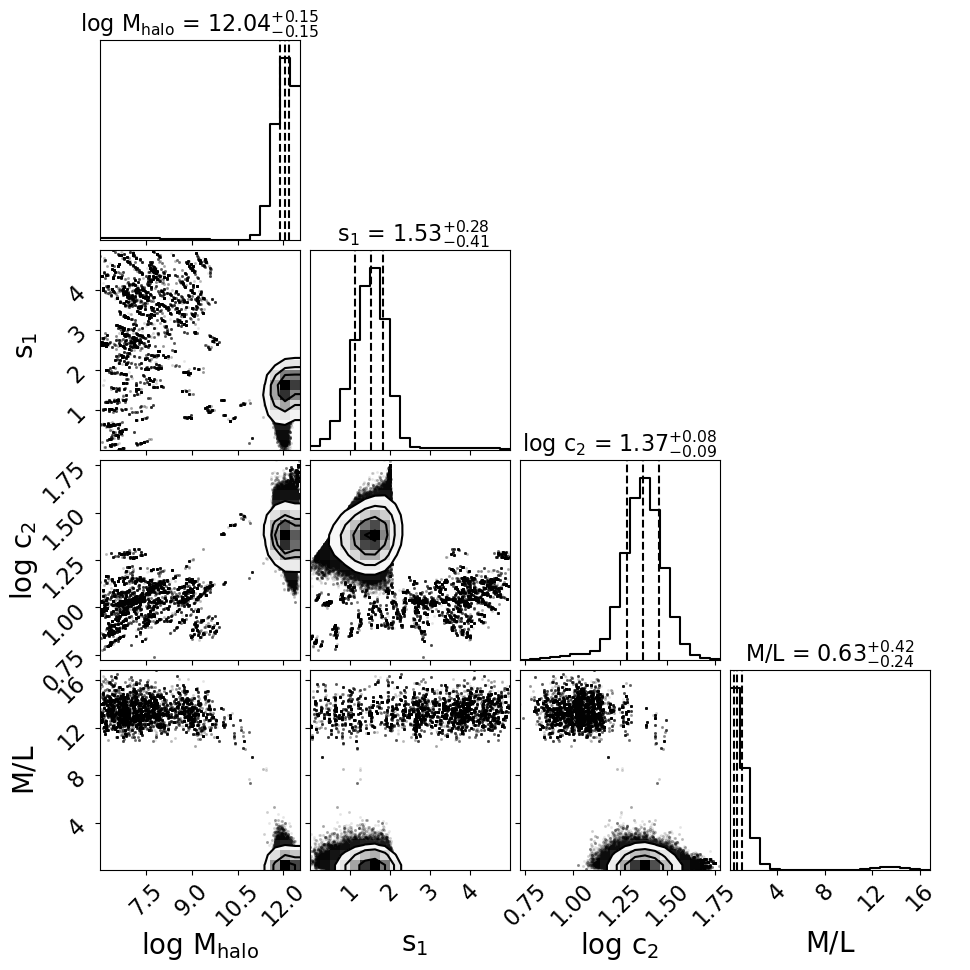}
  \includegraphics[width=7.cm]{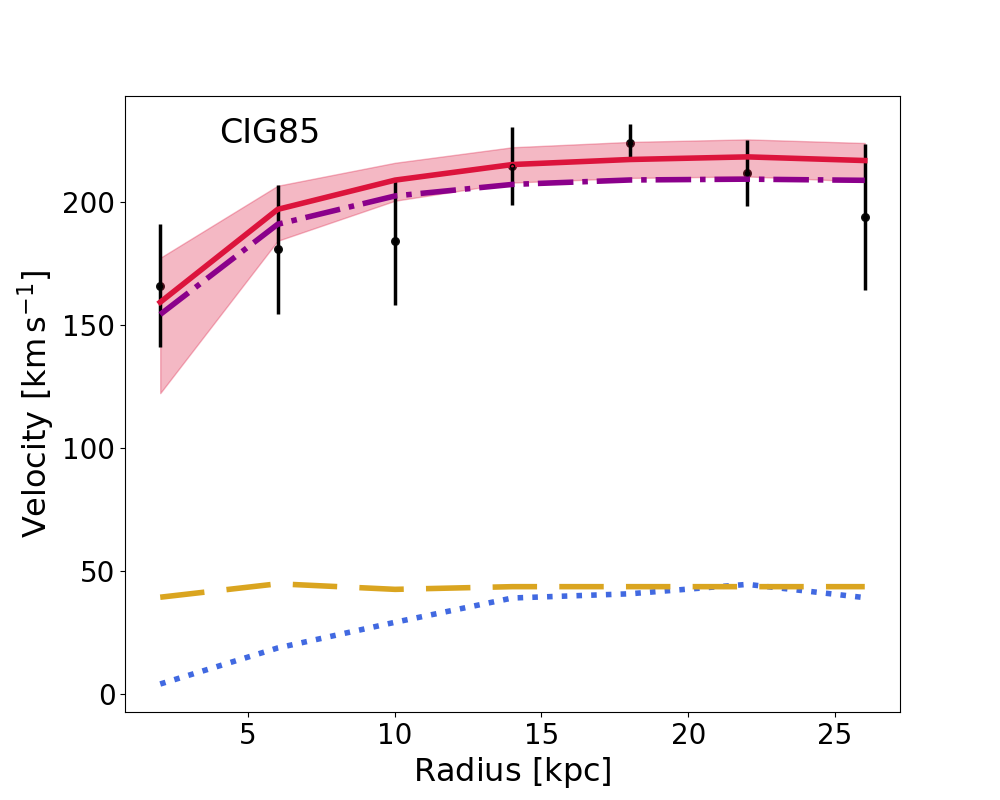}\\
   \includegraphics[width=6.cm]{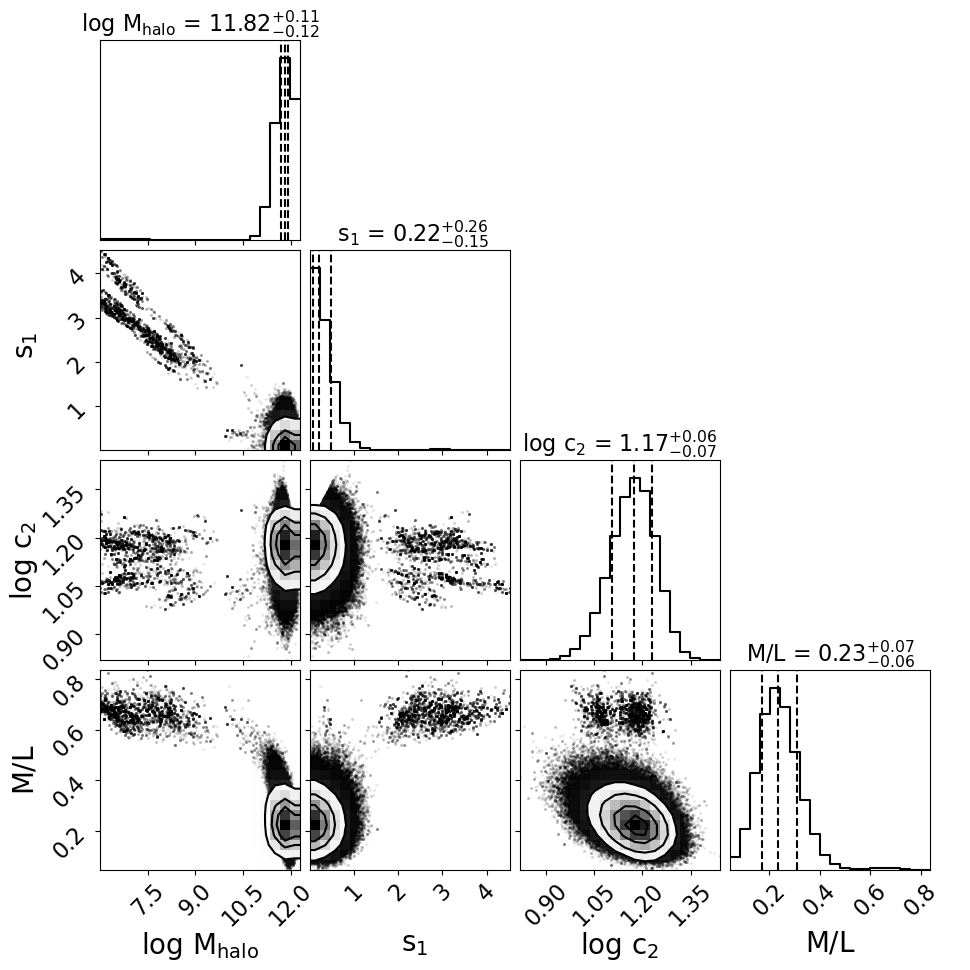}
   \includegraphics[width=7.cm]{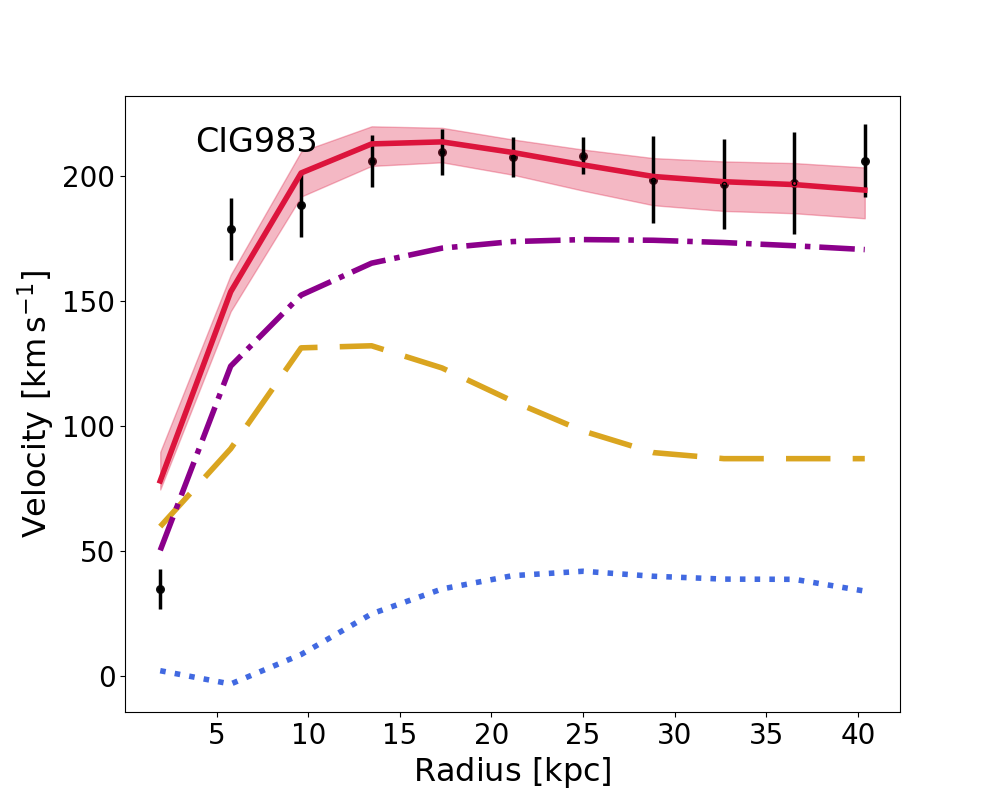}\\
      \caption[]{Continued.}
\end{figure*}

\clearpage
\section{Rotation curve fits for GHASP sample}  
\label{app:GHASP massmodel}

In Fig. \ref{fig:ghasp massmodel}, we present the model rotation curve fits (posterior distribution and rotation curve with its decomposition in different components) for GHASP sample using the Dekel-Zhao (DZ) parametrisation.

\begin{figure*}
\centering
   \includegraphics[width=6.cm]{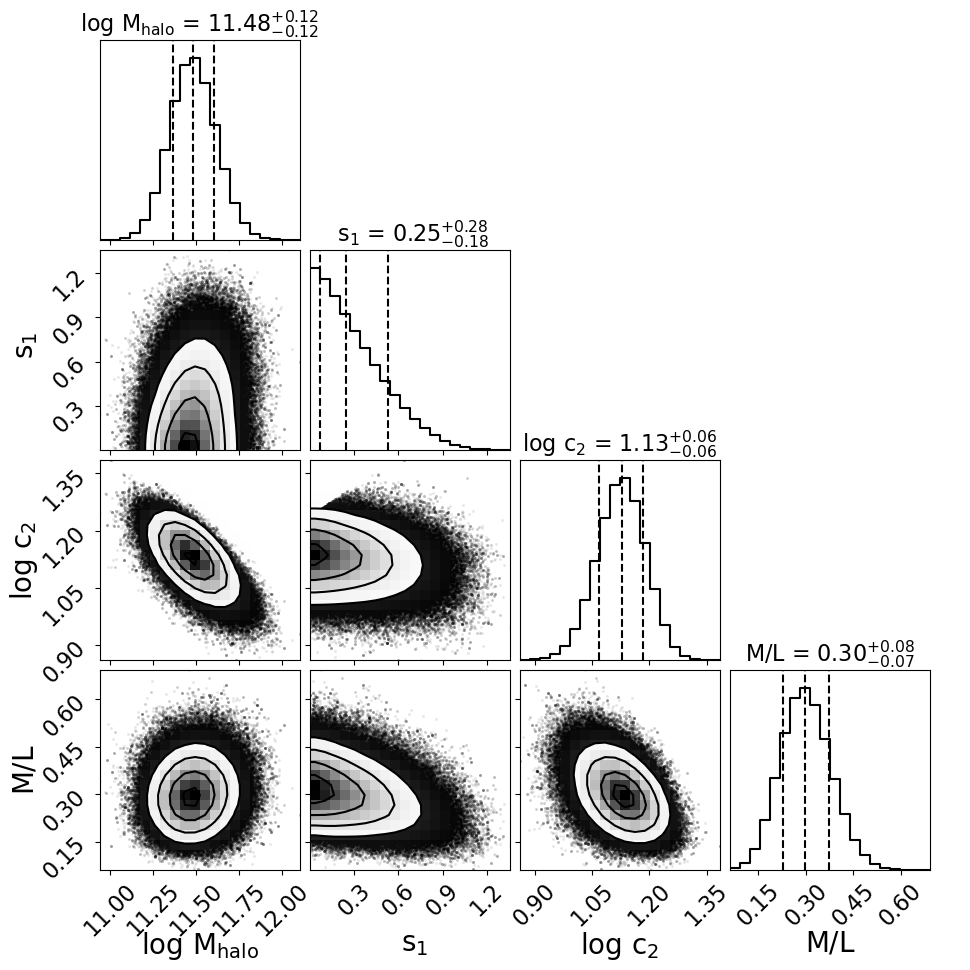}
 \includegraphics[width=7.cm]{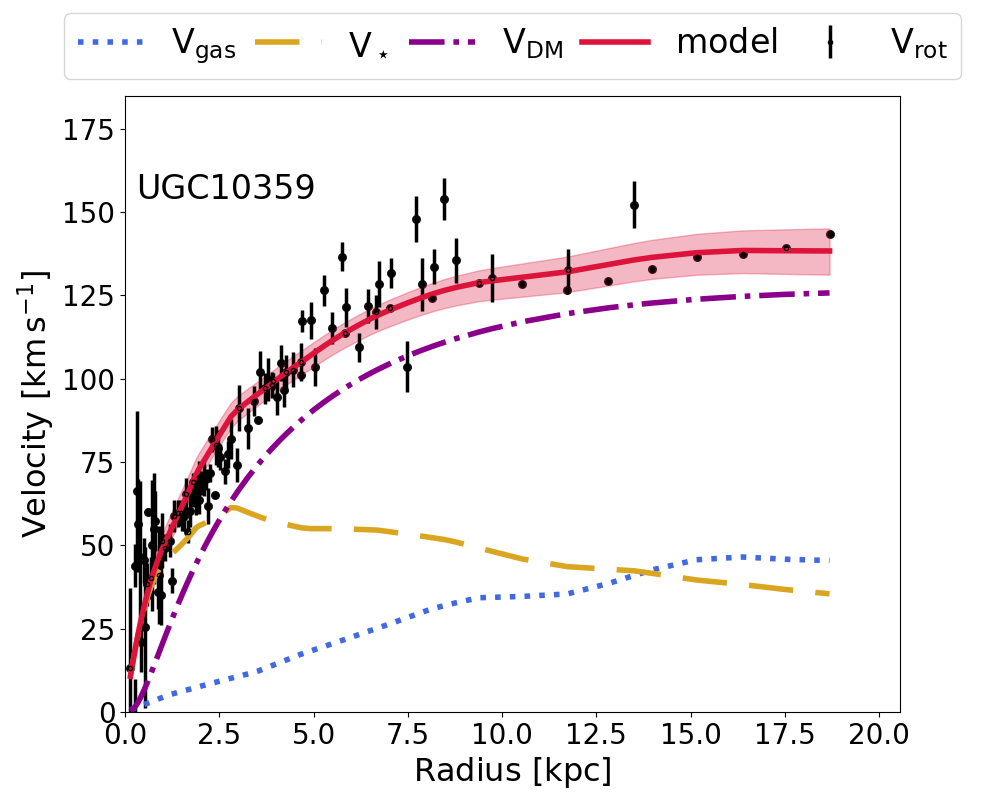}\\
 \includegraphics[width=6.cm]{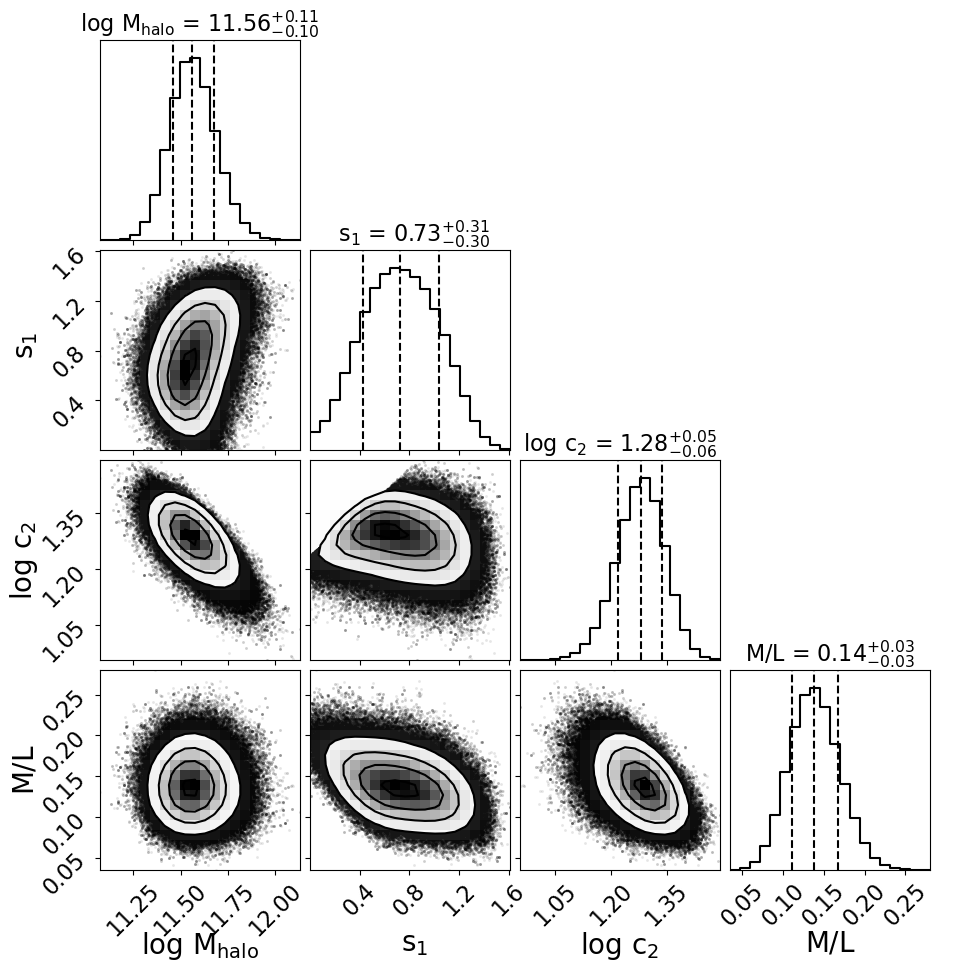}
 \includegraphics[width=7.cm]{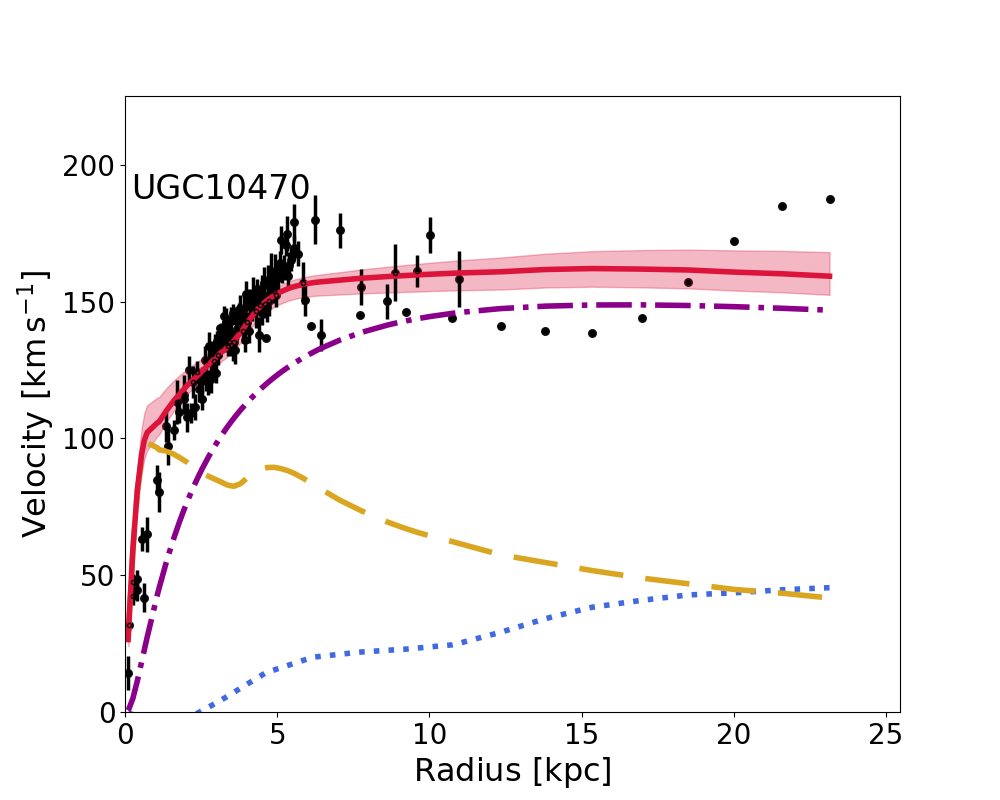}\\
  \includegraphics[width=6.cm]{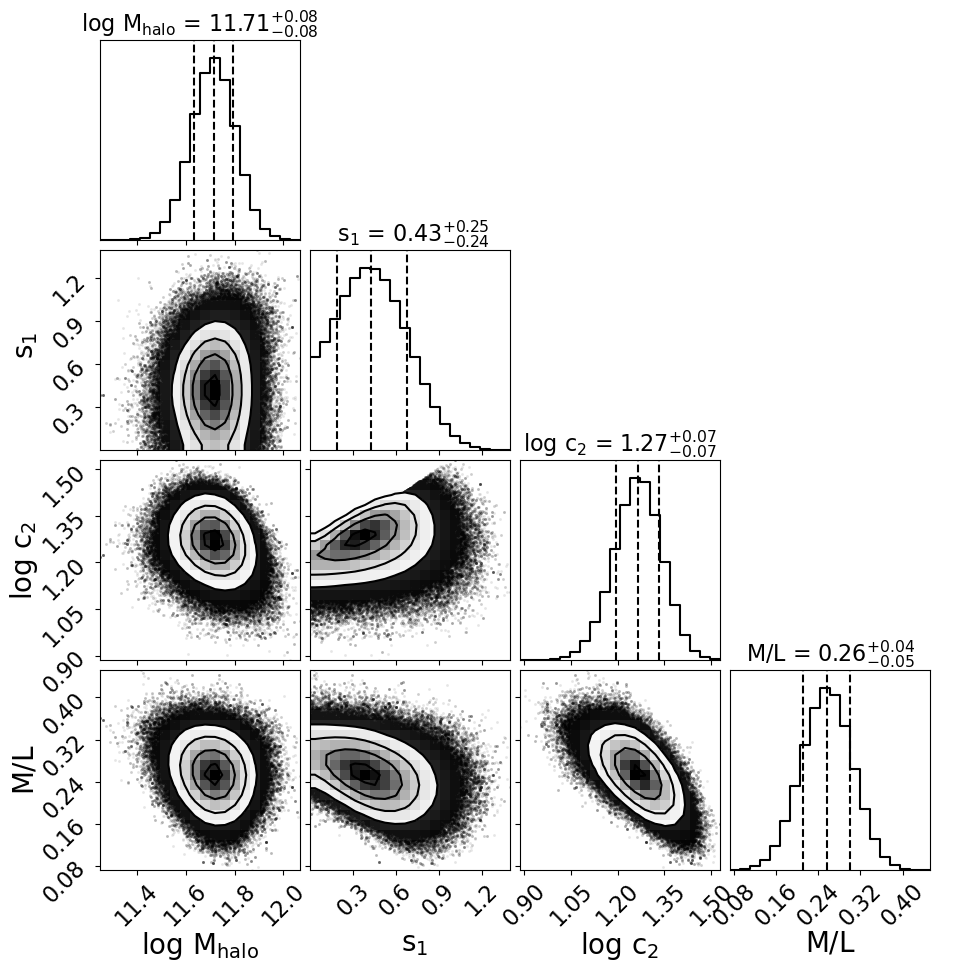}
  \includegraphics[width=7.cm]{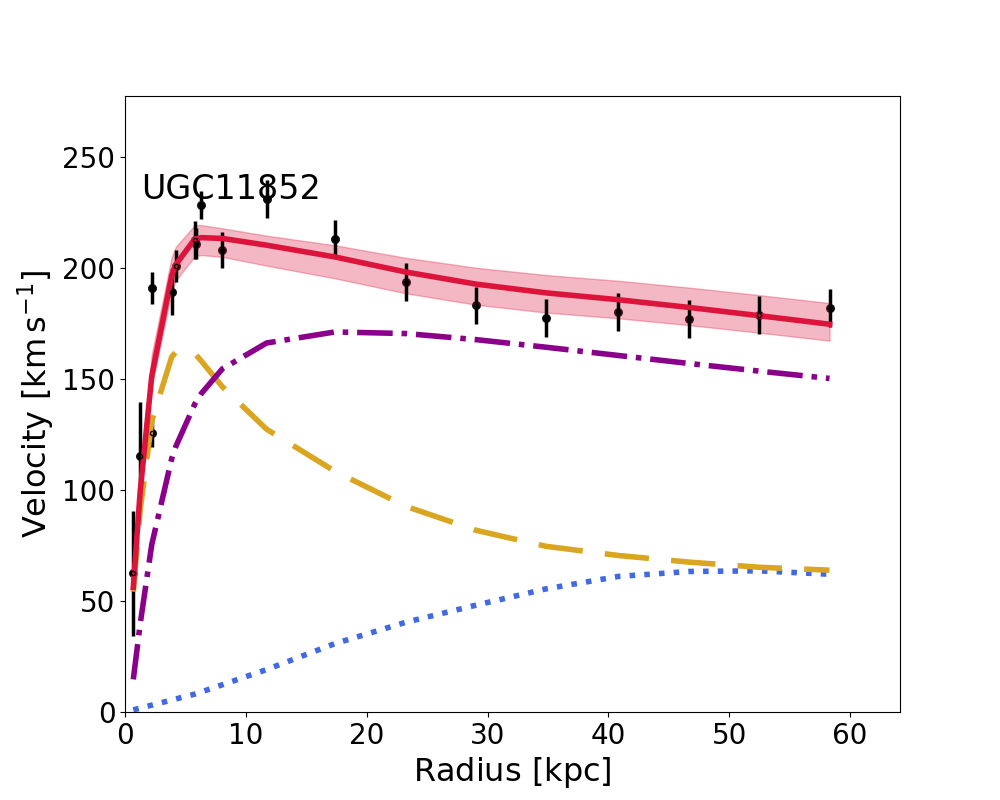}\\
  \includegraphics[width=6.cm]{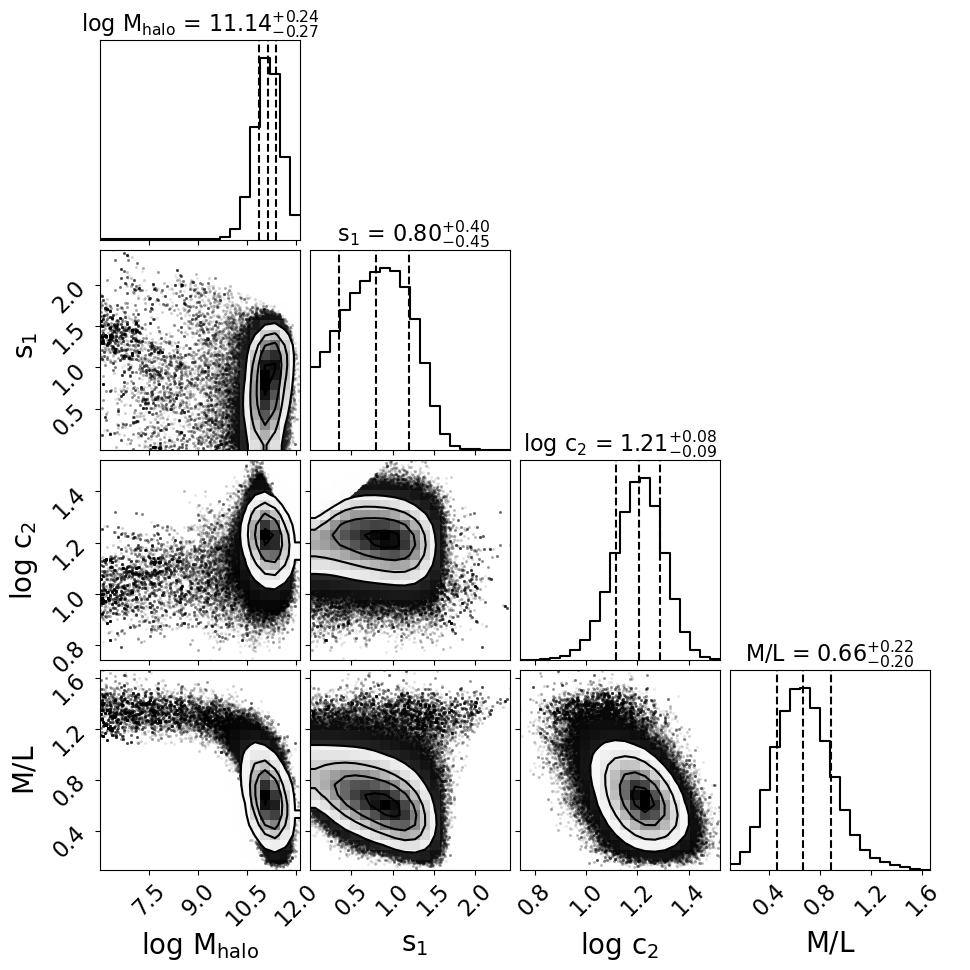}
  \includegraphics[width=7.cm]{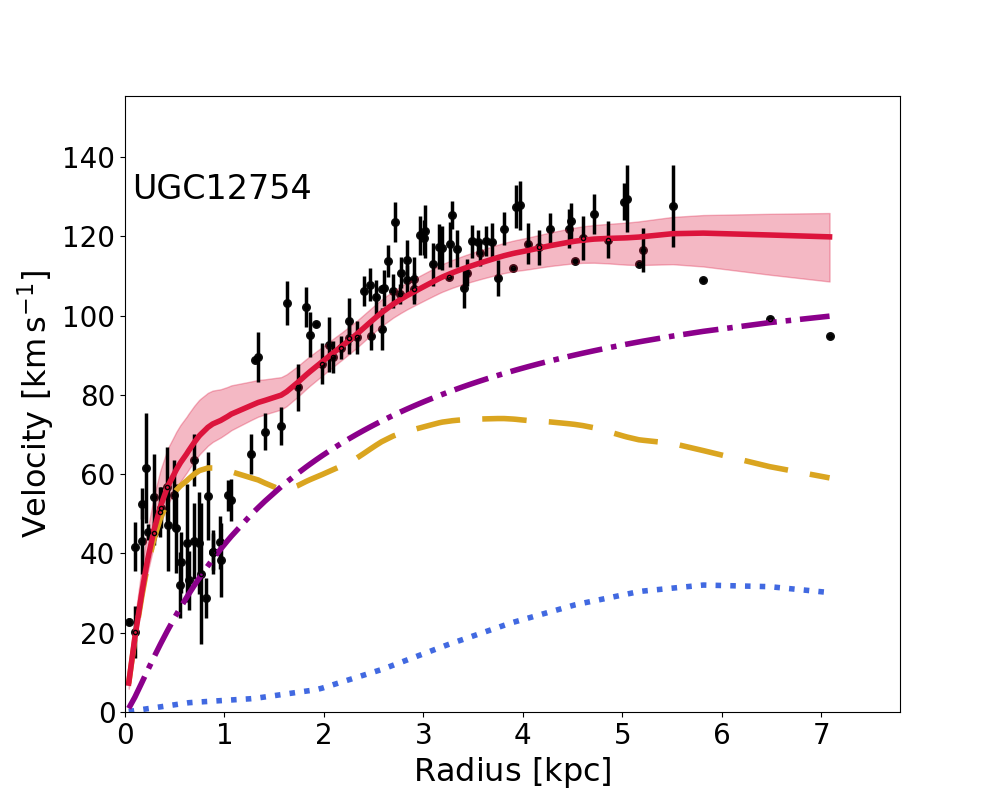}\\
     \caption{Galaxy from GHASP sample. The left panels show the posterior distributions of the model parameters, while the right panels present the corresponding mass models obtained using the DZ parametrisation}
 \label{fig:ghasp massmodel}
    \end{figure*}
  
\addtocounter{figure}{-1}
\begin{figure*}
    \centering
  \includegraphics[width=6.cm]{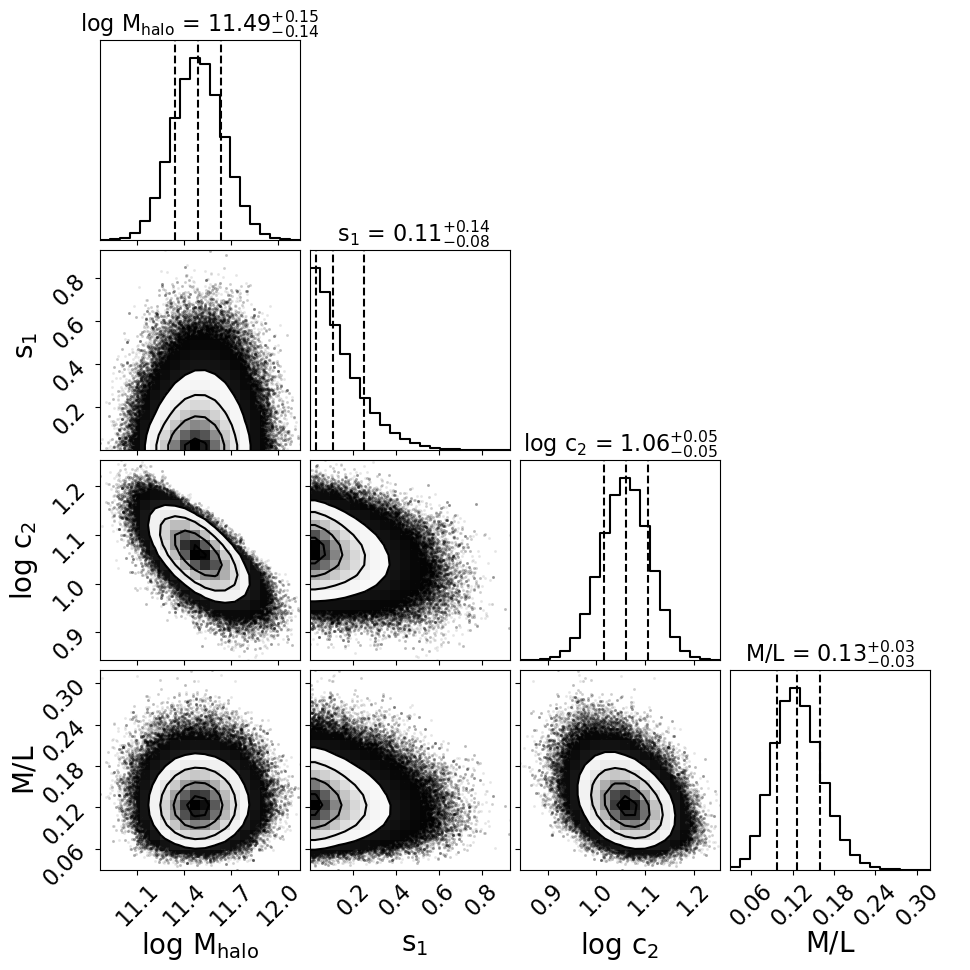}
  \includegraphics[width=7.cm]{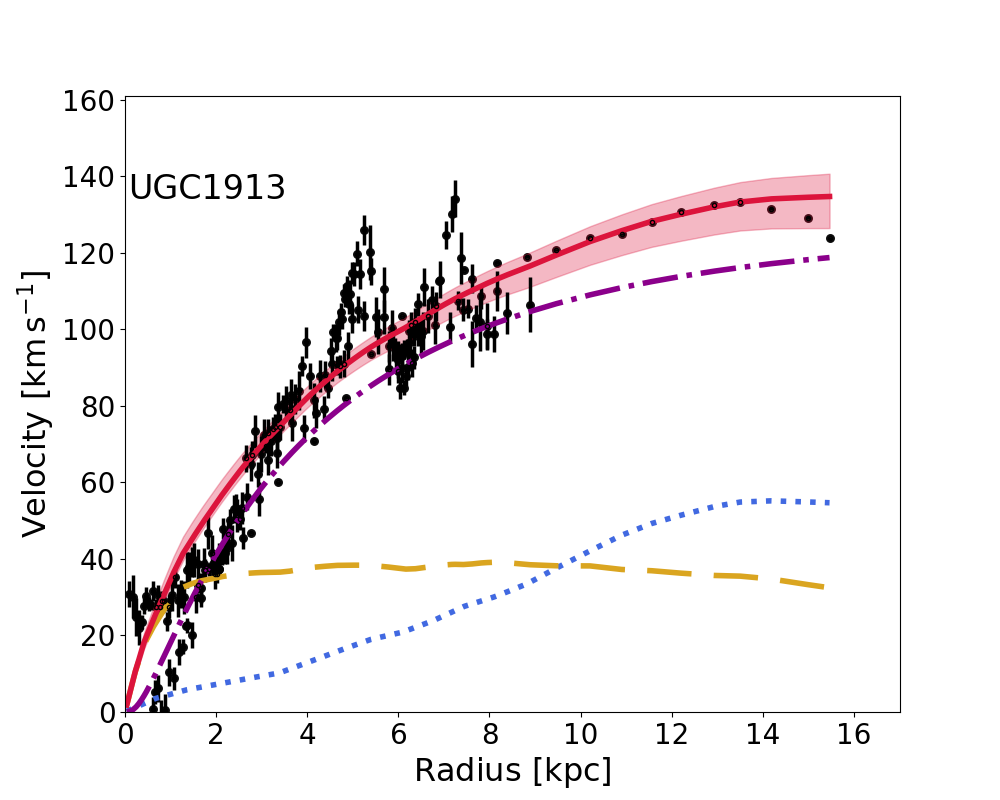}\\
  \includegraphics[width=6.cm]{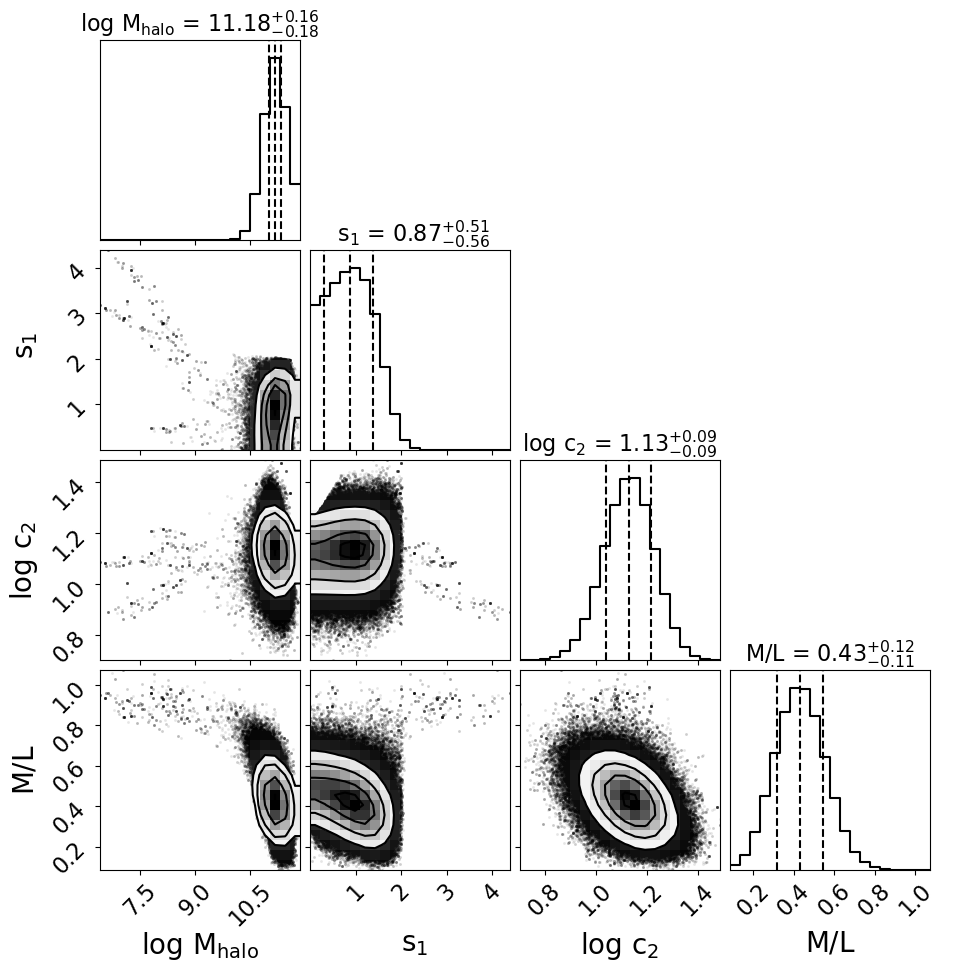}
  \includegraphics[width=7.cm]{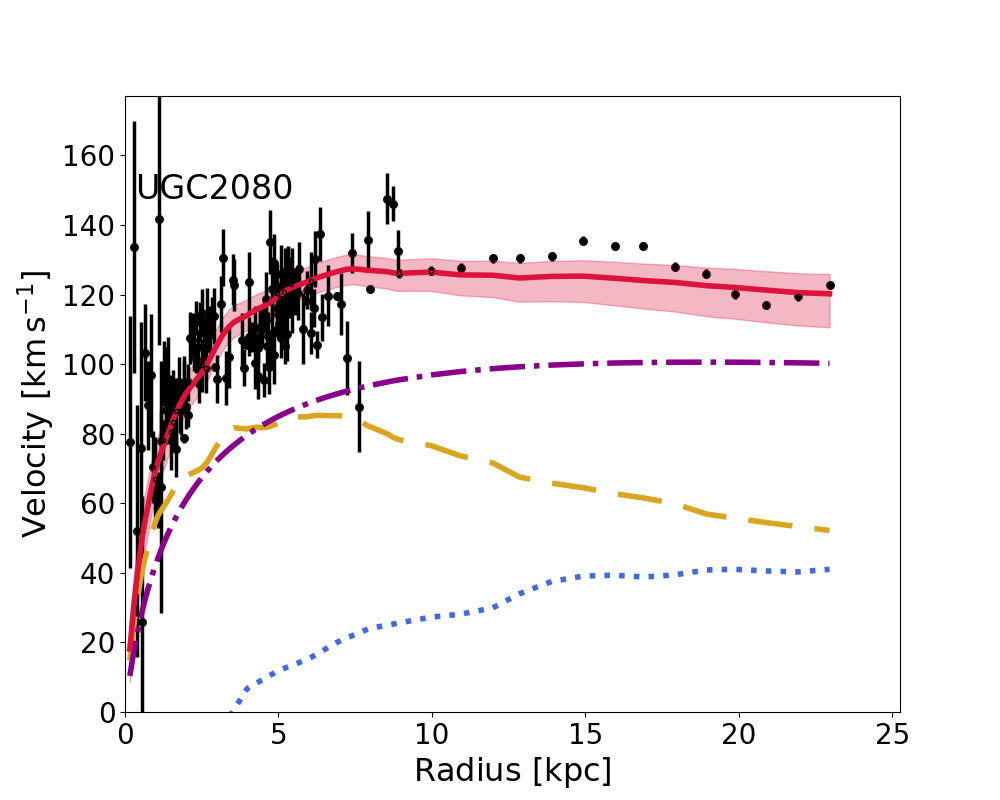}\\
  \includegraphics[width=6.cm]{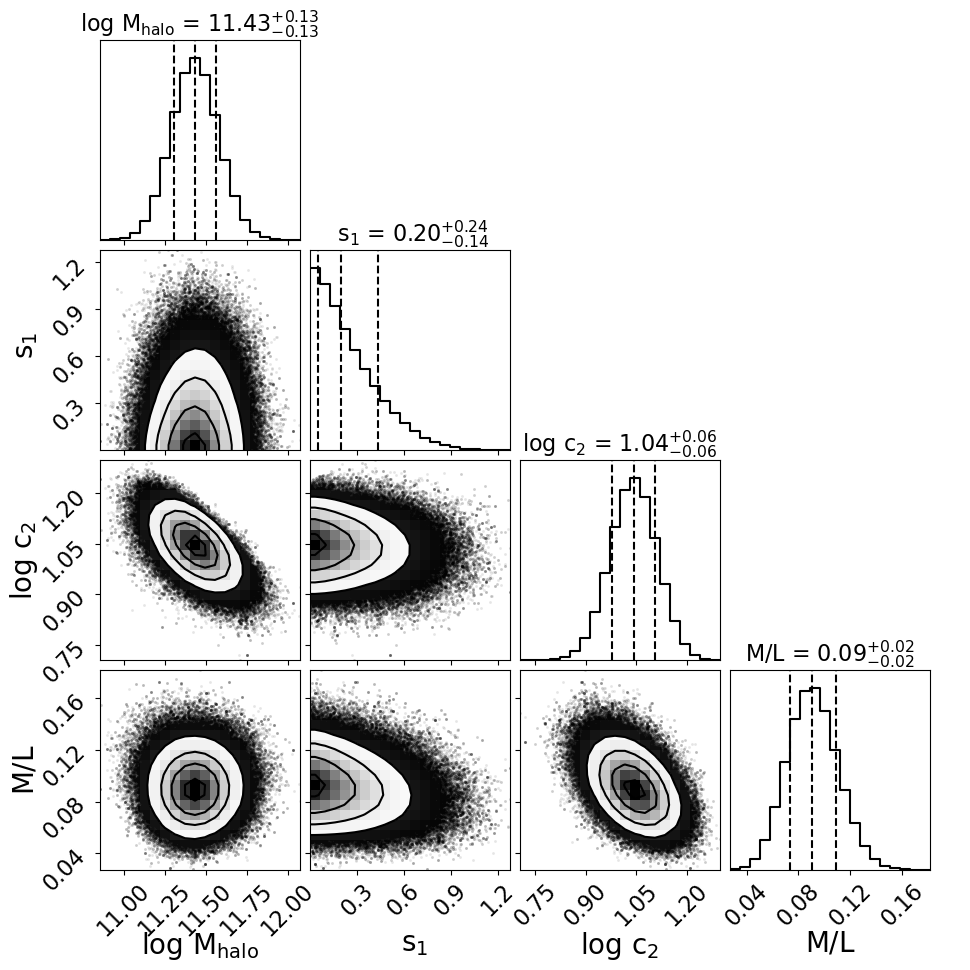}
  \includegraphics[width=7.cm]{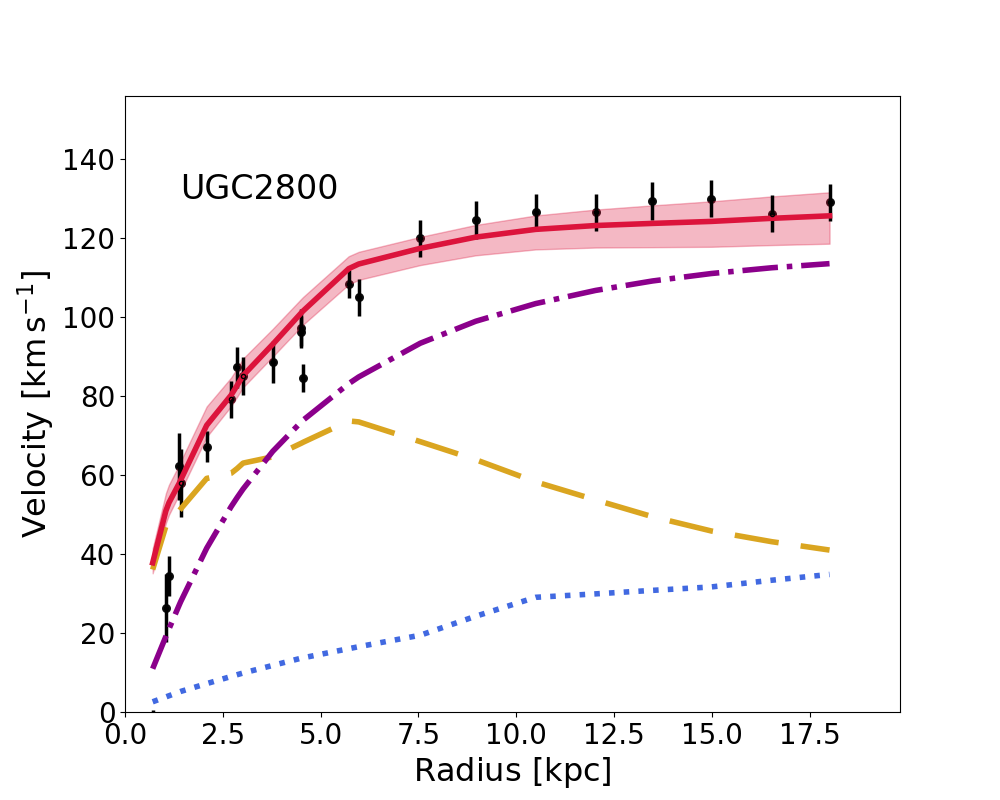}\\
  \includegraphics[width=6.cm]{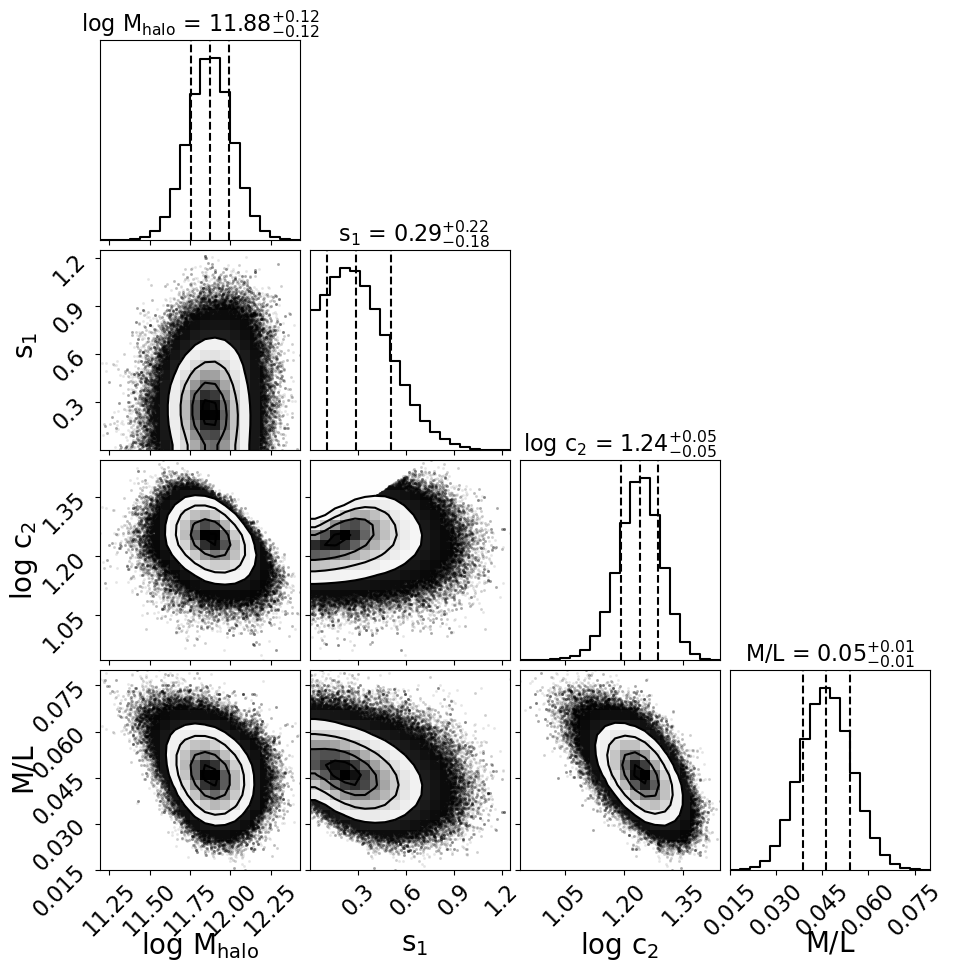}
  \includegraphics[width=7.cm]{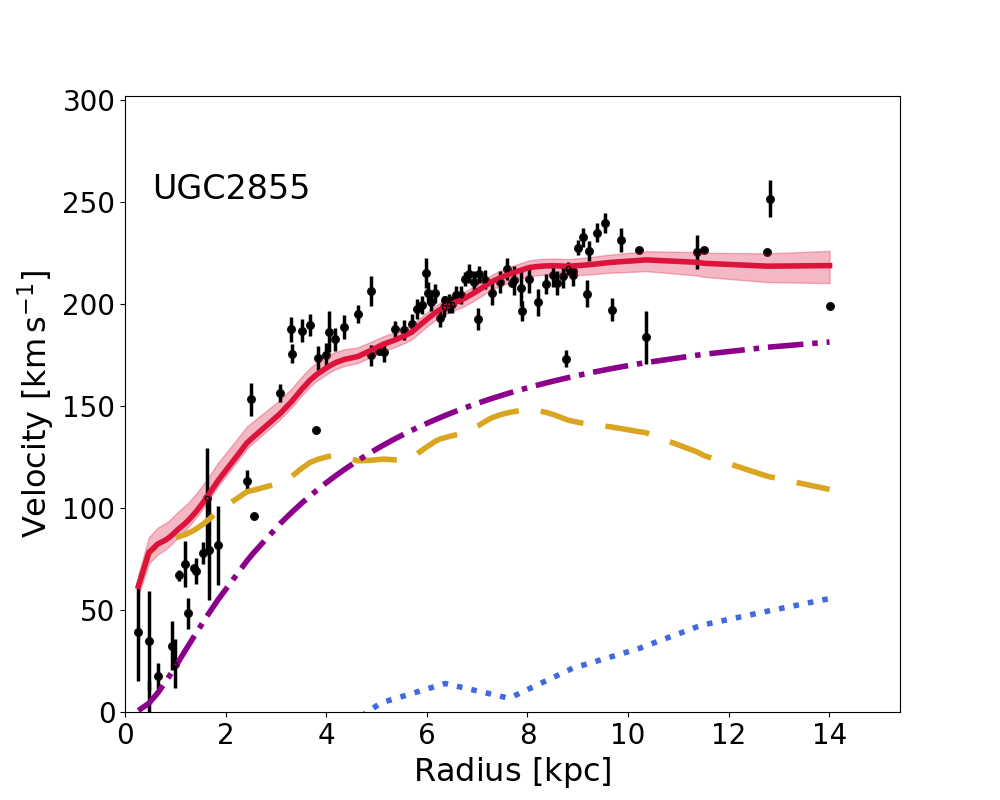}\\
      \caption[]{Continued.}
\end{figure*}
 
\addtocounter{figure}{-1}
\begin{figure*}
    \centering  
  \includegraphics[width=6.cm]{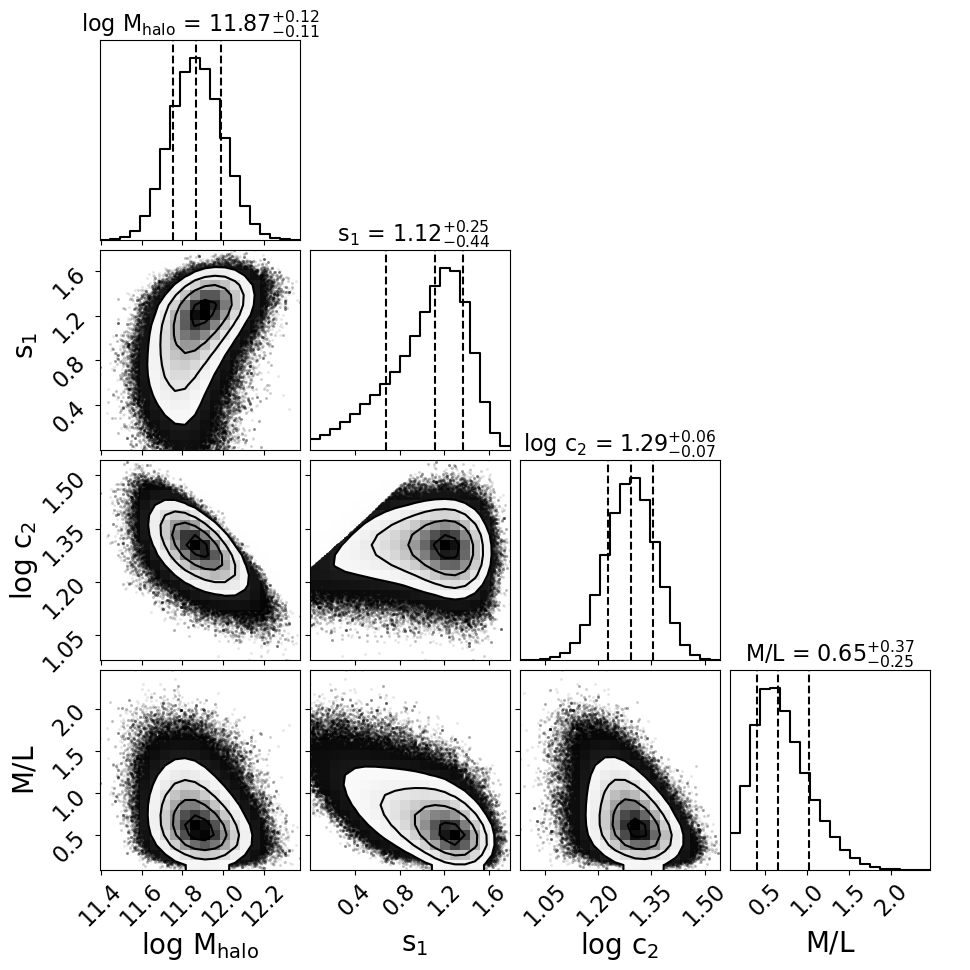}
  \includegraphics[width=7.cm]{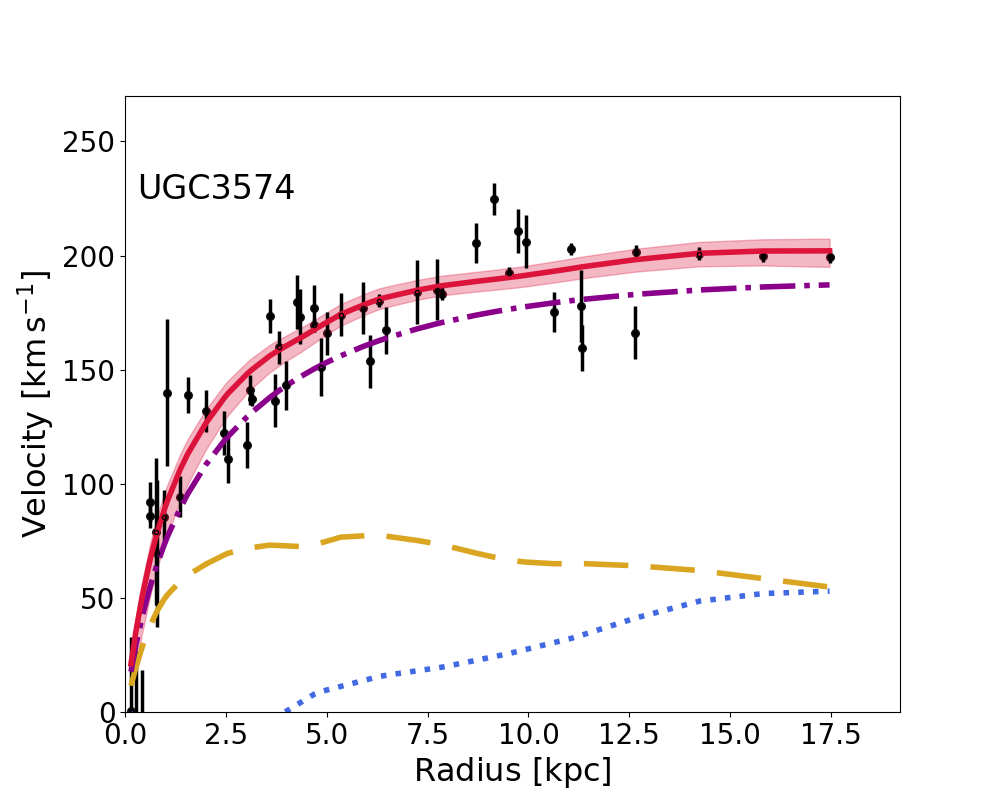}\\
  \includegraphics[width=6.cm]{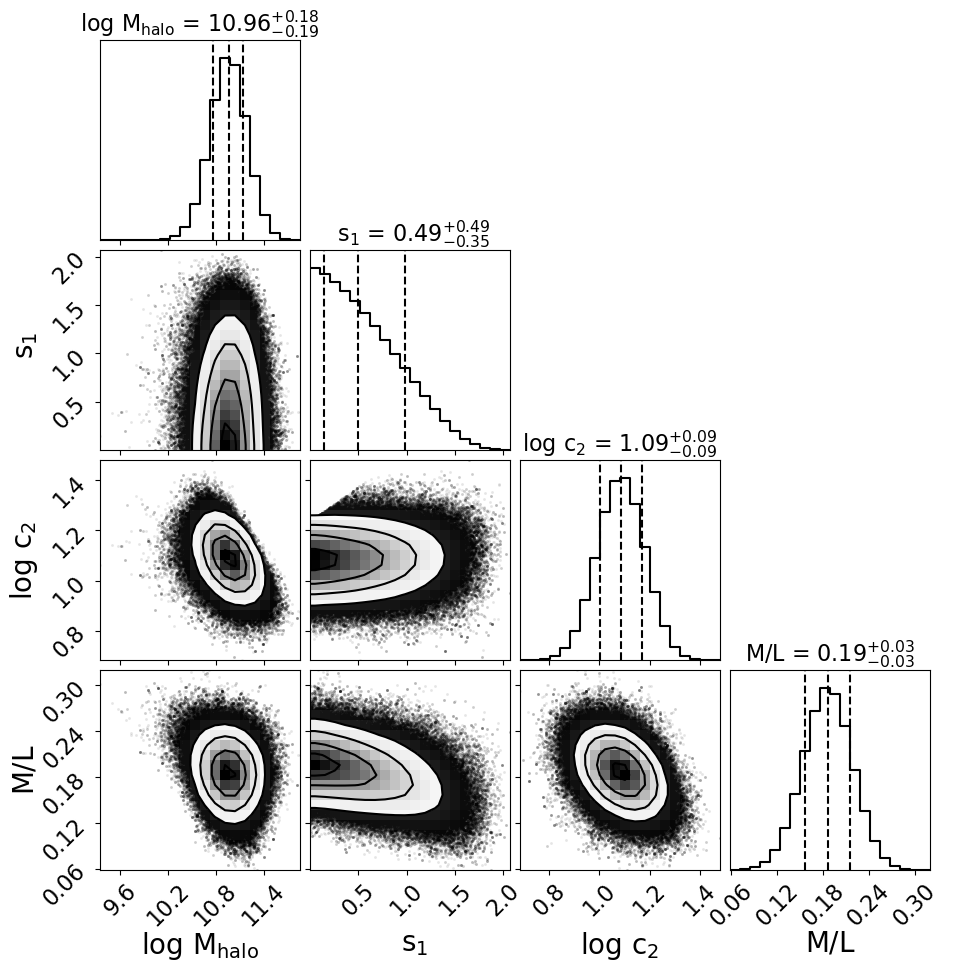}
  \includegraphics[width=7.cm]{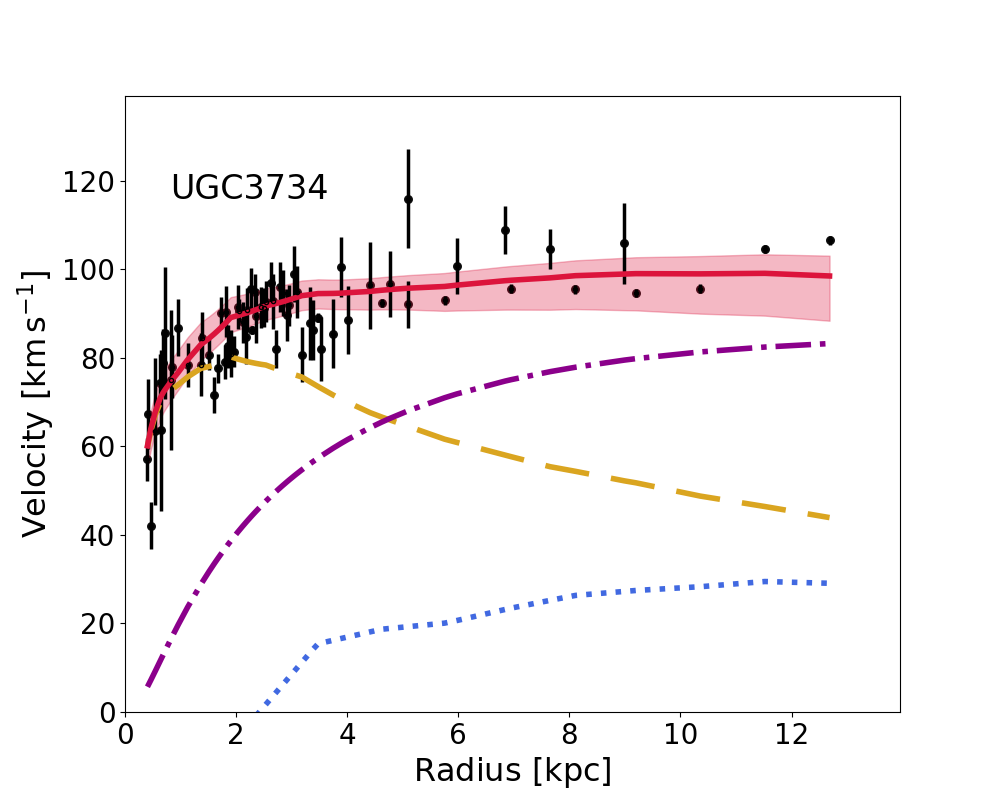}\\
  \includegraphics[width=6.cm]{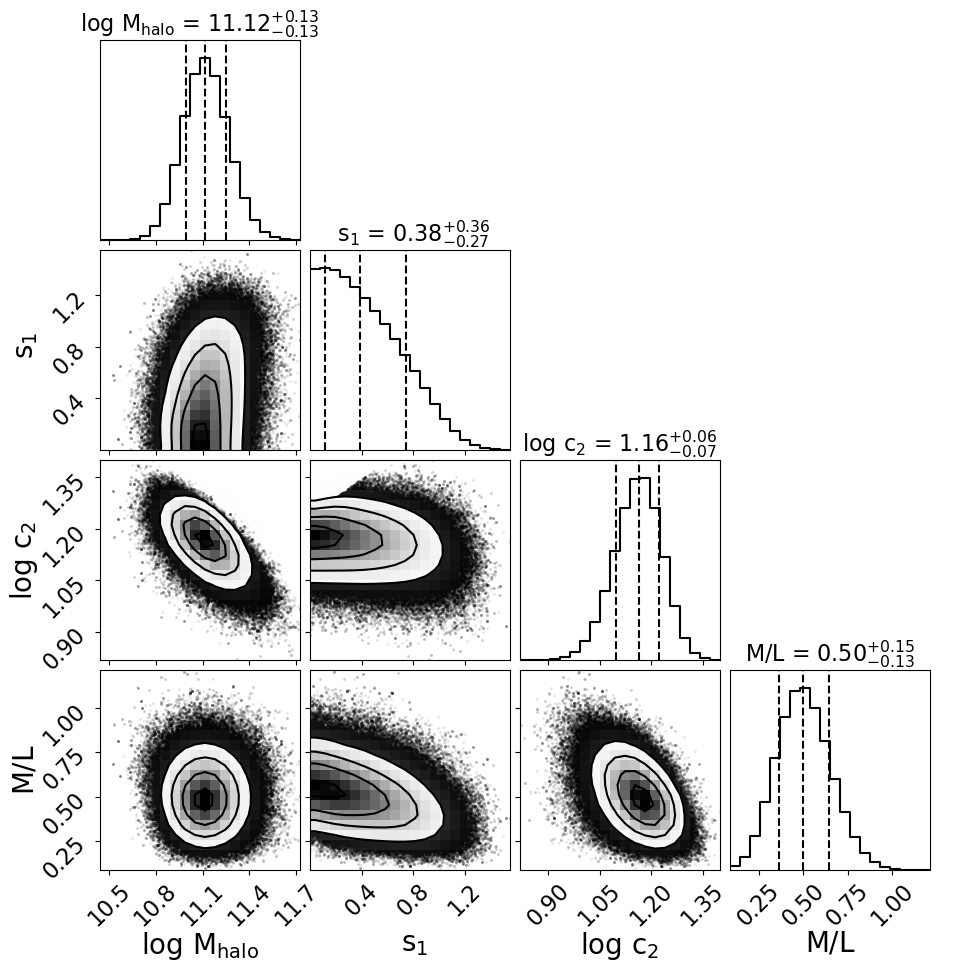}
  \includegraphics[width=7.cm]{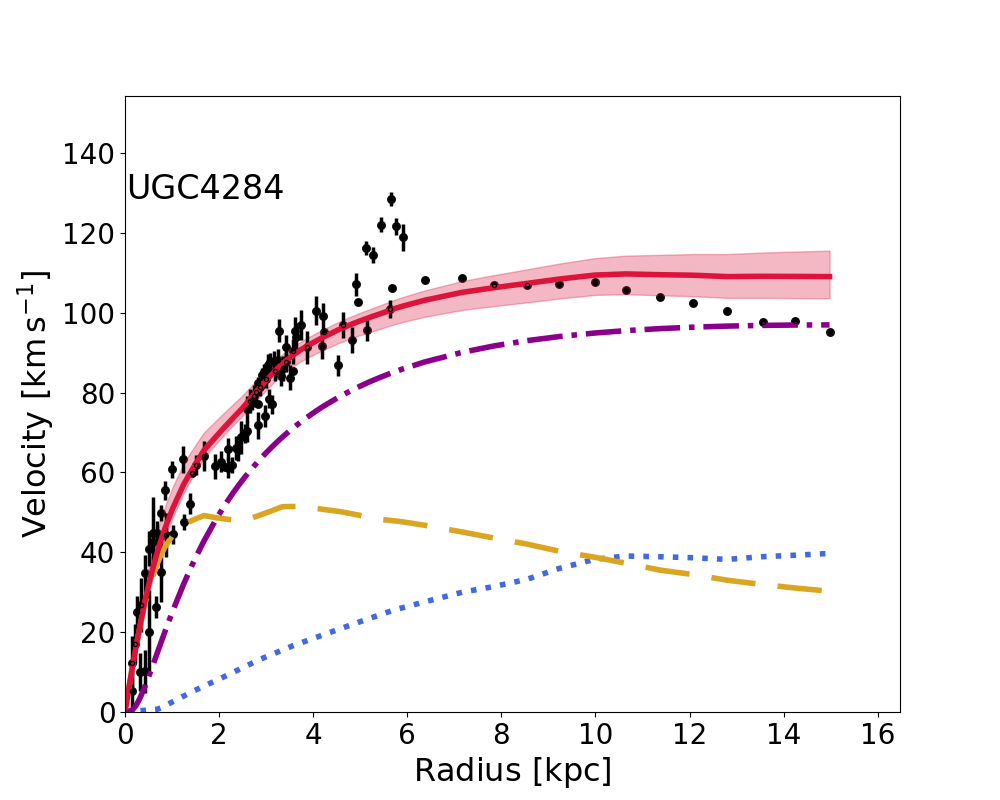}\\
  \includegraphics[width=6.cm]{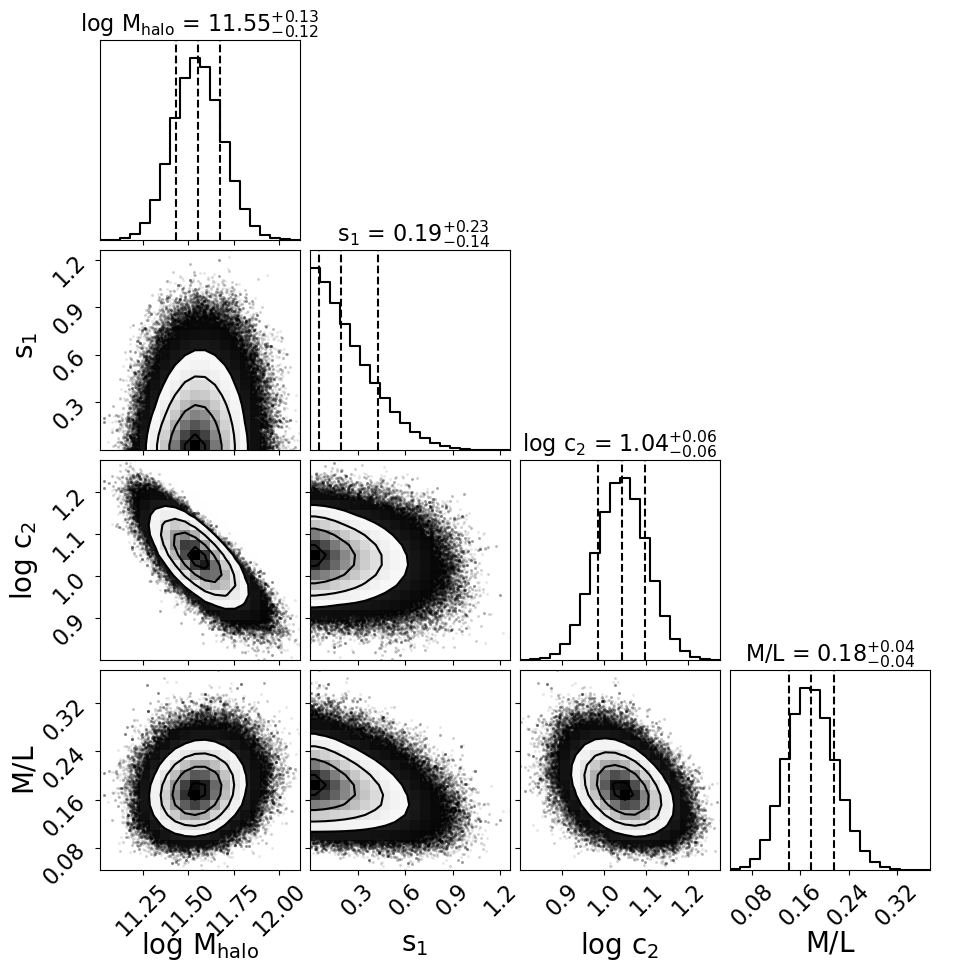}
  \includegraphics[width=7.cm]{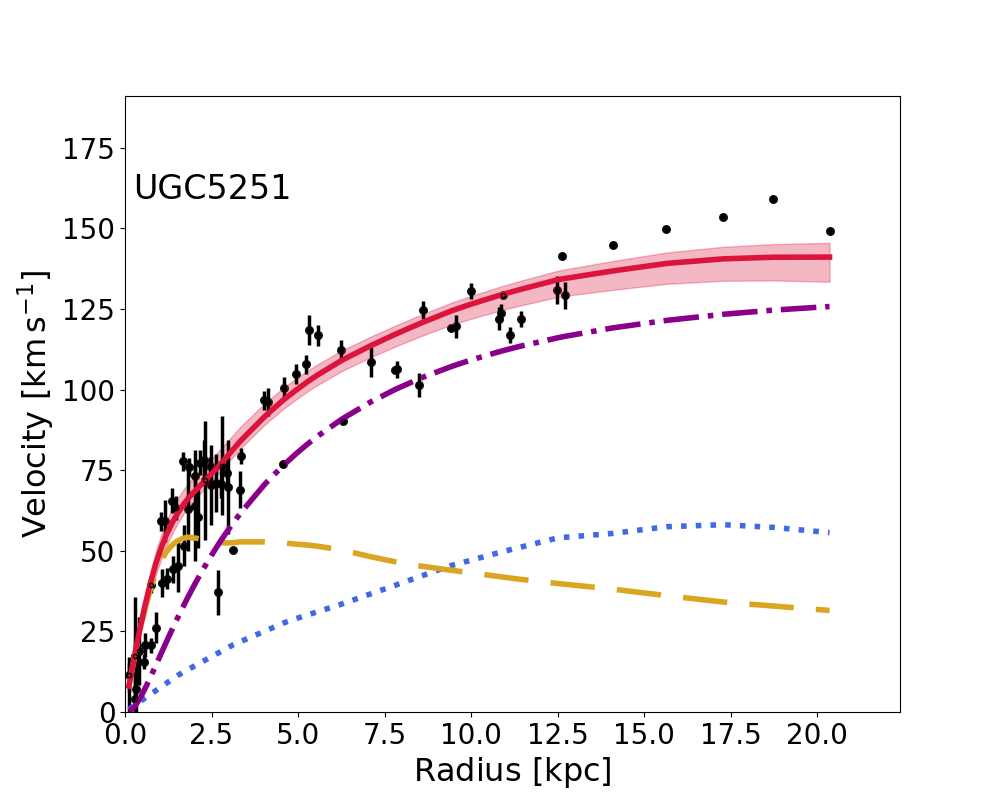}\\
      \caption[]{Continued.}
\end{figure*}
 
\addtocounter{figure}{-1}
\begin{figure*}
    \centering  
  \includegraphics[width=6.cm]{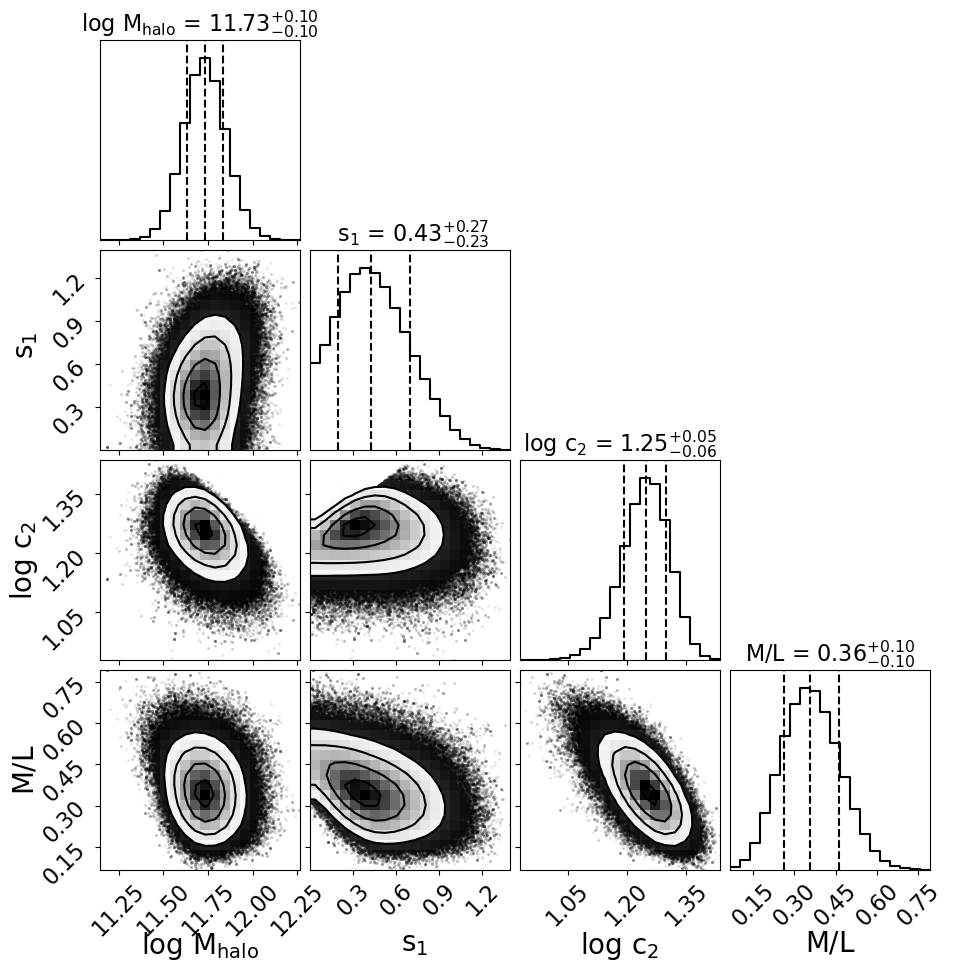}
  \includegraphics[width=7.cm]{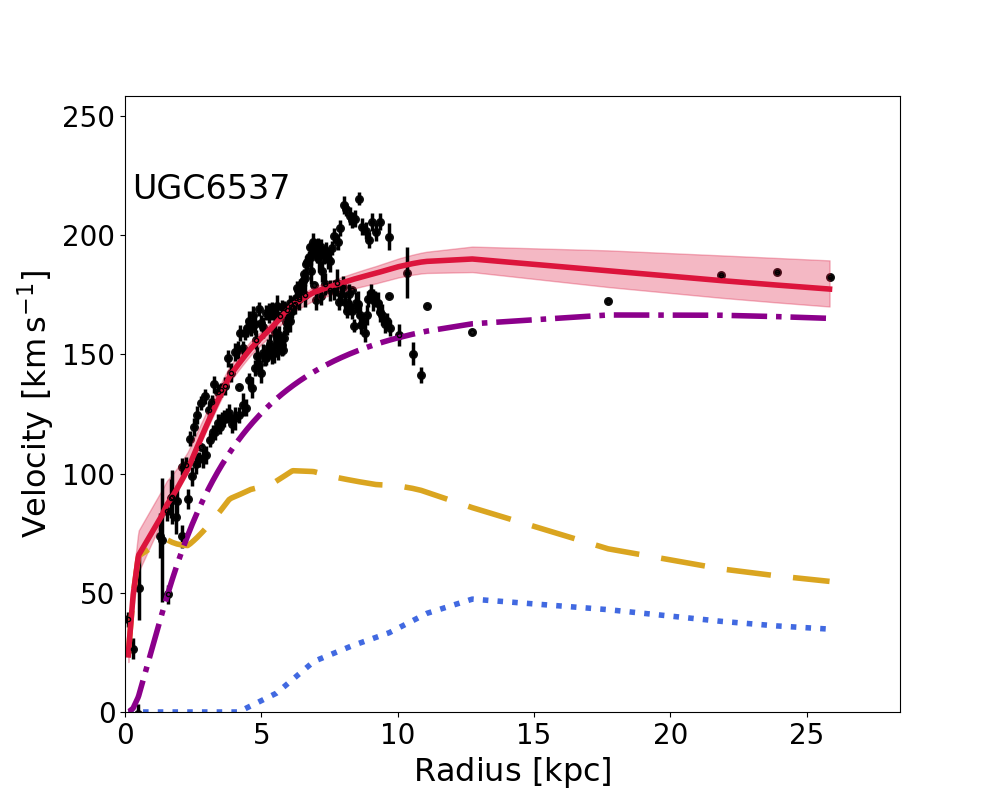}\\
  \includegraphics[width=6.cm]{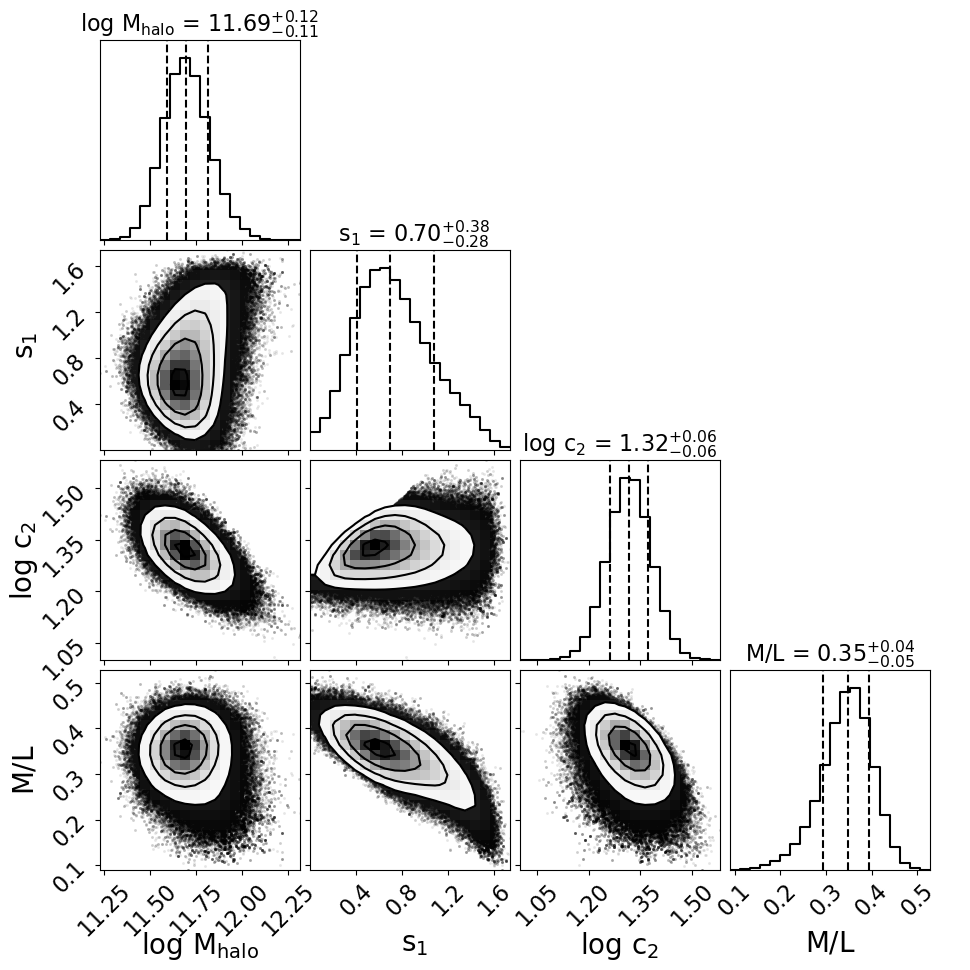}
  \includegraphics[width=7.cm]{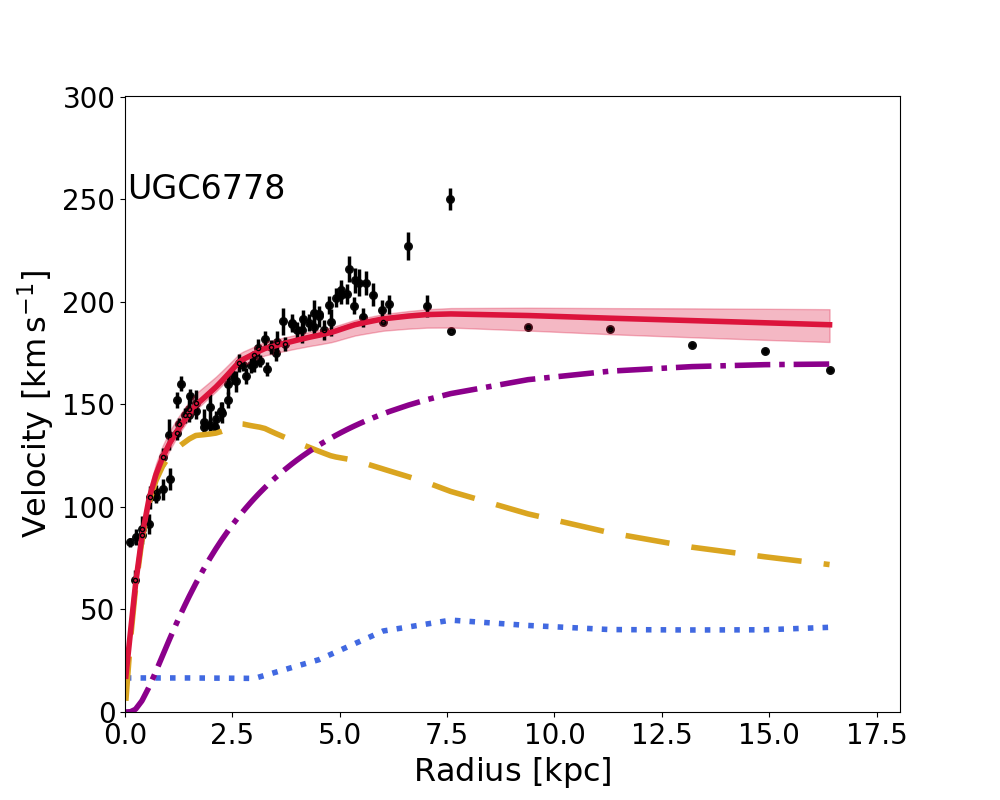}\\
   \includegraphics[width=6.cm]{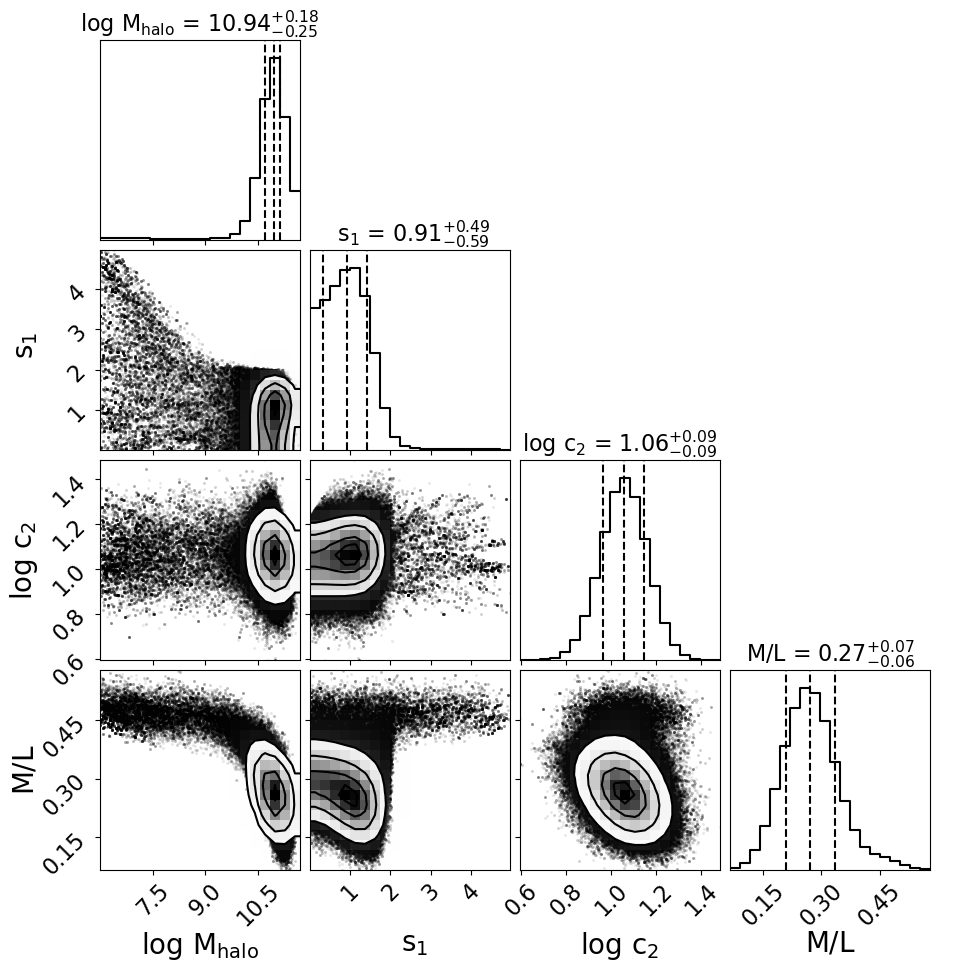}
 \includegraphics[width=7.cm]{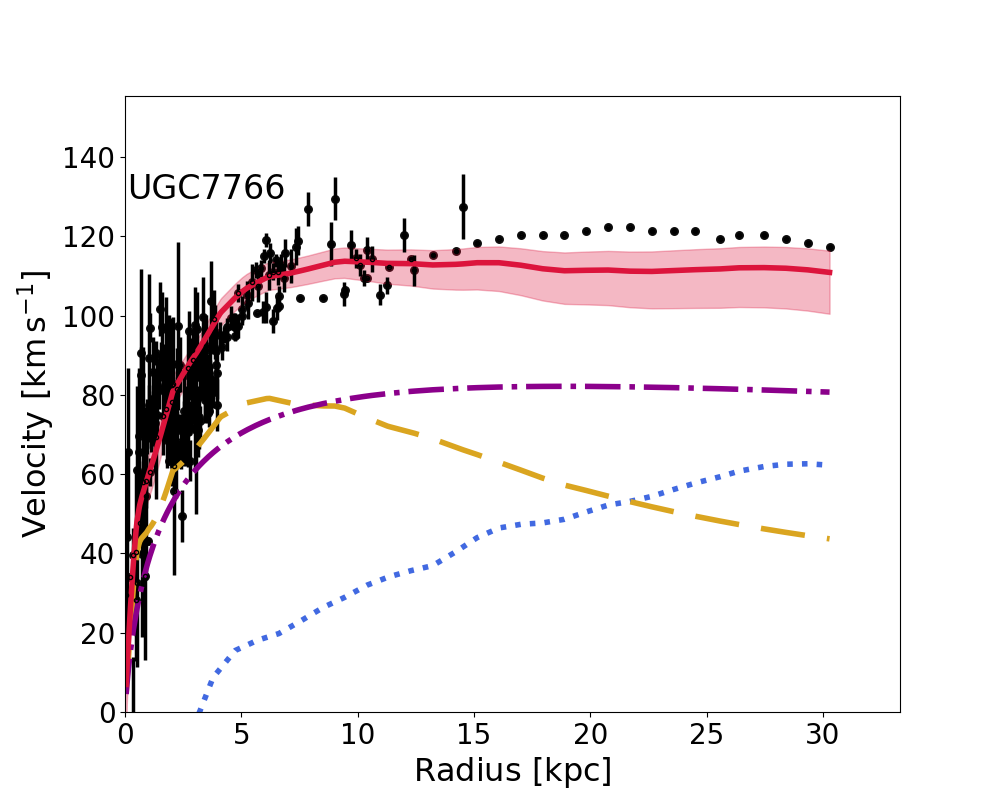}\\
  \includegraphics[width=6.cm]{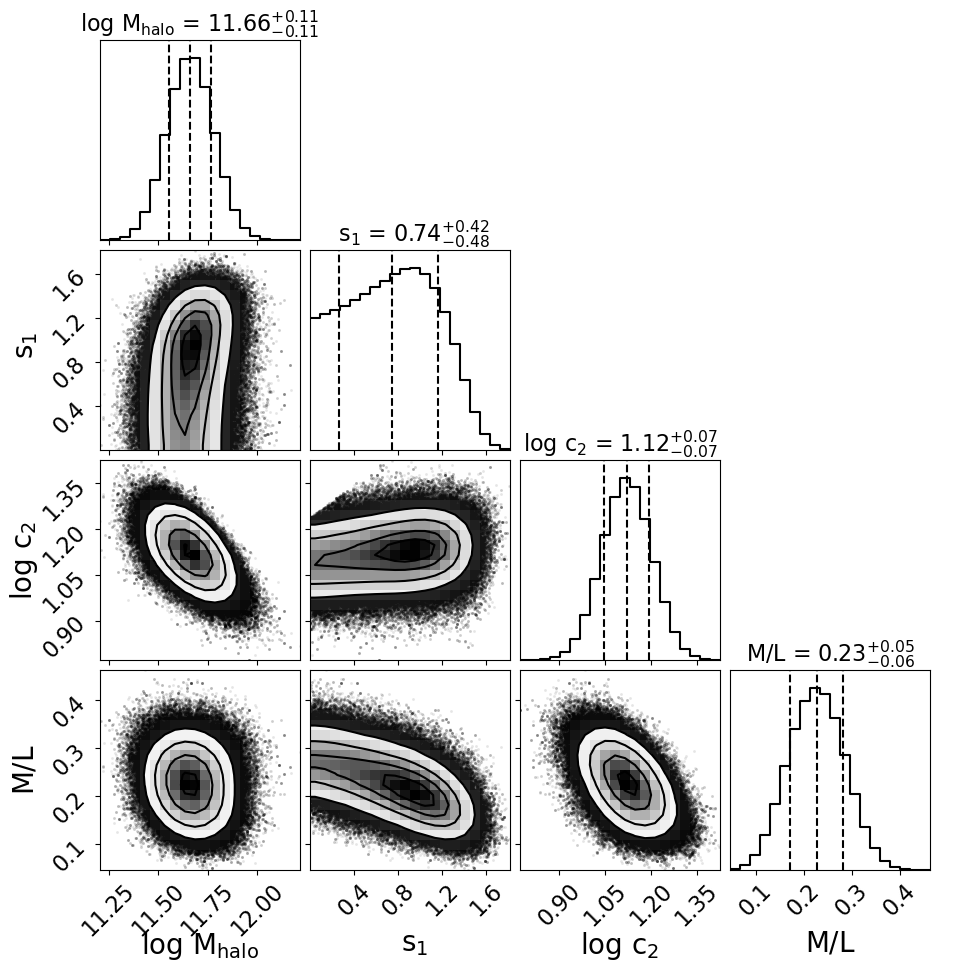}
  \includegraphics[width=7.cm]{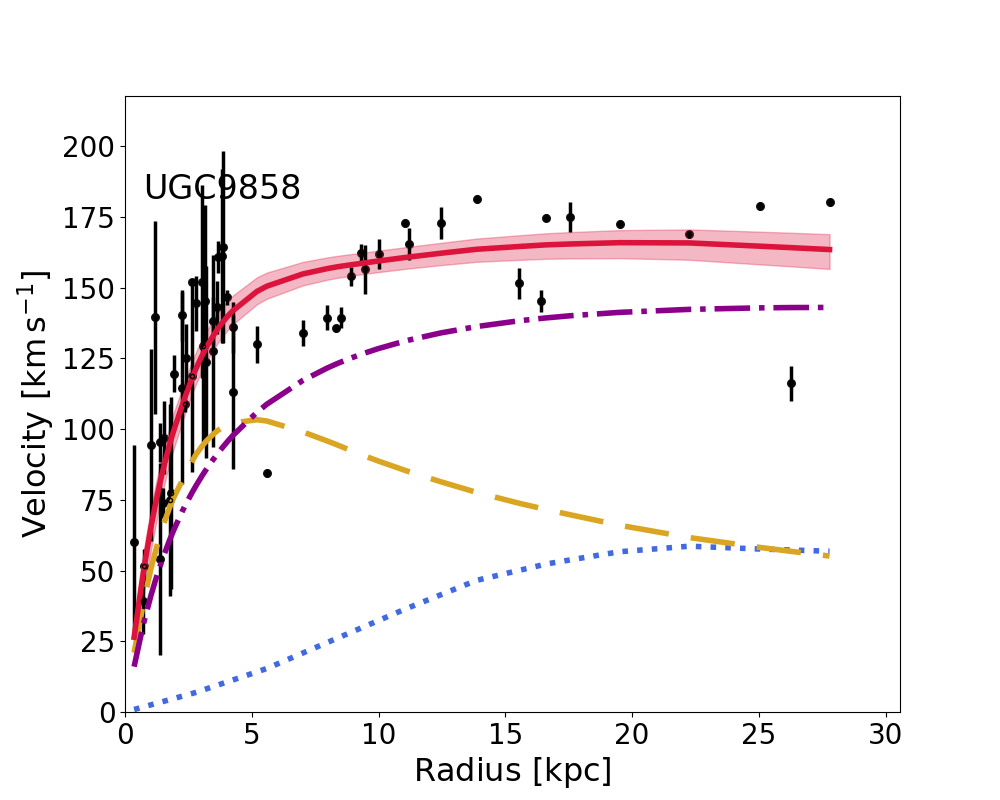}\\
      \caption[]{Continued.}
\end{figure*}

\addtocounter{figure}{-1}
\begin{figure*}
    \centering    
  \includegraphics[width=6.cm]{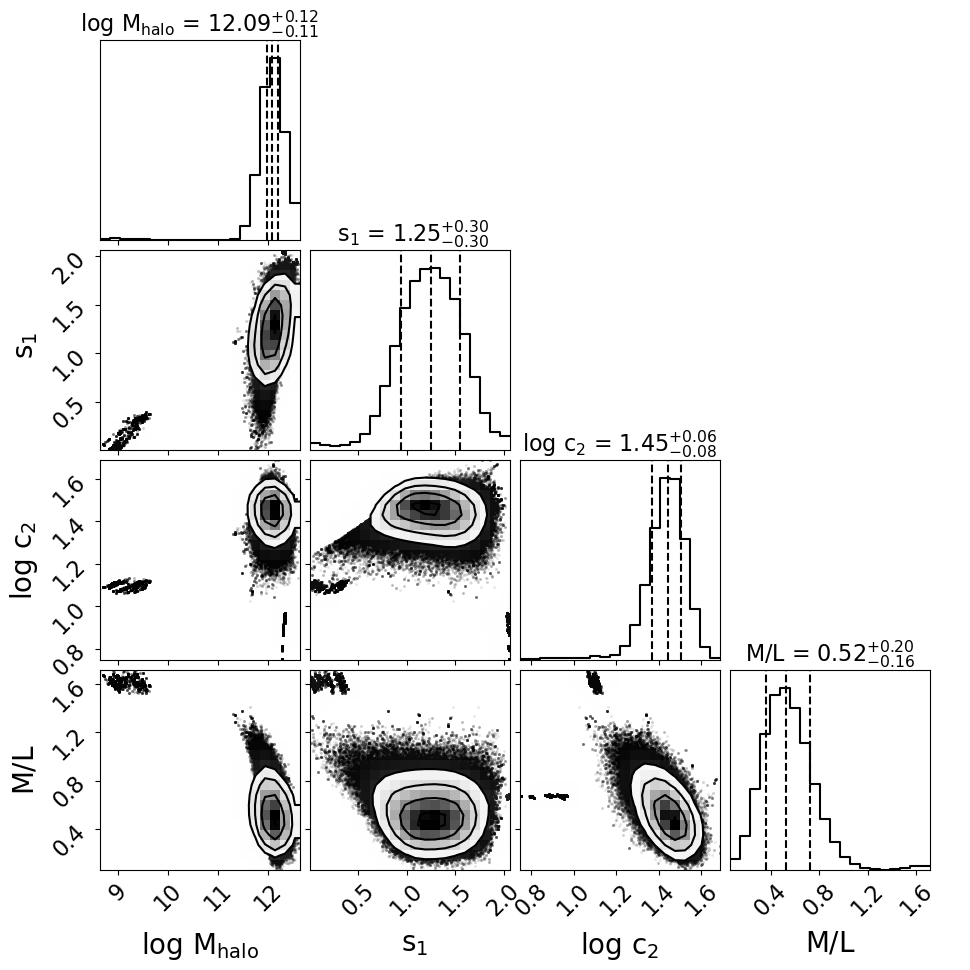}
  \includegraphics[width=7.cm]{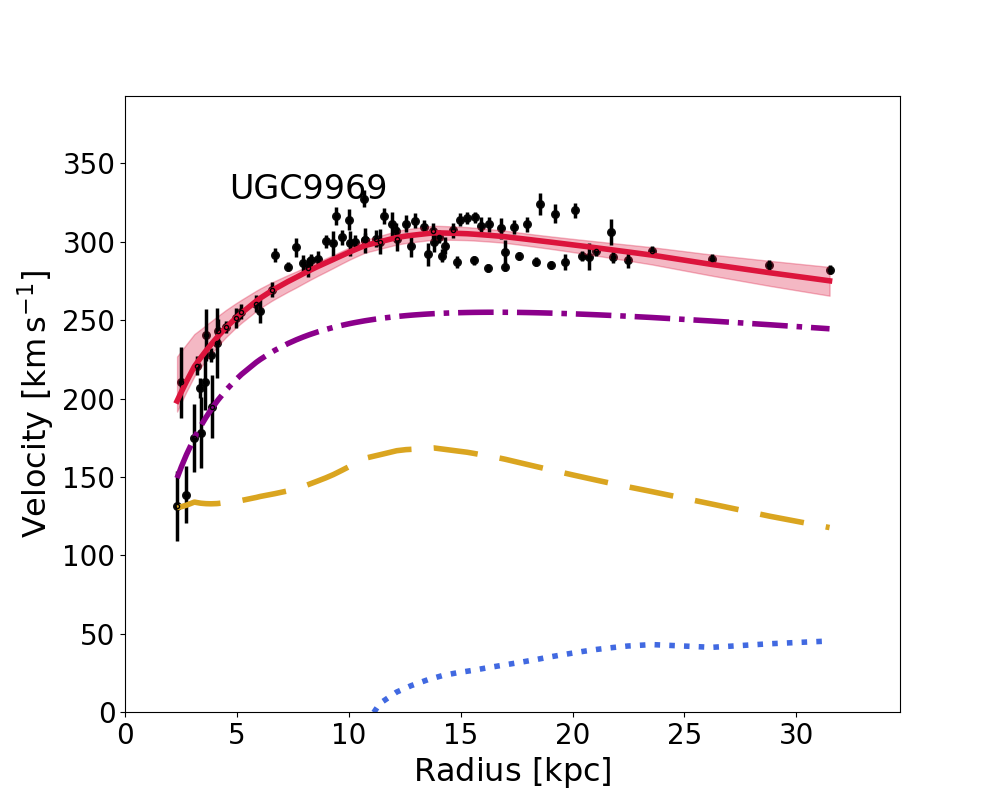}\\
      \caption[]{Continued.}
\end{figure*}

\end{appendix}

\end{document}